\documentclass[leqno]{amsart}

\usepackage[foot]{amsaddr}

\usepackage{color}
\usepackage{geometry}
\usepackage{subfiles}
\usepackage{subcaption}
\usepackage{graphicx}
\usepackage[section]{placeins}
\usepackage[linesnumbered,ruled,vlined]{algorithm2e}
\usepackage{hyperref}
\usepackage{stmaryrd}
\usepackage{amsmath,amssymb,amsfonts,amsthm,textcomp}
\usepackage{mathtools}
\usepackage{tensor}
\usepackage{multicol}
\usepackage{empheq}
\usepackage{tikz,pgfplots}
\usetikzlibrary{calc}
\usepackage{enumitem}

\usepackage{mathrsfs}
\DeclareMathOperator{\spn}{span}

\usepackage{xparse}
\usepackage{fancyhdr}

\usepackage{bigints}

\usepackage[most]{tcolorbox}

\newtcbtheorem{Summary}{\bfseries Box}{enhanced,drop shadow={black!50!white},
  coltitle=black,
  top=0.3in,
  attach boxed title to top left=
  {xshift=1.5em,yshift=-\tcboxedtitleheight/2},
  boxed title style={size=small,colback=black!50!white}
}{summary}

\NewDocumentCommand{\dgal}{sO{}m}{%
  \IfBooleanTF{#1}
    {\dgalext{#3}}
    {\dgalx[#2]{#3}}%
}

\NewDocumentCommand{\dgalext}{m}{%
  \sbox0{%
    \mathsurround=0pt 
    $\left\{\vphantom{#1}\right.\kern-\nulldelimiterspace$%
  }%
  \sbox2{\{}%
  \ifdim\ht0=\ht2
    \{\kern-.625\wd2 \{#1\}\kern-.625\wd2 \}%
  \else
    \left\{\kern-.7\wd0\left\{#1\right\}\kern-.7\wd0\right\}%
  \fi
}

\NewDocumentCommand{\dgalx}{om}{%
  \sbox0{\mathsurround=0pt$#1\{$}%
  \sbox2{\{}%
  \ifdim\ht0=\ht2
    \{\kern-.625\wd2 \{#2\}\kern-.625\wd2 \}%
  \else
    \mathopen{#1\{\kern-.7\wd0 #1\{}
    #2
    \mathclose{#1\}\kern-.7\wd0 #1\}}
  \fi
}

\graphicspath{
              {./00_Figuras_formulation/}
              {./06_Examples/}
              {./06_Examples/2D_MultiSeed/}
              {./06_Examples/2D_Experimental/}
              {./06_Examples/SingleSeed_dendrite/}
              {./06_Examples/MultipleSeed_dendrite/}
             }

\usepackage{mathtools}
\emergencystretch=\maxdimen
\newfont{\tenbfsl}{cmbxti9 scaled 1200}
\newfont{\tenbbb}{msbm10}
\newfont{\svnbbb}{msbm8}

\theoremstyle{remark}

\theoremstyle{definition}

\newcounter{syn}[section] 
\renewcommand{\thesyn}{\arabic{section}.\arabic{syn}}

\makeatletter
\def\threevdots{\mskip+4mu\vbox{\baselineskip2.25\p@ \lineskiplimit\z@
  \kern4.9\p@\hbox{.}\hbox{.}\hbox{.}}\mskip+3.8mu}
\makeatother

\begin{document}

\title[Dendrite formation in Li-metal batteries]{Dendrite formation in rechargeable lithium-metal batteries: Phase-field modeling using open-source finite element library}

\author{Marcos E. Arguello$^{\square,\ddagger,\star}$}
\email{m.arguello@postgrad.curtin.edu.au, m.arguello.19@abdn.ac.uk }
\author{Nicol\'as A. Labanda$^{\diamondsuit,\spadesuit}$}
\email{nlabanda@facet.unt.edu.ar}
\author{Victor M. Calo$^{\diamondsuit}$}
\email{victor.calo@curtin.edu.au}
\author{Monica Gumulya$^{\blacktriangledown}$}
\email{m.gumulya@curtin.edu.au}
\author{Ranjeet Utikar$^{\square}$}
\email{r.utikar@curtin.edu.au}
\author{Jos Derksen$^{\ddagger}$}
\email{jderksen@abdn.ac.uk}
\address{$^{\square}$ WA School of Mines, Mineral, Energy and Chemical Engineering, Curtin University, P.O. BOX U1987, Perth, WA 6845, Australia.}
\address{$^{\ddagger}$ School of Engineering, University of Aberdeen, Elphinstone Road, AB24 3UE Aberdeen, United Kingdom.}
\address{$^{\diamondsuit}$ School of Electrical Engineering, Computing and Mathematical Sciences, Curtin University, P.O. Box U1987, Perth, WA 6845, Australia}
\address{$^{\spadesuit}$ SRK Consulting, West Perth, Western Australia, Australia.}
\address{$^{\blacktriangledown}$ Occupation, Environment and Safety, School of Population Health, Curtin University, P.O. Box U1987, Perth, WA 6845, Australia}
\address[$^{\star}$]{Corresponding author: Marcos E. Arguello, m.arguello@postgrad.curtin.edu.au}

\date{\today}

\begin{abstract}
\noindent
We describe a phase-field model for the electrodeposition process that forms dendrites within metal-anode batteries. We derive the free energy functional model, arriving at a system of partial differential equations that describe the evolution of a phase field, the lithium-ion concentration, and an electric potential. We formulate, discretize, and solve the set of partial differential equations describing the coupled electrochemical interactions during a battery charge cycle using an open-source finite element library. The open-source library allows us to use parallel solvers and time-marching adaptivity. We describe two- and three-dimensional simulations; these simulations agree with experimentally-observed dendrite growth rates and morphologies reported in the literature.

\end{abstract}

\maketitle


\tableofcontents


\section{Introduction}
\label{section:Introduction}

Conventional lithium-ion batteries (LIB) that use intercalated graphite electrodes have dominated the rechargeable battery market over the past three decades; nevertheless, this battery technology is reaching its theoretical limit of 250~Wh/kg~\cite{ C3EE40795K, Fu2017}. Therefore, our society has a strong need for new chemistries and designs to achieve commercial implementation of ultra-high energy density batteries of around 500~Wh/kg~\cite{ Yanguang2017}. 

Currently, metallic lithium (Li) is the most prominent negative electrode material due to its combined high theoretical energy density (3860~mAh/g) and low reduction potential (-3.04~V vs. standard hydrogen electrode). However, despite recent successful operation reports for lithium metal batteries (LMBs), all reported short life cycles due to uneven and unstable lithium deposition leading to dendrite formation~\cite{ li2014review, LIU2018833}. Therefore, developing a stable rechargeable lithium metal anode is crucial to attaining higher energy density rechargeable technologies, such as Li-air, Li-S, and Li-flow batteries~\cite{ BAI20182434, Bruce2012}.

Lithium dendrites form during the battery charge cycle and can disconnect from the anode, generating “dead lithium” compounds that stop participating in the electrochemical reaction, resulting in loss of Coulombic efficiency~\cite{ Adams2018}. Ultimately, the uncontrolled growth of these lithium dendrites can cause the battery to short-circuit when they become large enough to contact the cathode electrode after penetrating through the separator~\cite{ ROSSO20065334, Dongping2015, JIAO2018110}. These critical shortcomings of dendrite formation during electrodeposition processes triggered efforts to control the dendritic pattern formation. Different strategies exist, including modifications in the electrode structure and porosity~\cite{ Fanlei2018, LiNan2019, Chanyuan2018}, pulsed charge-discharge cycle~\cite{ YANG2014900, Mayers2012, Aryanfar2014}, electrode surface morphology modification (roughness and wavenumber)~\cite{ Tikekar_2014}, solvent and electrolyte composition~\cite{ SUNDSTROM1995599, Li2017, Zheng2017, Qian2015, Suo2013, KIM2018517, Cheng2016}, ionic mass transport and electrolyte management~\cite{ Yang2005, Aoxuan2019, TAN201667, Crowther_2008, Wlasenko2010, Li2018,Iverson2008}, controls of internal temperature and pressure~\cite{ GIREAUD20061639, AKOLKAR201484, YAN2018193}, and stable interfacial coatings~\cite{ Yayuan2017, ZhuBin2017}.

The formation of dendrites in Li anodes has been noted to vary considerably from one system to the other. Bai et al.~\cite{ C6EE01674J} noted the formation of mossy/whiskers-like morphologies at current densities lower than the intrinsic diffusion-limited current density (reaction limited regime, where lithium grows from the base). On the other hand, at higher current densities (greater than the diffusion-limited current density, i.e. transport or diffusion limited regime, where lithium grows from the tip), fractal dendritic morphology or finger-like structures are obtained. Transition from base-controlled mossy structure to tip-controlled dendrite growth was also observed experimentally when under limiting current condition. Further, various 2D numerical simulations have found the formation of spike-like or tree-like structures where dendrite growth was found to occur on multiple branches and in multiple directions. These types of formation, however, have not been observed widely experimentally due to instabilities at the branch structures and mass transfer rate limitations at these regions~\cite{ LIU2019100003}.

The experimental investigation of the lithium dendrite formation in rechargeable metal batteries is challenging~\cite{ cheng2017toward}. Thus, the combined insights from experiment and simulation enhance our understanding of the mechanisms of dendrite formation and growth in lithium anodes~\cite{ LIU2019100003, JANA2017552, TAN2017155}. Within these computational models, the thermodynamic ones include several surface nucleation models based on density functional theory (DFT) that explore the mechanisms of lithium electrodeposition and dendrite suppression. Considering lithium dendrite growth as an inherent property of the metal, these models invoke surface energies and diffusion barriers as keys parameters in determining the specific metal structures growing on electrodes~\cite{ Markus2014, LING2012270, ozhabes2015stability}. Dynamic models describe the electrochemical propagation of dendrites; for example, the space-charge (migration-limited) models violate the electroneutrality at the dendrite tips leading to high overpotentials and the subsequent formation of ramified electrodeposits~\cite{ PhysRevA.42.7355}. Furthermore, stress and inelastic deformation models simulate the dendrite’s surface-tension mitigated growth~\cite{ Barton1962, Monroe_2004}. Additionally, film growth models apply classical film-growth theory within the lithium anode battery~\cite{ Wang2016}, and diffusion-limited aggregation models (DLA) develop fractal shapes using random-walk statistics~\cite{ Aryanfar2014}. The latter successfully modeled extremely low currents without considering surface energy.

Moreover, several phase-field (diffuse-interface) models simulate the morphologic evolution of lithium electrodeposits as a result of reaction-driven phase transformation within the metal anode batteries and describe the dendrite morphologies observed experimentally for low and high applied current densities~\cite{ PhysRevE.69.021603, PhysRevE.69.021604, SHIBUTA2007511, OKAJIMA2010118, PhysRevE.86.051609, Bazant2013, doi:10.1063/1.4905341, ELY2014581, Zhang_2014, CHEN2015376, PhysRevE.92.011301, doi:10.1021/acsenergylett.8b01009, YURKIV2018609, MU2019100921, jana2019electrochemomechanics, https://doi.org/10.1002/advs.202003301,CHEN2021229203, Zhang_2021}. The phase-field method describes and predicts the mesoscale microstructural evolution by implicitly tracking the boundaries and interfaces using an auxiliary function (i.e., the order parameter). These models rigorously incorporate interfacial energy, interface kinetics, and curvature-driven phase boundary movement. The evolution of the phase-field variables satisfies local equilibrium~\cite{ Ruurds84} and free energy minimization~\cite{ STEINBACH1996135}, leading to nonlinear partial differential equations (PDE’s). Different phase-field models of the electrochemical dendrite deposition describe the phase-field evolution using the Cahn-Hilliard equation~\cite{ GARCIA200411, HAN20044691}, the Allen-Cahn equation~\cite{ ALLEN1972423}, or a modified non-linear Allen-Cahn reaction equation~\cite{ Bazant2013}. In~\cite{ wang2020application}, the authors present a comprehensive review of the existing phase-field simulations in rechargeable batteries. 

Despite continuous efforts, there are several open issues related to the evolution of dendritic patterns in lithium metal electrodes. Most published models solve two-dimensional problems to rationalize the dendritic patterns and suppress their growth~\cite{ PhysRevE.92.011301, doi:10.1021/acsenergylett.8b01009, YURKIV2018609, MU2019100921, https://doi.org/10.1002/advs.202003301, Ruurds84}. Nevertheless, the lithium dendrite morphologies are inherently three dimensional~\cite{ jana2019electrochemomechanics, ding2016situ, TATSUMA20011201}. Thus, simulation tools to study and understand the three-dimensional effects triggering pattern formation are essential. Its application includes enhancing our understanding of the differences between 2D and 3D simulations on the free space effect on the Li-ion diffusion to the anode surface, the surface anisotropy, the electric field distribution, the nuclei distribution, the interaction between neighboring dendrites; and assessing control strategies to prevent the dendrite formation (e.g., 3D porous current collectors hosts with lithiophilic sites)~\cite{ yang2015accommodating, xu2021dendrite}.

Nevertheless, few attempts to simulate three-dimensional lithium dendrite growth exist in the literature. For instance,~\cite{ natsiavas2016effect} develops a 3D growth model for the electrode–electrolyte interface in lithium batteries in the presence of an elastic prestress; the model assesses the linearized stability of planar interface growth. Later, Jang et al. performed stochastic modeling of the three-dimensional effect of applied voltage and diffusion constant on the growth of lithium dendrites~\cite{ jang2021effect}. Recently, Yangyang et al.~\cite{ https://doi.org/10.1002/advs.202003301} presented a phase-field model to study the three-dimensional effect of the exchange current density on the electrodeposition behavior of lithium metal; however, without focusing on dendritic morphologies. The scarcity of three-dimensional phase-field results is evidence of the complexity and large computational cost involved in solving the highly non-linear set of equations describing coupled electrochemical interactions during a battery charge cycle. 
 
In this paper, we present a 3D phase-field framework to describe the dendritic electrodeposition of lithium. This work constitutes a step toward the physics-based, quantitative models to rationalize hazardous three-dimensional dendritic patterns needed to achieve the commercial realization of Li-metal batteries. Herein, we use several computational efficiency improvements to deliver the 3D simulations at a reasonable cost, such as time step adaptivity strategy, mesh rationalization, parallel computation, and balanced phase-field interface thickness to mesh resolution ratio.

We organize the paper as follows: Section~\ref{section:Formulation} presents the basic equations describing the lithium-battery dendrite growth process. Then, we derive the weak variational statement and its numerical implementation in Section~\ref{section:weakvar}. We describe the time-adaptive strategy and the parameters that deliver convergence in Section~\ref{section:TimeInt} and detail its implementation in Section~\ref{section:Algebraic}. We discuss numerical simulations of lithium-battery dendrites growth in Section~\ref{section:NumExp}, where we validate a two-dimensional model in terms of dendrite propagation rates and spatial distribution analysis of the system's variables in comparison with phase-field simulation results reported in the literature. Furthermore, we simulate three-dimensional spike-like lithium structures that grow under high current density~\cite{ jana2019electrochemomechanics, ding2016situ, TATSUMA20011201} (fast battery charge); these structure's growth is dangerous for battery operation. We perform single and multiple nuclei numerical experiments to study the 3D distribution of the electric field and the lithium-ion concentration to understand the mechanism behind tip-growing lithium morphologies better. Finally, we draw conclusions in Section~\ref{section:concl}.

\section{Governing partial differential equations}
\label{section:Formulation}

We model a battery cell composed of a solid pure Lithium metal anode and a binary liquid electrolyte. The variables of interest are the lithium concentration, $\zeta$, and the electric potential, $\phi$. Under an electro potential difference $\Delta \phi$, the battery charges and the electrolyte dissociates in $\text{Li}^+$ cation and $\text{A}^-$ anion species; Faradic reactions occur, the current passes through the electrode-electrolyte interface, the $\text{Li}^+$ cation gains an electron and deposits on the anode surface ($\text{Li}^++e^-\rightarrow \text{Li}$). Thus, the physical processes involved in the electrochemical deposition of Lithium are charge and mass transport. The Butler-Volmer kinetics~\eqref{eq:currentoverpot} is a standard phenomenological model that describes the charge and mass transport at the electrode-electrolyte interface~\cite{ 1967IV, Allen2001}, known as the current-overpotential equation:
\begin{equation}
  i= \displaystyle i_0\left(e^{-\frac{\alpha
        n F\eta}{ RT}}-e^\frac{\left(1-\alpha\right)
      n F\eta}{RT}\right) \ ,
  \label{eq:currentoverpot}
\end{equation}
where $i$ is the current density and $i_0$ is the exchange current density, identified as an intrinsic kinetic parameter and assumed constant in this case~\cite{ https://doi.org/10.1002/advs.202003301}. The applied overpotential $\eta\left(\zeta,\phi\right)$ is the total free energy change per charge transferred~\cite{ Bazant2013}. The first and second term in brackets represent the oxidation and reduction reactions, respectively, whereas $\alpha$ is the charge transfer coefficient that characterizes the forward and reverse reactions~\cite{ Kuznetsov99}. $R,\ T,\ n,$ and $F$ represent the gas constant, temperature, valence, and Faraday’s constant, respectively. The lithium electrodeposition rate depends on the applied current density via a Faradic reaction~\cite{ NISHIKAWA201184},
\begin{equation}
v=\frac{\partial\lambda}{\partial t}=\frac{i}{F  n  \text{C}_m^s} \ ,
\label{eq:faradaicreactionrate}
\end{equation}
where $\lambda$ represents the electrodeposited film thickness over a time $t$, and $\text{C}_m^s$ is the site density of Lithium metal (inverse of molar volume) (e.g., see~\cite{ Bazant2013, Arguello2022} for further details). 

\subsection{Phase-field}

We introduce a continuous phase-field variable $\xi$ that corresponds to a dimensionless lithium atom concentration $\widetilde{\zeta}=\zeta/\text{C}_m^s$, with $\zeta$ normalized against $\text{C}_m^s$. Thus, $\xi=1$ and $0$ represent the pure electrode and electrolyte phases. Superscripts “s” and “l” represent the solid-electrode and liquid-electrolyte phases. We assume the electrode phase is a pure solid and, thus, neglect the presence of a solid-electrolyte interface (SEI). Consequently, neither species nor charge can be stored at the electrode-electrolyte interface. Therefore, $\xi$ is a non-conserved order parameter in our model~\cite{ doi:10.1063/1.4905341}, satisfying the Allen-Cahn reaction (ACR) model: $\frac{\partial\xi}{\partial t}=\mathcal{R}\left(\frac{\delta \Psi}{\delta \zeta_i}\right)$, where $\Psi$ is the total free functional, expressed as the summation of the Helmholtz free energy density $\text{f}_{\text{ch}}$, surface energy density $\text{f}_{\text{grad}}$, and electrostatic energy density $\text{f}_{\text{elec}}$~\cite{ Bazant2013, PhysRevE.69.021604, GARCIA200411, HAN20044691}
\begin{equation}
\Psi=\int_{V}\left[\text{f}_{\text{ch}}\left(\xi,\zeta_i\right)+\frac{1}{2}\kappa\left(\xi\right)\left(\mathrm{\nabla}\xi\right)^2+\text{f}_{\text{elec}}\left(\xi,\zeta_i,\phi\right)\right]dV \ ,
\label{eq:gibbsfreeen}
\end{equation}
where $\zeta_i$ represents the set of concentrations of all the ionic species, e.g., lithium cation $\zeta_+$, and anion $\zeta_-$, respectively. $\text{f}_{\text{elec}}=\rho\phi$ is the electrostatic energy density, where $\rho=\sum_{i}{n _iF \zeta_i}$ is the charge density. The Helmholtz free energy density relative to the standard state is~\cite{ Bazant2013, YURKIV2018609},
\begin{equation}
\text{f}_{\text{ch}}=W{\xi}^2\left(1-\xi\right)^2+C_0 R T \left({\widetilde{\zeta}}_+ \ln{\widetilde{\zeta}}_++{\widetilde{\zeta}}_- \ln{\widetilde{\zeta}}_-\right) \ ,
\label{eq:helmholtzfreeenergy}
\end{equation}
with ion concentrations, ${\widetilde{\zeta}}_+=\zeta_+/C_0$ and ${\widetilde{\zeta}}_-=\zeta_-/C_0$, being normalised against the initial bulk concentration of Lithium in the electrolyte $C_0$. We model the equilibrium states using a double-well function $g\left(\xi\right)=W{\xi}^2\left(1-\xi\right)^2$, where $W/16$ acts as a barrier between the states~\cite{ CHEN2015376}. 

\begin{figure}
\includegraphics[height = 5cm]{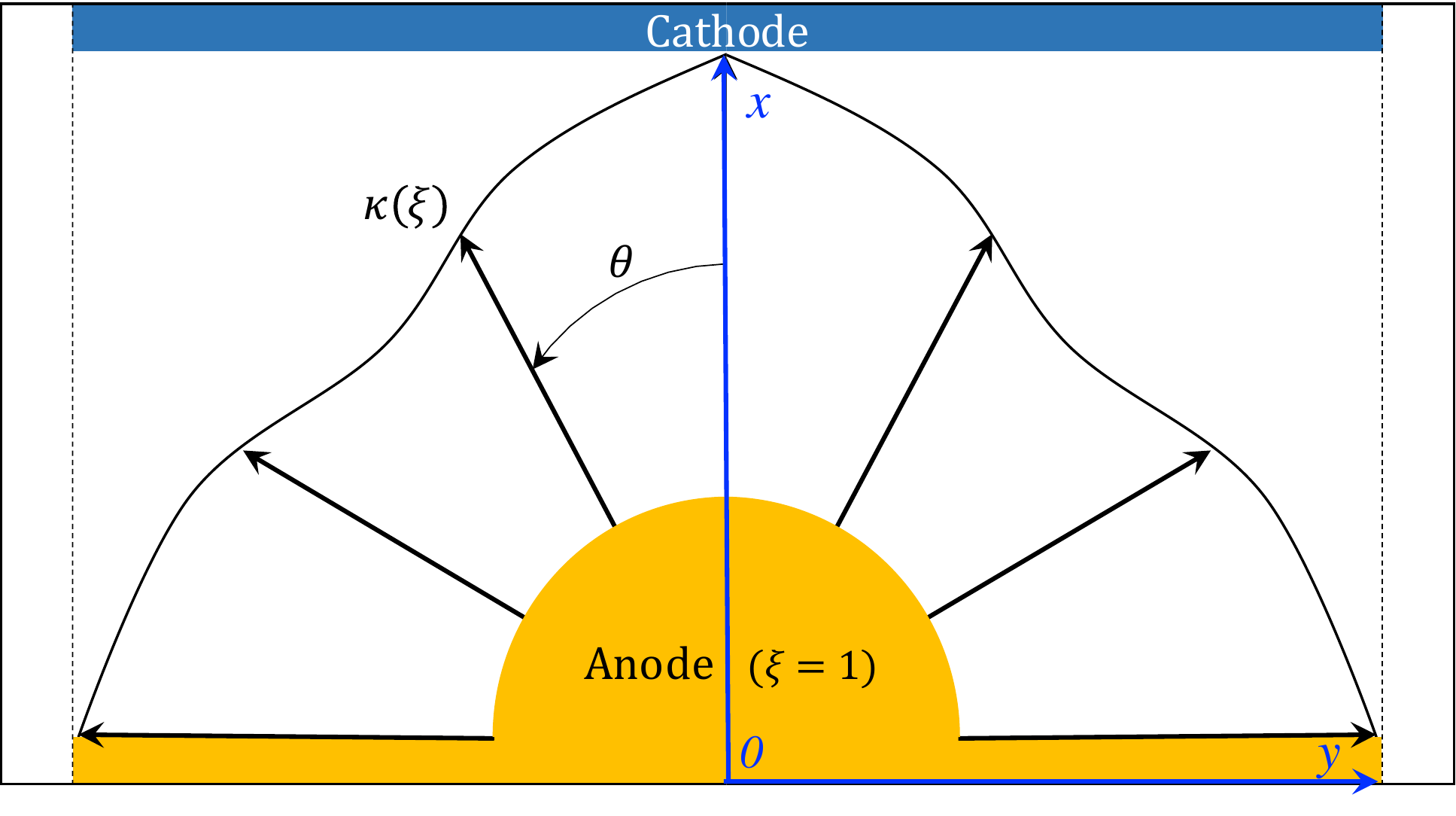}
\caption{2D schematic of Lithium electrodeposit, with four-fold surface energy anisotropy $\kappa\left(\xi\right)$ as a function of $\theta$, defined as the angle between the surface normal and the crystallographic orientation “$x$”.}
\label{fig:kappa2D}
\end{figure}

The surface energy anisotropy plays an important role in the morphology of the electrodeposit~\cite{ PhysRevE.92.011301}, which we implement using a standard approach~\cite{ KOBAYASHI1993410, Zhang_2014} with $\kappa\left(\xi\right)=\kappa_0\left[1+\delta_{\text{aniso}}\cos\left(\omega\theta\right)\right]$, where $\kappa_0$ relates to the Lithium surface tension $\gamma$; $\delta_{\text{aniso}}$ and $\omega$, are the strength and mode of  anisotropy, respectively~\cite{ TRAN201948, ZHENG202040}. $\theta$ is the angle between the surface normal and the crystallographic orientation of the lithium dendrite; aligned with the “$x$” direction of the cell stack for simplicity, as Figure~\ref{fig:kappa2D} shows. We use four-fold anisotropy ($\omega=4$) in agreement with lithium body centred cubic crystal structure~\cite{ PhysRevE.92.011301}. We make use of trigonometric identities to arrive to $\cos\left(4\theta\right)=8n_1^4-8n_1^2+1$, where $n_1=\nabla_x\xi/||\nabla\xi||$. Therefore, the 2D gradient coefficient is,
\begin{equation}
\kappa\left(\xi\right)=\kappa_0\left[1+\delta_{\text{aniso}}\left(8n_1^4-8n_1^2+1\right)\right]\ .
\label{eq:KAPPA2D}
\end{equation}
Figure~\ref{fig:kappa2D} sketches~\eqref{eq:KAPPA2D}, showing its preferred growth directions of a lithium electrodeposit ($\theta=0^{\circ},90^{\circ},270^{\circ}$) due to four-fold surface energy anisotropy, tending to grow fractal or branched dendrites. In 3D, we use version of~\eqref{eq:KAPPA2D} (four-fold anisotropy) derived by William et al. ~\cite{ GEORGE2002264, TAKAKI201321} to simulate crystal growth
\begin{equation}
\kappa\left(\xi\right)=\kappa_0\left(1-3\delta_{\text{aniso}}\right)\left[1+\frac{4\delta_{\text{aniso}}}{1-3\delta_{\text{aniso}}}\left(n_1^4+n_2^4+n_3^4\right)\right],
\label{eq:KAPPA3D}
\end{equation}
where $n_2=\frac{\nabla_y\xi}{||\nabla\xi||}$ and $n_3=\frac{\nabla_z\xi}{||\nabla\xi||}$, with “$y$” and “$z$” defined orthogonal to “$x$” as depicted in Figure~\ref{fig:BC_Planar3D}

Following~\cite{ CHEN2015376}, we calculate the variational derivative of the free energy functional~\eqref{eq:gibbsfreeen}. Thus, the phase-field Butler-Volmer equation matches the velocity of the sharp interface limit of the phase-field equation, with the current-overpotential~\eqref{eq:currentoverpot} as follows~\cite{ PhysRevE.86.051609, PhysRevE.92.011301, PhysRevE.64.021604} 
\begin{equation}
\frac{\partial\xi}{\partial t}=-L_\sigma\left[\frac{\partial g\left(\xi\right)}{\partial\xi}-\kappa\left(\xi\right)\nabla^2\xi\right]-L_\eta\frac{\partial h\left(\xi\right)}{\partial\xi}\left[e^{\left(\frac{\left(1-\alpha\right)n F \phi}{R T}\right)}-{\widetilde{\zeta}}_+\ e^{\left(\frac{-\alpha n F \phi}{R T}\right)}\right] \ ,
\label{eq:linealPF}
\end{equation}
where we adjust the interface mobility $L_\sigma$ value (constant for each simulation) for the selected voltage charging the battery. These mobility values balance the phase-field interface energy and the electrochemical reaction contribution~\cite{ Arguello2022}.  The electrochemical reaction kinetic coefficient is~\cite{ doi:10.1021/acsenergylett.8b01009} 
$$L_\eta=\frac{\gamma i_0}{n F \kappa\text{C}_m^s},$$
where the quantities of Li surface energy $\gamma$ and the phase-field interfacial thickness $\delta_{PF}$ relate to the model parameters according to $\delta_{PF}=\frac{2\kappa_0}{3\gamma}$~\cite{ Cahn1959}. The interpolation function $h\left(\xi\right)$ smooths the diffuse interface in the current implementation; we use a sigmoid interpolation function~\cite{ CHAI2017335, Arguello2022}, which has better computational efficiency than the popular choice of polynomial interpolation function~\cite{ doi:10.1146/annurev.matsci.32.101901.155803}:
\begin{equation}
h(\xi)=\frac{e^{\varsigma\left(\xi-\frac{1}{2}\right)}}{1+e^{\varsigma\left(\xi-\frac{1}{2}\right)}} \ ,
\end{equation}
where $\varsigma$ is a parameter that determines the interface thickness of the interpolation function; we use $\varsigma=20$, to adjust for interpolation between $\xi=0$ and $\xi=1$.

\subsection{Diffusion-migration process}

A diffusion-migration equation describes the motion of charged chemical species (lithium-ion) in the fluid electrolyte. The temporal evolution of ${\widetilde{\zeta}}_+$, satisfies the modified Nernst-Planck diffusion equation that describes the flux of Li-ions under the influence of a concentration gradient $\nabla {\zeta}_+$ and an electric field $\mathrm{\nabla\phi}$~\cite{ PhysRevA.42.7355}.  We add the term $\frac{\text{C}_m^s}{C_0}\frac{\partial\xi}{\partial t}$ to the Nernst-Planck diffusion equation to account for the eliminated lithium ion from the electrolyte solution, due to electrodeposition on the solid phase (metal electrode). Thus, the diffusion equation becomes:
\begin{equation}
\frac{\partial{\widetilde{\zeta}}_+}{\partial t}=\nabla\cdot\left(D^{\text{eff}}\left(\xi\right)\ \mathrm{\nabla}{\widetilde{\zeta}}_++D^{\text{eff}}\left(\xi\right)\frac{n F }{R T}{\widetilde{\zeta}}_+\mathrm{\nabla\phi}\right)-\frac{\text{C}_m^s}{C_0}\frac{\partial\xi}{\partial t} \ ,
\label{eq:Diffusion_Eq}
\end{equation}
where we interpolate the effective diffusivity by $D^{\text{eff}}\left(\xi\right)=D^sh\left(\xi\right)+D^l\left[1-h\left(\xi\right)\right]$, with $D^s$ and $D^l$ as the electrode and electrolyte diffusivities, respectively. 

\subsection{Electrostatic Potential}

We account for the electrostatic potential distribution $\phi$ using the charge continuity equation~\cite{ PhysRevE.69.021604}
\begin{equation}
\frac{\partial\rho}{\partial t}=-\nabla\cdot\vec{i} \ ,
\end{equation}
where $\vec{i}$ is the current density and $\rho=\Sigma_i n _iF  \zeta_i$ is the charge density. Experimental observations support the assumption that space-charge effects do not affect the stability of electrodeposits~\cite{ Elezgaray1998LinearSA}. Therefore, we ignore the effects of the double-layer structure and assume electroneutrality, that is,
$${\widetilde{\zeta}}_+={\widetilde{\zeta}}_-,$$
where we use Ohm’s law in the continuity equation, $\vec{i}=\sigma  \vec{E}$, and $\sigma$ is the conductivity and $\vec{E}=-\nabla\phi$ is the electric field. Thus, we arrive at a Poisson equation, including a source term to represent the charges that enter or leave the system due to the electrochemical reaction~\cite{ CHEN2015376}
\begin{equation}
\mathrm{\nabla}\cdot\left[\sigma^{\text{eff}}\left(\xi\right)\ \mathrm{\nabla\phi}\right]= n F \text{C}_m^s\frac{\partial\xi}{\partial t} \ .
\end{equation}

We interpolate the effective conductivity using 
$$\sigma^{\text{eff}}\left(\xi\right)=\sigma^sh\left(\xi\right)+\sigma^l\left[1-h\left(\xi\right)\right],$$
where $\sigma^s$ and $\sigma^l$ are the electrode and electrolyte phase conductivity, respectively.

We collect the system of equations that models the physical process in the following box.
\begin{Summary}{Dendrite growth model}{firstsummary}  
Lithium-battery dendrite growth process based on phase-field theory. \textit{Find $\Xi=\left(\xi, {\widetilde{\zeta}}_+,\phi\right)$ fulfilling}
\begin{equation*}
  \left\{
    \begin{aligned}
      \displaystyle \frac{\partial\xi}{\partial t} &= \displaystyle  -L_\sigma\left[\frac{\partial g\left(\xi\right)}{\partial\xi}- \nabla \cdot \left( \kappa\left( \xi \right) \nabla \xi \right)\right]-L_\eta\frac{\partial h\left(\xi\right)}{\partial\xi}\left[e^{\left(\frac{\left(1-\alpha\right)n F \phi}{R T}\right)}-{\widetilde{\zeta}}_+\ e^{\left(\frac{-\alpha n F \phi}{R T}\right)}\right],
      && \text{ in }V \times I  \\
      \displaystyle \frac{\partial{\widetilde{\zeta}}_+}{\partial t}   &= \displaystyle  \nabla\cdot\left[ D^{\text{eff}}\left(\xi\right)\ \mathrm{\nabla}{\widetilde{\zeta}}_+ +D^{\text{eff}}\left(\xi\right)\frac{n F }{R T}{\widetilde{\zeta}}_+\mathrm{\nabla\phi}\right]  -  \frac{\text{C}_m^s}{C_0}\frac{\partial\xi}{\partial t} ,
      &&  \text{ in }V \times I   \\
      \displaystyle n F \text{C}_m^s\frac{\partial\xi}{\partial t}  &= \displaystyle \mathrm{\nabla}\cdot\left[\sigma^{\text{eff}}\left(\xi\right)\ \mathrm{\nabla\phi}\right]   ,
      &&  \text{ in }V \times I   \\ 
      \xi &= \xi_{D}  ,
      && \text{ on } \partial V_{D} \times I   \\ 
      {\widetilde{\zeta}}_{+} &= \widetilde{\zeta}_{+D}  ,
      &&  \text{ on } \partial V_{D} \times I  \\
      \phi &= \phi_{D}    ,
      &&\text{ on } \partial V_{D} \times I   \\ 
      \nabla \xi   \cdot \boldsymbol{n} &= 0   ,
      &&  \text{ on }  \partial V_{N}  \times I  \\
      \nabla {\widetilde{\zeta}}_+ \cdot \boldsymbol{n}  &= 0 ,
      &&  \text{ on }  \partial V_{N} \times I \\
      \nabla \phi   \cdot \boldsymbol{n} &= 0  ,
      &&  \text{ on }  \partial V_{N} \times I  \\ 
      \xi \left( \boldsymbol{x} , t_0 \right) &= \xi_{0}   ,
      &&   \text{ in }V  \\
      {\widetilde{\zeta}}_{+} \left( \boldsymbol{x} , t_0 \right) &= {\widetilde{\zeta}}_{+0}  ,
      &&   \text{ in }V  \\
      \phi \left( \boldsymbol{x} , t_0 \right) &= \phi _{0}   ,
      &&  \text{ in }V 
\end{aligned}\right. 
\end{equation*}
where $V$ is the problem domain with boundary $\partial V = \partial V_{N} \cup \partial V_{D}$, the subscript $N$ and $D$ related to the Neumann and Dirichlet parts, with outward unit normal $\boldsymbol{n}$, and $I$ is the time interval.
\end{Summary}

\section{Weak variational formulation and space-time discretization}
\label{section:weakvar}

\subsection{Weak formulation}

We now state the weak variational formulation~\cite{ hughes2012finite, pardo2021modeling} of the system of equations as: \textit{Find  $\Xi=\left(\xi, \widetilde{\zeta}_{+}, \phi \right)$ such that
  $\forall V=\left( v,w,p\right)$}
\begin{equation}
  \begin{aligned}
      \underbrace{\left\langle v , \dot{\xi}\right\rangle_{V} 
       +\left\langle w , \dot{{\widetilde{\zeta}}_+} \right\rangle_{V} +  \left\langle w , \frac{\text{C}_m^s}{C_0} \dot{\xi}\right\rangle_{V}
         +     \left\langle p , \dot{\xi} \right\rangle_{V}}_{M\left(V,\dot{\Xi}\right)}
      = \underbrace{a\left( v , \xi,  \widetilde{\zeta}_{+}, \phi \right)
       +  b\left( w , \xi, \widetilde{\zeta}_{+}, \phi \right) 
      + c\left( p , \xi, \phi  \right)}_{A\left(V,\Xi\right)}
\end{aligned}
\label{eq:weakform}
\end{equation}
%
where  $\dot{\bullet} = \frac{\partial \bullet}{\partial t}$ denotes the time derivative of the variable $\bullet = \xi, \widetilde{\zeta}_{+}$,  $\left\langle u , w\right\rangle_{V} = \int_{V} w u  \ dV$ expresses the inner product, and the functions on the right-hand side are
\begin{equation}\label{eq:ops}
    \begin{aligned}
       a\left( v , \Xi\right)
      = & - \int_{V} L_\sigma\left[\frac{\partial g\left(\xi\right)}{\partial\xi} v +  \kappa\nabla \xi \cdot \nabla v \right]   \ dV - \int_{V}  L_\eta\frac{\partial h\left(\xi\right)}{\partial\xi}\left[e^{\left(\frac{\left(1-\alpha\right)n F \phi}{R T}\right)}-{\widetilde{\zeta}}_+\ e^{\left(\frac{-\alpha n F \phi}{R T}\right)}\right] v \ dV  \   ,\\
        b\left( w , \Xi\right)
      =& - \int_{V} \left[ D^{\text{eff}}\left(\xi\right)\ \mathrm{\nabla}{\widetilde{\zeta}}_+ \cdot \nabla w  + D^{\text{eff}}\left(\xi\right)\frac{n F }{R T}{\widetilde{\zeta}}_+\mathrm{\nabla\phi} \cdot \nabla w \right] \ dV \ , \\ 
        c\left( p , \Xi  \right)
      = & - \frac{1}{n F \text{C}_m^s} \int_{V} \sigma^{\text{eff}}\left(\xi\right) \mathrm{\nabla\phi} \cdot \nabla p \ dV.
    \end{aligned}
\end{equation}
We use standard finite element spaces where each function and its gradient are square integrable.

\subsection{Time semi-discretization}

We use a second-order backward-difference (BDF2) time marching scheme with an adaptive time step size. BDF2 is an implicit time marching method that requires the solution at two previous time instants; the initial step uses a first-order backward-difference method (BDF1). BDF2 has second-order accuracy and numerically damps the highest frequencies of the solution, unlike the conservative Crank-Nicolson method~\cite{ hughes2012finite}. Liao et al.~\cite{ 10.1093/imanum/draa075} showed that BDF2 is an effective time integrator for the phase-field crystal model, especially when coupled with an adaptive time-step strategy.

We discretize the time interval into $t_0<t_1<...<t_n<...<t_f$  and define the time-step size as $\Delta t_n=t_n-t_{n-1}$, approximate $\xi \left(t_n\right)$, $\dot{\xi } \left( t_n \right)$, ${\widetilde{\zeta}}_+  \left( t_n \right)$ and $\dot{{\widetilde{\zeta}}}_+  \left( t_n \right)$, respectively, as $\xi_n$, $\dot{\xi}_n$, ${\widetilde{\zeta}}_{+n}$ and $\dot{\widetilde{\zeta}}_{+n}$, and express the time increments as $\Delta \bullet = \bullet_{n+1} -  \bullet_{n}$ for $\bullet=\xi,\dot{\xi},{\widetilde{\zeta}}_{+},\dot{\widetilde{\zeta}}_{+}$. We use a second-order approximation of the time derivative at $t_{n+1}$ as follows~\cite{ CELAYA20141014},
\begin{equation} \label{eq:BDF2}
\frac{\partial \bullet_{n+1}}{\partial t} = \left( \frac{1+2\omega_{n+1}}{1+\omega_{n+1}} \right) \frac{\bullet_{n+1} -  \frac{\left(1+\omega_{n+1} \right)^2}{1+2\omega_{n+1}}  \bullet_{n} + \frac{\omega^{2}_{n+1}}{1+2\omega_{n+1}} \bullet_{n-1}}{\Delta t_{n+1}}, \ \text{  with  } \ \omega_{n+1}=\frac{\Delta t_{n+1}}{\Delta t_{n}}, \ \text{  and  } \ \bullet = \xi, {\widetilde{\zeta}}_{+} \ .
\end{equation}
Then using this definition and letting $V_v=\left(v,0,0\right)$, we define the $\xi$-residual as
\begin{equation}\label{eq:R_xi}
  \begin{aligned}
    0=R_{\xi}\left(V_v,\Xi_{n+1}\right)
    &=M\left(V_v,\dot{\Xi}_{n+1}\right)-A\left(V_v,\Xi_{n+1}\right)\\
    &=\left\langle v , {\xi}_{n+1} \right\rangle_{V}
    + \underbrace{\left\langle v ,   \frac{\left(1+\omega_{n+1} \right)^2}{1+2\omega_{n+1}}  \xi_{n} - \frac{\omega^{2}_{n+1}}{1+2\omega_{n+1}} \xi_{n-1} \right\rangle_{V}}_{\ell_{\xi} \left( v \right)\text{ (known at $t_{n}$)}}
    - \beta_{n+1}  \Delta t_{n+1} \  a\left( v , ,\Xi_{n+1}\right)
  \end{aligned}
\end{equation}
with $\beta_{n+1}  = \frac{1+\omega_{n+1}}{1+2\omega_{n+1}}$.  We now approximate $a\left( v , \Xi_{n+1}\right)$ as a Taylor series expansion from $\Xi_{n}$ to obtain
\begin{equation}\label{eq:aprime}
  \begin{aligned}
    a\left( v , \Xi_{n+1}\right)
    &=  a\left( v , \Xi_{n}\right)
    + a'_{\xi} \left( v ,  \Delta \xi; \Xi_{n}\right)
    +  a'_{\widetilde{\zeta}_{+}} \left( v , \Delta \widetilde{\zeta}_{+} ;\Xi_{n}\right)
    + \ a'_{\phi} \left( v , \Delta \phi ; \Xi_{n}\right) 
  + \mathcal{O}\left(\Delta t_{n+1}^2\right)
  \end{aligned}
\end{equation}
where $\mathcal{O}\left(\Delta^2\right)$ represents neglected higher-order terms in the expansion and 
\begin{equation}
  {f}'_{\bullet}\left( v, \Delta \bullet ; \bullet_{n} \right)=\dfrac{d}{d\epsilon}\left.f\left( {v} ,
      \bullet_{n} +\epsilon \Delta \bullet \
    \right)\right|_{\epsilon=0} \ 
\end{equation}
represents the directional Gâteaux derivative of the functional $f$ in the direction $\bullet$. Combining~\eqref{eq:R_xi} and~\eqref{eq:aprime}, we obtain a linear equation system to solve, that is
\begin{equation}\label{eq:linealization1}
  \begin{aligned}
    0&= R_{\xi}\left(V_v,\Xi_{n}\right)
    + \left\langle v , \Delta {\xi} \right\rangle_{V}
    + \beta_{n+1}  \Delta t_{n+1} \left[
      a'_{\xi} \left( v , \Delta \xi ;  \Xi_{n}   \right)  +
      a'_{\widetilde{\zeta}_{+}} \left( v ,
        \Delta \widetilde{\zeta}_{+} ;  \Xi_{n}   \right)
      + \ a'_{\phi} \left( v ,  \Delta \phi ;  \Xi_{n} \right) \right]
  \end{aligned}
\end{equation}
Similarly, we define the weighting functions $V_w=(0,w,0)$ and $V_p=(0,0,p)$; the residuals $\mathcal{R}_{\bullet}$ with $\bullet = \widetilde{\zeta}_{+}, \phi$ and linearize the resulting residuals to obtain the linearized system of equations to update the Newton iteration. We collect the linearized system of equations that models the physical process in the following box.
\begin{Summary}{Linearized equation system}{Linearization}  
  Discrete linearized equations for lithium-battery dendrite growth process based on phase-field theory. \textit{Find}
    $\Xi_{n+1}=\left(\xi_n, {\widetilde{\zeta}}_{+n},\phi_n\right)+\left(\Delta \xi,\Delta \phi,\Delta \widetilde{\zeta}_{+}\right)$
    \textit{ such that}
    \begin{equation}\label{eq:linealization}
      \left\{
        \begin{aligned}
          0&= R_{\xi}\left(V_v,\Xi_{n}\right)
          + \left\langle v , \Delta {\xi} \right\rangle_{V}\\
          &\qquad\qquad+ \beta_{n+1}  \Delta t_{n+1} \left[
            a'_{\xi} \left( v , \Delta \xi ;  \Xi_{n}   \right)  +
            a'_{\widetilde{\zeta}_{+}} \left( v ,
              \Delta \widetilde{\zeta}_{+} ;  \Xi_{n}   \right)
            + \ a'_{\phi} \left( v ,  \Delta \phi ;  \Xi_{n} \right) \right]\\          
          0 &=  \mathcal{R}_{{{\widetilde{\zeta}}}_{+}} \left( V_w , \Xi_{n} \right) 
          + \left\langle w , \Delta {{\widetilde{\zeta}}}_{+}
            + \frac{\text{C}_m^s}{C_0} \Delta {\xi} \right\rangle_{V}\\
          &\qquad\qquad+ \beta_{n+1}  \Delta t_{n+1} \left[
            b'_{\xi} \left( w , \Delta \xi ;  \Xi_{n}   \right)   +
            b'_{\widetilde{\zeta}_{+}} \left( w ,  \Delta \widetilde{\zeta}_{+} ;  \Xi_{n}\right) +
            b'_{\phi} \left( w ,    \Delta \phi ;  \Xi_{n} \right) \right]
          \\
          0&= \mathcal{R}_{\phi} \left( V_p , \Xi_{n} \right)
          + \left\langle p , \Delta {\xi}  \right\rangle_{V}
          + \beta_{n+1}  \Delta t_{n+1} \left[
            c'_{\xi} \left( p , \Delta \xi ;  \Xi_{n}   \right)
            +  c'_{\phi} \left( p , \Delta \phi ;  \Xi_{n}  \right) \right]
        \end{aligned}\right. 
    \end{equation}
    We only expand $c$ in the directions $\Delta \xi$ and $\Delta \phi$ as it is independent of $ \Delta \widetilde{\zeta}_{+}$, see~\eqref{eq:ops}$_3$.

\end{Summary}

For completeness, we summarize the Gâteaux derivatives in~\eqref{eq:linealization} in the following box.
\begin{Summary}{Gâteaux derivatives}{Gateaux}  
\begin{equation} \label{eq:gateaux}
  \begin{aligned}
    a'_{\xi} \left( v , \Delta \xi ;  \Xi_{n}  \right) &=
     \bigintss_{V} L_\sigma\left[
      \left.\frac{\partial g^{2}\left(\xi\right)}{\partial\xi^{2}} \right\vert_{\xi_n} \Delta \xi \ v
      + \left( \kappa\left(\xi_n \right)\nabla \Delta \xi 
        +  \left.\frac{\partial\kappa\left(\xi \right)}{\partial\xi}\right\vert_{\xi_n}
        \Delta \xi \ \nabla \xi_{n} \right)\cdot \nabla v \right]   \ dV\\
    &  + \bigintss_{V}  L_\eta
    \left.\frac{\partial h^{2}\left(\xi\right)}{\partial\xi^{2}}\right\vert_{\xi_n}
    \left[e^{\left(\frac{\left(1-\alpha\right)n F \phi_n}{R T}\right)}-{\widetilde{\zeta}}_{+n}\ e^{\left(\frac{-\alpha n F \phi_n}{R T}\right)}\right] \Delta\xi \ v \ dV  \\
    a'_{\widetilde{\zeta}_{+}} \left( v , \Delta \widetilde{\zeta}_{+} ;  \Xi_{n} \right)
    &= \bigintss_{V}  L_\eta
    \left.\frac{\partial h\left(\xi\right)}{\partial\xi} \right\vert_{\xi_n}
    e^{\left(\frac{-\alpha n F \phi_n}{R T}\right)} \Delta {\widetilde{\zeta}}_+ v \ dV  \\
    a'_{\phi} \left( v , \Delta \phi ;  \Xi_{n}   \right)
    &=  \bigintss_{V}  L_\eta
    \left.\frac{\partial h\left(\xi\right)}{\partial\xi} \right\vert_{\xi_n} \frac{n F }{R T}
    \left[
      \left(1-\alpha\right) e^{\left(\frac{\left(1-\alpha\right)n F \phi_{n}}{R T}\right)} + \alpha {\widetilde{\zeta}}_{+n}   \ e^{\left(\frac{-\alpha n F \phi_{n}}{R T}\right)}
    \right] \Delta \phi \ v \ dV \\
    b'_{\xi} \left( w , \Delta \xi ;  \Xi_{n}   \right)
    &= \bigintss_{V}
    \left.\frac{ \partial D^{\text{eff}}\left(\xi\right) }{\partial\xi}\right\vert_{\xi_{n}}
    \Delta \xi \  \left( \mathrm{\nabla}{\widetilde{\zeta}}_{+n} 
      +   \frac{n F }{R T}\ {\widetilde{\zeta}}_{+n}\mathrm{\nabla\phi}_n \right)
    \cdot \nabla w \ dV  \\
    b'_{\widetilde{\zeta}_{+}} \left( w , \Delta \widetilde{\zeta}_{+} ;  \Xi_{n} \right)
    &= \bigintss_{V} D^{\text{eff}}\left(\xi_{n} \right)
    \left( \ \mathrm{\nabla} \Delta {\widetilde{\zeta}}_+
      +  \frac{n F }{R T}\Delta{\widetilde{\zeta}}_{+} \mathrm{\nabla\phi}_{n} \right)
    \cdot \nabla w \ dV  \\
    b'_{\phi} \left( w , \Delta \phi ,  \Xi_{n} \right)
    &= \bigintss_{V}   D^{\text{eff}}\left(\xi_{n}\right)
    \frac{n F }{R T}{\widetilde{\zeta}}_{+n}
    \mathrm{\nabla \Delta \phi} \cdot \nabla w   \ dV  \\
    c'_{\xi} \left( p , \Delta \xi ;  \Xi_{n}  \right)
    &=  \frac{1}{n F \text{C}_m^s} \bigintss_{V}
    \left.\frac{\partial \sigma^{\text{eff}} \left(\xi\right)}{\partial \xi}\right\vert_{\xi_{n}}
    \Delta \xi \ \mathrm{\nabla\phi}_{n} \cdot \nabla p \ dV  \\
    c'_{\phi} \left( p , \Delta \phi ;  \Xi_{n}\right)
    &=  \frac{1}{n F \text{C}_m^s} \bigintss_{V} \sigma^{\text{eff}}\left(\xi_{n}\right)
    \mathrm{\nabla \Delta \phi} \cdot \nabla p \ dV 
  \end{aligned}
\end{equation}
\end{Summary}

\subsection{Space discretization}

We discretize the domain $V \subset\mathbb{R}^3$ with Dirichlet boundary condition $\partial V_D$ and Neumann boundary condition $\partial V_N$ using finite elements. We express the domain as the union of non-overlapping elements, $K_i$; thus, 
$V_h=\bigcup_{i=1}^{M}K_i .$ We define continous piecewise polynomial functions over the discrete domain. In particular, we use linear functions over simplexes (i.e., triangles in 2D, tetrahedra in 3D). Since the three variables share the spatial distribution of the boundary conditions, we use the shape functions  $N_A\left(\boldsymbol{x}\right)$ for each degree of freedom $\bullet_A^n$ at $t_n$ satisfying  the Dirichlet boundary conditions to discretize each degree of freedom $\bullet=X,Z,Y$ corresponding to the variables $\xi, {\widetilde{\zeta}}_+,\phi$, respectively; thus, we have
\begin{equation} \label{eq:matricesdisc}
\begin{aligned}
  {\xi}^{h}_n  = \sum_{i=1}^{S} N_A \left( \boldsymbol{x} \right)   X_{A}^n
  && {\widetilde{\zeta}_{+n}}^h    = \sum_{i=1}^{S} N_A \left( \boldsymbol{x} \right)  Z_{A}^n
  && {\phi}_n^{h}  = \sum_{i=1}^{S} N_A  \left( \boldsymbol{x} \right) Y_{A}^n \\
 \Delta {\xi}^{h}  = \sum_{i=1}^{S} N_A \left( \boldsymbol{x} \right)  \Delta X_A
  &&\Delta {\widetilde{\zeta}_{+}}^h    = \sum_{i=1}^{S} N_A \left( \boldsymbol{x} \right)  \Delta Z_{A}
  && \Delta{\phi}^{h}  = \sum_{i=1}^{S} N_A  \left( \boldsymbol{x} \right) \Delta Y_{A}
  \end{aligned}
\end{equation}
where $S$ corresponds to the total number of unknowns in each solution variable. We define the weighting spaces using  test spaces using the same functions but restricting those with support on $\partial V_D$ to be zero, thus
\begin{equation} \label{eq:matricesdisc2}
  \begin{aligned}
    {v}^{h} \in \spn \left\lbrace  N_B\right\rbrace^{W}_{B=1}
    &&  {w}^{h} \in \spn \left\lbrace  N_B  \right\rbrace^{W}_{B=1}
    && {p}^{h} \in \spn \left\lbrace  N_B
  \right\rbrace^{W}_{B=1} 
  \end{aligned}
\end{equation}
where $W$ corresponds to the total number of weighting functions for each variable.

Using these spatial discretizations, we obtain the fully discrete algebraic problem:
\begin{equation} \label{eq:algsystem}
  \begin{aligned}
    \underbrace{
      \begin{Bmatrix}
        \Delta \mathcal{R}_{\xi} \\
        \Delta \mathcal{R}_{{\widetilde{\zeta}_{+}}} \\
        \Delta \mathcal{R}_{\phi}
      \end{Bmatrix}
    }_{\Delta  \mathcal{R}}
    &=  
    \underbrace{
      \left( 
        \begin{aligned} 
          \begin{bmatrix}
            \mathbb{M}_{\xi\xi} & \textbf{0} & \textbf{0} \\
            \mathbb{M}_{{\widetilde{\zeta}_{+}}\xi}
            & \mathbb{M}_{{\widetilde{\zeta}_{+}}{\widetilde{\zeta}_{+}}}
            & \textbf{0} \\
            \mathbb{M}_{\phi\xi} & \textbf{0} & \textbf{0}
          \end{bmatrix}  
          +
          \beta_{n+1}  \Delta t  
          \begin{bmatrix}
            \mathbb{K}_{\xi\xi} & \mathbb{K}_{\xi{\widetilde{\zeta}_{+}}}
            & \mathbb{K}_{\xi\phi} \\
            \mathbb{K}_{{\widetilde{\zeta}_{+}}\xi}
            & \mathbb{K}_{{\widetilde{\zeta}_{+}}{\widetilde{\zeta}_{+}}}
            & \mathbb{K}_{{\widetilde{\zeta}_{+}}\phi}  \\
            \mathbb{K}_{\phi\xi} & \textbf{0}  & \mathbb{K}_{\phi\phi}
          \end{bmatrix}  
        \end{aligned}
      \right)
    }_{\mathbb{J}}
    \cdot 
    \underbrace{
      \begin{Bmatrix}
        \Delta  \boldsymbol{X} \\
        \Delta  \boldsymbol{Z} \\
        \Delta  \boldsymbol{Y}
      \end{Bmatrix}}_{\Delta \boldsymbol{X}}.
  \end{aligned}
\end{equation}
\begin{Summary}{Matrix blocks}{matrices}  
The mass matrix blocks are
\begin{equation} \label{eq:matricesM}
  \begin{aligned}
    \mathbb{M}^{\xi\xi}_{AB}
    =   \mathbb{M}^{{\widetilde{\zeta}_{+}}{\widetilde{\zeta}_{+}}}_{AB}
    = \mathbb{M}^{\phi\xi}_{AB} = \displaystyle \sum^{M}_{k=1}  \int_{K_k} N_A N_B \ dK_k
    && \text{ and }
    &&\mathbb{M}_{AB}^{{\widetilde{\zeta}_{+}}\xi}  = \sum^{M}_{k=1}  \int_{K_k} \frac{\text{C}_m^s}{C_0} N_A N_B \ dK_k 
  \end{aligned}
\end{equation}
while stiffness matrix blocks are
\begin{equation} \label{eq:matricesK}
  \left\{
  \begin{aligned}
    \mathbb{K}_{AB}^{\xi\xi}  =
    & \sum^{M}_{k=1}  \bigintss_{K_k} L_\sigma
    \left[
      \left.\frac{\partial g^{2}}{\partial\xi^{2}} \right\vert_{\xi_n^h}
      N_A \ N_B
      + \left( \kappa\left( \xi_n^h \right)\nabla N_A
        +  \left.\frac{\partial\kappa}{\partial\xi}\right\vert_{\xi_n^h}
        N_A \ \nabla \xi_n^h
        \right)\cdot \nabla N_B \right]   \ d{K_k} \\
      +&  \sum^{M}_{k=1}  \bigintss_{K_k}  L_\eta
      \left.\frac{\partial h^{2}}{\partial\xi^{2}}\right\vert_{\xi_n^h}
      \left[
        e^{\left(\frac{\left(1-\alpha\right) n F \phi_n^h}{R T}\right)}
        -{\widetilde{\zeta}}_{+n}^h\ e^{\left(\frac{-\alpha n F \phi_n^h}{R T}\right)}
      \right]
      N_A \ N_B \ d{K_k}  \\
      \mathbb{K}_{AB}^{\xi{\widetilde{\zeta}_{+}}}  =
      & \sum^{M}_{k=1}  \bigintss_{K_k}  L_\eta
      \left.\frac{\partial h}{\partial\xi} \right\vert_{\xi_n^h}
      e^{\left(\frac{-\alpha n F  \phi_n^h}{R T}\right)} N_A N_B \ d{K_k}   \\
      \mathbb{K}_{\xi\phi} =
      & \sum^{M}_{k=1}  \bigintss_{K_k}   L_\eta
      \left.\frac{\partial h}{\partial\xi}\right\vert_{\xi_n^h}
      \frac{n F }{R T} \left[
        \left(1-\alpha\right) e^{\left(\frac{\left(1-\alpha\right)n F  \phi_n^h}{R T}\right)}
        + \alpha {\widetilde{\zeta}}_{+n}^h   \
        e^{\left(\frac{-\alpha n F  \phi_n^h}{R T}\right)}\right] N_A \ N_B \ d{K_k} \\
      \mathbb{K}_{AB}^{{\widetilde{\zeta}_{+}}\xi} =
      & \sum^{M}_{k=1}  \bigintss_{K_k}
      \left.\frac{ \partial D^{\text{eff}}}{\partial\xi}\right\vert_{\xi_n^h}
      N_A \  \left[ \mathrm{\nabla}{\widetilde{\zeta}}_{+n}^h 
        +   \frac{n F }{R T}\ {\widetilde{\zeta}}_{+n}^h \mathrm{\nabla \phi}_n^h 
      \right] \cdot \nabla N_B \ d{K_k} \\
      \mathbb{K}_{{\widetilde{\zeta}_{+}}{\widetilde{\zeta}_{+}}} =
      & \sum^{M}_{k=1}  \bigintss_{K_k} D^{\text{eff}}
      \left( \xi_n^h \right) \left[ \ \mathrm{\nabla} N_A
        +  \frac{n F }{R T} N_A \mathrm{\nabla{\phi}}_n^h  \right]
      \cdot \nabla N_B \ d{K_k}   \\ 
      \mathbb{K}_{AB}^{{\widetilde{\zeta}_{+}}\phi} =
      & \sum^{M}_{k=1}  \bigintss_{K_k}  D^{\text{eff}}
      \left( \xi_n^h \right)
      \frac{n F }{R T}\ {\widetilde{\zeta}}_{+n}^h \
      \mathrm{\nabla N_A} \cdot \nabla N_B   \ dK_k  \\ 
      \mathbb{K}_{AB}^{\phi\xi}  =
      & \frac{1}{n F \text{C}_m^s}\ \sum^{M}_{k=1}  \bigintss_{K_k}
      \left.\frac{\partial \sigma^{\text{eff}} }{\partial \xi}\right\vert_{\xi_n^h}
      N_A \ \mathrm{\nabla{\phi}_n^h} \cdot \nabla N_B \ dK_k \\
      \mathbb{K}_{AB}^{\phi\phi}  =
      & \frac{1}{n F \text{C}_m^s}\ \sum^{M}_{k=1}  \bigintss_{K_k}
      \sigma^{\text{eff}}\left( \xi_n^h \right)\
      \mathrm{\nabla} N_A \cdot \nabla N_B   \ dK_k
  \end{aligned}\right.
\end{equation}
\end{Summary}

\section{Time-adaptive strategy}
\label{section:TimeInt}

Time step adaptivity is highly useful in this problem, where the time step requirements vary significantly at different simulation stages. For example, initially, the simulation requires small time steps to achieve convergence during the development of the phase-field interface and thereon, their size grows a few orders of magnitude (i.e., from $\Delta t = 10^{-6}\left[s\right]$ up to $\Delta t = 10^{-1}\left[s\right]$) depending on the parameters (e.g.,  mesh size, electrodeposition rate). 

Following~\cite{ Vignal:2017, LABANDA2022114675}, we express the local truncation error for BDF2 using Taylor expansions as follows
\begin{equation}
  {\tau}^{\text{BDF2}}\left(t_{n+1}\right)
  = \frac{ \Delta t^{2}_{n+1} \left( \Delta t_{n}
      +  \Delta t_{n-1}  \right)}{6} \dddot{u} \left(t_{n+1}\right)
  + \mathcal{O} \left( \Delta t^{4}  \right) \ ,
  \label{eq:truncation}
\end{equation}
where $u = \xi, {\widetilde{\zeta}_{+}}$, since $\phi$'s time derivative does not appear explicitly in the formulation. We use the solutions $u_{n+1}, u_{n}, u_{n-1}$ and $u_{n-2}$ from the BDF2 scheme to estimate the truncation error of~\eqref{eq:truncation} using the third-order backward difference formula (BDF3)
\begin{equation}
  \dddot{u} \left(t_{n+1}\right) \approx \frac{1}{\Delta t^{2}_{n+1}}\left[ \frac{u_{n+1}
      -u_{n}}{\Delta t_{n+1}}
    - \left( 1 + \frac{\Delta t_{n+1}}{\Delta t_{n}} \right) \frac{u_{n}
      -u_{n-1}}{\Delta t_{n}}
    + \frac{\Delta t_{n+1}}{\Delta t_{n} \Delta t_{n-1}} \left( u_{n-1}
      - u_{n-2}\right)\right].
  \label{eq:backwarddifferenceformula}
\end{equation}
Thus, we estimate BDF2's local truncation error by substituting~\eqref{eq:backwarddifferenceformula} into~\eqref{eq:truncation} to obtain
\begin{equation}
  {\tau}^{\text{BDF2}}\left(t_{n+1}\right) \approx \frac{ \Delta t_{n} +  \Delta t_{n-1}  }{6}\left[ \frac{u_{n+1}-u_{n}}{\Delta t_{n+1}} - \left( 1 + \frac{\Delta t_{n+1}}{\Delta t_{n}} \right) \frac{u_{n}-u_{n-1}}{\Delta t_{n}} + \frac{\Delta t_{n+1}}{\Delta t_{n} \Delta t_{n-1}} \left( u_{n-1} - u_{n-2}\right)\right],
  \label{eq:local}
\end{equation}
Finally, we compute the weighted local truncation error as an error indicator~\cite{ hairer2010solving}
\begin{equation}
  E_{u} \left(t_{n+1}\right) = \sqrt{\frac{1}{N} \sum^{N}_{i=1} \left(\frac{ {\tau}^{\text{BDF2}}_{i} \left(t_{n+1}\right) }{\rho_{abs} + \rho_{rel} \max \left( \vert u_{n+1} \vert_{i} , \vert u_{n+1} \vert_{i} + \vert {\tau}^{\text{BDF2}} \left(t_{n+1}\right) \vert_{i} \right) }\right)^{2}},
  \label{eq:WLE}
\end{equation}
where $\rho_{abs}$ and $\rho_{rel}$ are user-defined parameters that define the absolute and relative tolerances, respectively. In our examples we set these parameters to $\rho_{abs} = \rho_{rel} = 10^{2}$. The time error is computed using the maximum time error of time-dependent variables
\begin{equation}\label{eq:WLE2}
E  \left(t_{n+1}\right) = \max \left( E_{\xi} \left(t_{n+1}\right) , E_{\widetilde{\zeta}_{+}} \left(t_{n+1}\right) \right) ,
\end{equation}
and the time-step adaptivity simply follows \cite{GOMEZ20084333,LANG1995223}
\begin{equation}
  \Delta t^{k+1}_{n+1} \left(t_{n+1}\right)
  = \boldsymbol{F} \left( E \left(t_{n+1}\right) ,\Delta t^{k}_{n+1}, \text{tol} \right)
  = \rho_{tol} \left( \frac{\text{tol}}{E \left(t_{n+1}\right)}\right)^{\frac{1}{2}} \Delta t^{k}_{n+1} ,
  \label{eq:DT}
\end{equation}
where $k$ is the time-step refinement level and $\rho_{tol} $ is a safety factor parameter set to $0.9$ in our simulations. We summarize the time-adaptive scheme in Algorithm~\ref{Alg1Adap}, where we define two tolerances $\text{tol}_{max}$ and $\text{tol}_{min}$ that limit the range of reduction or increments of the time-step size.

\section{Implementation details}
\label{section:Algebraic}

This section briefly discusses the implementation aspects and details the step-by-step calculations. We perform all the numerical experiments using the open-source FEniCS environment~\cite{ alnaes2015fenics} using the FIAT package~\cite{ Kirby2004} to integrate numerically and assemble the matrices and vectors. We use four-node quadrilateral elements in 2D and eight-node hexahedral elements in 3D,; in all cases, we use consistent Gaussian quadrature. We use a passing interface package MPI4py~\cite{ 9439927, DALCIN20111124, DALCIN2008655, DALCIN20051108} for parallelization and solve nonlinear using SNES combined with BiCGStab for each linear system, including a Nonlinear Additive Schwarz methods (NASM) for parallel solution~\cite{ petsc2021}. Table~\ref{table:nonlin} summarizes the parameters we use.
\begin{table}[h]
\caption{Numerical parameters summarize} 
\centering 
\begin{tabular}{l c c} 
\hline\hline 
Description & Symbol & Value  \\ [0.5ex] 
\hline 
Max. iteration number for SNES & $\text{it}_{\text{max}}$ & $8$  \\ 
Relative tolerance for SNES & $\text{tol}$ &  $10^{-8}$ \\
Max. iteration number for Krylov & $\text{it}_{\text{Kr}}$ & $1000$  \\ 
Relative tolerance for Krylov & $\text{tol}_{\text{Kr}}$ &  $10^{-22}$ \\
Max. tolerance for time-adaptive scheme & $\text{tol}_{\text{max}}$ & $10^{-5}$   \\
Min. tolerance for time-adaptive scheme & $\text{tol}_{\text{min}}$ & $10^{-7}$   \\
Safety factor for time-adaptive scheme & $\rho_{tol}$ & $0.9$      \\
Relative scale factor for time-error computation & $\rho_{rel}$ &  $10^{2}$   \\
Absolute scale factor for time-error computation & $\rho_{abs}$ &  $10^{2}$ \\ [1ex] 
\hline 
\end{tabular}
\label{table:nonlin} 
\end{table}

We compute the time increment by gathering the unknown vectors at the master core, estimating the error for the current time step, adapting the time step size, and broadcasting its value to the other cores. This straightforward implementation is practical given the small number of processors used. We initialize the simulations using a first-order backward difference formula (BDF1) until $\bullet_{n-2}$ is different from null. We perform the simulations using a laptop with a 2.4 GHz processor with 8-core Intel Core i9 and 16 GB 2667 MHz DDR4 RAM, obtaining satisfactory results using a regular computer. Algorithm~\ref{Alg1Adap} sketches the time marching scheme, where $rank$ is the core number, and $master$ is the master core used as the communicator.   
\begin{algorithm} 
  \SetAlgoLined
  \KwData{$\bullet_{n}$, $\bullet_{n-1}$ , $\bullet_{n-2}$, $\dot{\bullet}_{n}$,  $\Delta t_{n+1}$, $\Delta t_{n}$, $\Delta t_{n-1}$}
  \KwResult{updated variables $\bullet_{n+1}$, $\dot{\bullet}_{n+1}$, $\Delta t_{n+1}$}
  Initialize $\bullet_{n} = \bullet_{0}$, with $\bullet = \xi, {\widetilde{\zeta}_{+}}, \phi$ \;
  \While{$t_{n+1} \leq t_f$}{
        Solve non-linear problem~\eqref{eq:algsystem} \;
        Gather $\Delta \boldsymbol{X}$ in master core and calculate temporal error ${\tau}^{\text{BDF2}}\left(t_{n} + \Delta t_{n+1}\right)$ with~\eqref{eq:local} \;
        \If{$\text{rank}$ == $\text{master}$}{
          Calculate $E_{u} \left(t_{n} + \Delta t_{n+1} \right)$ with~\eqref{eq:WLE} \;
          Take the maximum error with~\eqref{eq:WLE2} \;
          \eIf{$ E \left(t_{n} + \Delta t_{n+1} \right) \leq \text{tol}_{\text{max}}$}{
            Update current time step $t_{n+1} \leftarrow t_{n} + \Delta t_{n+1}$\;
            Update $\bullet_{n+1}$ and $\dot{\bullet}_{n+1}$ \;
            Update $\bullet_{n-2} \leftarrow \bullet_{n-1}$, $\bullet_{n-1} \leftarrow \bullet_{n}$, $\bullet_{n} \leftarrow \bullet_{n+1}$ and $\dot{\bullet}_{n} \leftarrow \dot{\bullet}_{n+1}$\;
            \If{$E \left(t_{n} + \Delta t_{n+1} \right) < \text{tol}_{\text{min}}$}{
              Increase delta step for next time increment $\Delta t_{n+1} \leftarrow  \boldsymbol{F} \left( E \left(t_{n+1}\right) ,\Delta t_{n+1}, \text{tol}_{\text{min}} \right)$ using~\eqref{eq:DT} \;}
              Broadcast to rest of ranks all variables $\Delta t_{n+1}$, $\bullet_{n-2}$, $\bullet_{n-1}$ and $\bullet_{n}$ \;
          }{
            Reduce time-step size$\Delta t_{n+1} \leftarrow \boldsymbol{F} \left( E \left(t_{n+1}\right) ,\Delta t_{n+1}, \text{tol}_{\text{max}} \right)$ using~\eqref{eq:DT} \;
            Broadcast to rest of ranks only $\Delta t_{n+1}$ \;
          }
          }
  }
  \caption{Dendrite growth model based on phase-field theory with time adaptivity}
  \label{Alg1Adap}
\end{algorithm}

\section{Simulations of lithium-metal dendrite formation}
\label{section:NumExp}

We present several studies that evaluate the performance of the proposed electrodeposition model for metal anode battery simulations. These studies consist of 2D and 3D simulations of dendritic electrodeposition of lithium metal during battery charge status.

\subsection{System layout and properties}

\begin{figure}
\begin{subfigure}[b]{\linewidth}
    \centering%
	{\includegraphics[height = 6.5cm]{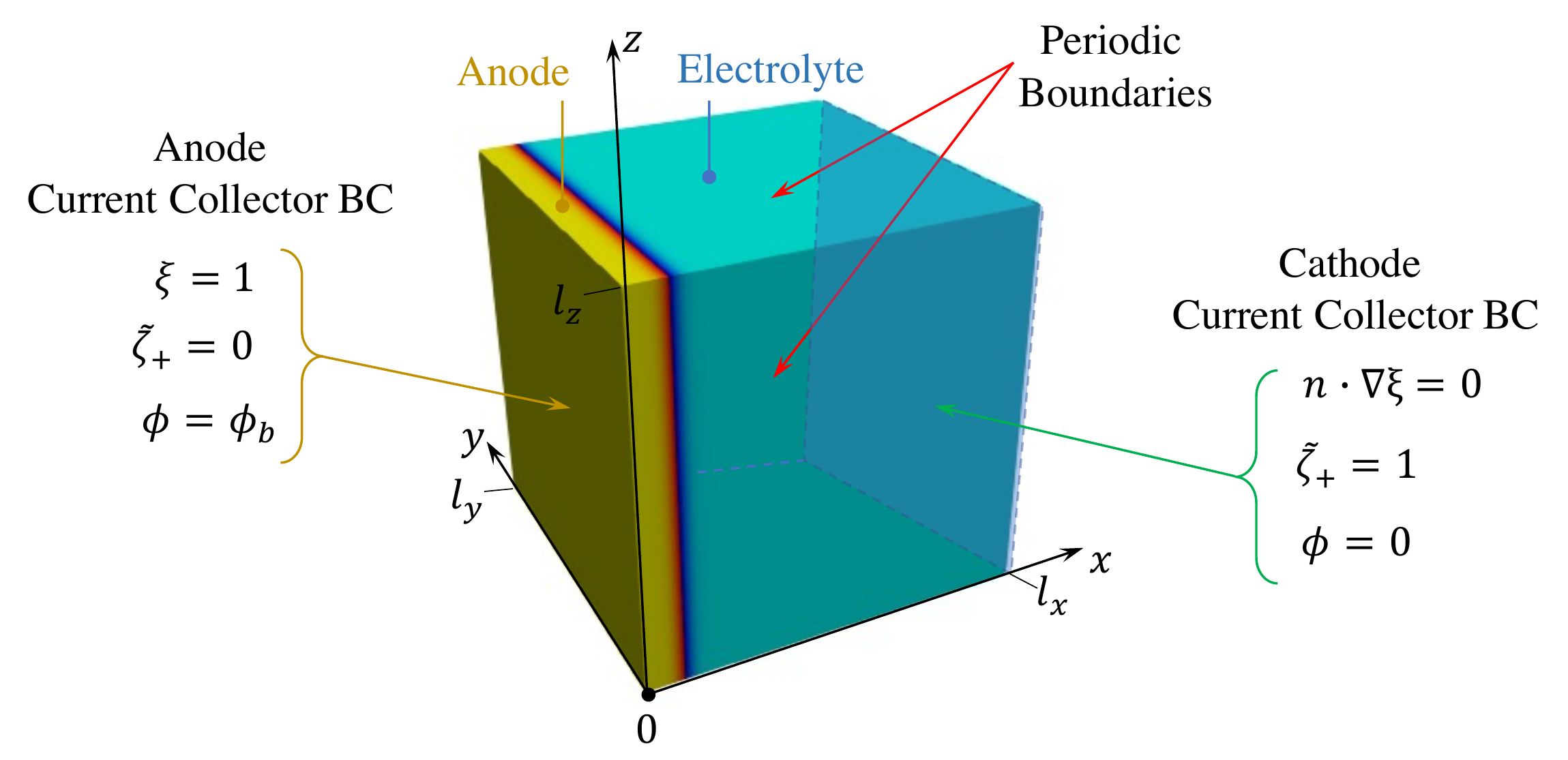}}
	\caption{Planar electrode.}
	\label{fig:BC_Planar3D}
\end{subfigure} 
\bigbreak
\begin{subfigure}[b]{\linewidth}
    \centering%
    {\includegraphics[height = 6.5cm]{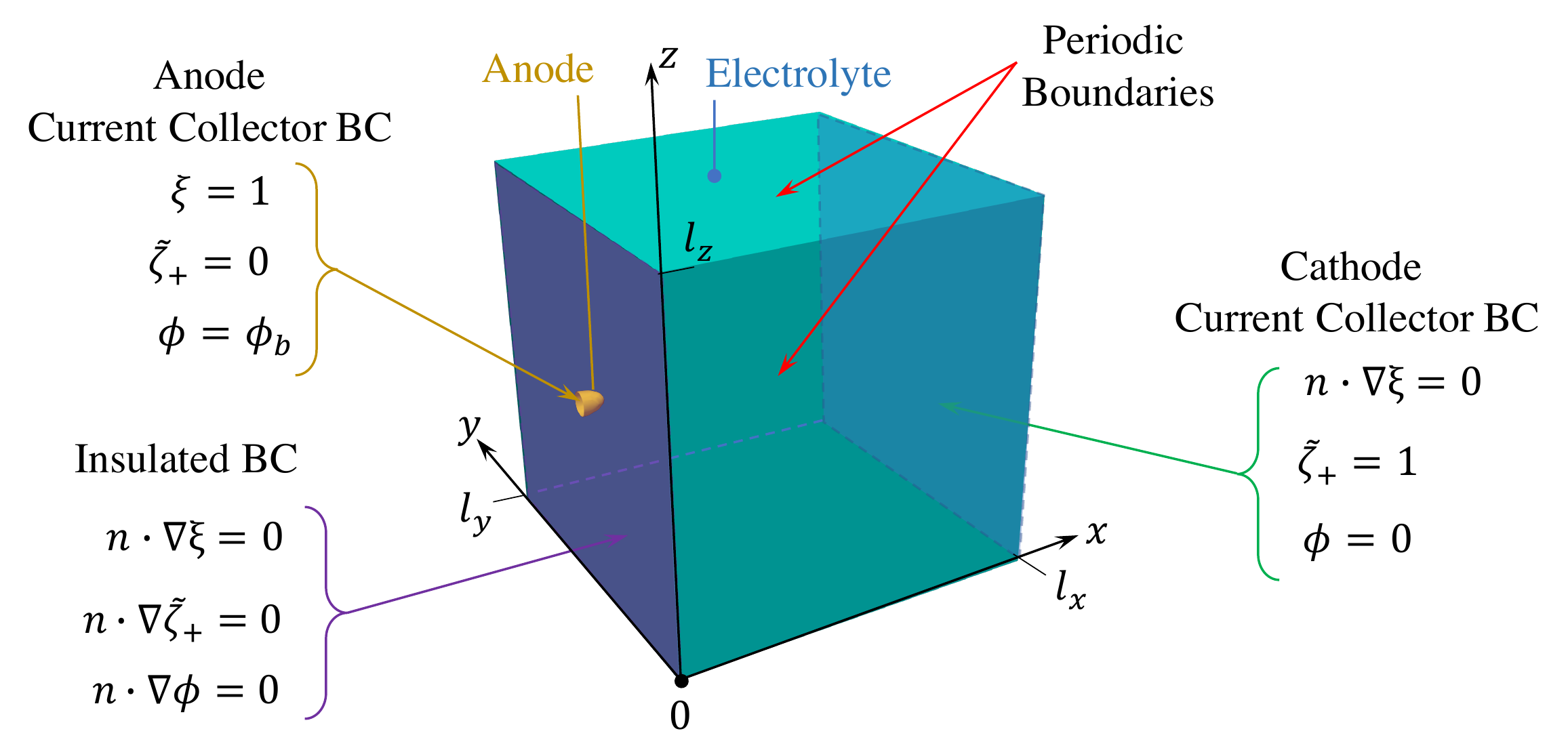}}
    \caption{Artificial nucleation.}
    \label{fig:BC_Seed3D}
\end{subfigure}
\caption{Boundary conditions for planar electrode (\subref{fig:BC_Planar3D}) and artificial nucleation (\subref{fig:BC_Seed3D}) simulations.}
\label{fig:3DSystemLayout_BC}
\end{figure}

Generally, the computational domain for a battery simulation comprises the anode and cathode regions as well as the space between the electrodes filled with electrolyte~\cite{ Trembacki_2019}. However, in most phase-field simulations of metal electrodeposition, including those we perform, the cathode region reduces to a current collector boundary condition on the electrolyte side of the domain (see Figure~\ref{fig:3DSystemLayout_BC}). We model a battery cell with a traditional sandwich architecture, undergoing a recharging process under fixed applied electro potential. We represent this cell as an $l_x\times l_y$ rectangular domain in 2D and as an $l_x\times l_y\times l_z$ hexagonal domain in 3D. The initial perturbation (dendrite nuclei) significantly impacts the simulation outcome. These nuclei are usually part of the problem's initial conditions. Some models include initial seeds (artificial deformation on the electrode surface) combined with surface anisotropy to propagate the dendrite shape~\cite{ OKAJIMA2010118, doi:10.1063/1.4905341, JANA2017552}. Alternatively, other models combine surface anisotropy with Langevin noise at the phase-field interface to generate a random nucleation pattern~\cite{ PhysRevE.92.011301, YURKIV2018609, MU2019100921, Ruurds84}. In our experience, the shape of the initial seed and the noise levels have a major influence on the resulting dendrite morphology.

We study two different initial structures. The first structure consists of a 5$\mu m$-thick metal anode ($\text{l}_{0}=5\left[\mu m\right]$), made up of pure Lithium, separated from the liquid electrolyte by a smooth interface, as  Figure~\ref{fig:3DSystemLayout_BC}-\subref{fig:BC_Planar3D} depicts. The initial condition drives from the equilibrium solution for a one-dimensional transition zone between solid ($\xi=1$) and liquid ($\xi=0$), where our variables $\left(\xi, \widetilde{\zeta}_{+}, \phi \right)$ vary in the “$x$” spatial direction according to $\xi\left(x\right)=\frac{1}{2}\left[1-\text{tanh}\left(x\sqrt{\frac{W}{2\kappa_{0}}}\right)\right]$~\cite{ doi:10.1146/annurev.matsci.32.101901.155803}. 

In the second structure we assume that artificial nucleation regions, ellipsoidal protrusions (seeds) with semi-axes $r_x$, $r_y$, $r_z$, and center $\left(l_{0_x},l_{0_y},l_{0_z}\right)$, exist at the surface of the anode undergoing electrodeposition (see Figure~\ref{fig:3DSystemLayout_BC}-\subref{fig:BC_Seed3D}). This is a widely-used strategy in phase-field simulations of electrodeposition~\cite{ Zhang_2014, CHEN2015376, YURKIV2018609, MU2019100921}; since it reduces the computational cost as the lithium metal is only able to electrodeposit on the same nuclei, enhancing dendrite growth and allowing for detailed study of its morphology. For the artificial nucleation case, we modify the initial condition formula, replacing “$x$” by $\left[\left(\frac{x-l_{0_x}}{r_x}\right)^2+\left(\frac{y-l_{0_y}}{r_y}\right)^2+\left(\frac{z-l_{0_z}}{r_z}\right)^2-1\right]$ within the hyperbolic tangent argument, to account for a smooth transition between the solid seed (lithium metal anode) and the surrounding liquid electrolyte region.

We assume the cell's electrolyte to be 1M $\text{LiPF}_\text{6}$ EC/DMC 1:1 volume ratio solution~\cite{ doi:10.1021/acsenergylett.8b01009}; we compute the electrolyte's site density ($C_m^l$) using its density ($1.3 \left[g/cm^3 \right]$) and molar mass ($90 \left[g/mol \right]$) and electrode's site density $C_m^s$ using the density ($0.534 \left[g/cm^3 \right]$) and molar mass ($6.941 \left[g/mol\right]$) of pure Lithium~\cite{ doi:10.1021/acsenergylett.8b01009}.

We model the anode boundary as a Dirichlet boundary condition $\xi=1$ at $x=0$ for the phase-field parameter (solid electrode phase). In contrast, we use a no-flux Neumann boundary condition at $x=l_x$ (cathode) to allow the electrodeposition process to take place ($\xi$ changing from 0 to 1) when the reaction front approaches the short-circuit condition~\cite{ Arguello2022}.
For the Li-ion concentration, we apply Dirichlet boundary conditions setting $\widetilde{\zeta}_{+}=0$ and $\widetilde{\zeta}_{+}=1$ at the anode ($x=0$) and cathode ($x=l_x$) boundaries, respectively. These boundary conditions allow the Li-ion flux into the battery (electrolyte side), ensuring that the Li deposited at the electrode-electrolyte interface equals the amount of $Li^+$ supplied to the electrolyte. Thus, these boundary conditions avoid the quick Li-ion depletion and keep the electrodeposition process running for the full simulation~\cite{ Arguello2022}. 
Finally, we apply periodic boundary conditions on the remaining faces; these are $\left(x,0,z\right)$, $\left(x,l_y,z\right)$, $\left(x,y,0\right)$, and $\left(x,y,l_z\right)$. Figure ~\ref{fig:3DSystemLayout_BC} summarizes the boundary conditions we apply.

Additionally, Table ~\ref{table:material_parameters} presents the phase-field model parameters. The normalization constants for length, time, energy and concentration scales are set as $h_0=1\left[\mu m\right]$, $t_0=1\left[s\right]$, $E_0=2.5\times10^6 \left[J/m^3\right]$, and $C_0=1\times10^3\left[mol/m^3\right]$, respectively~\cite{ doi:10.1021/acsenergylett.8b01009}. Table ~\ref{table:material_parameters} also shows that although the electrode and electrolyte materials can exhibit $Li$/$Li^+$ dependent conductivities and diffusivities, their values are set to constants across each phase for simplicity~\cite{ doi:10.1021/acsenergylett.8b01009, PhysRevE.92.011301}.

\begin{table}[h]
\caption{Simulation Parameters} 
\centering 
\begin{tabular}{l c c c c} 
\hline\hline 
Description & Symbol & Real Value & Normalized & Source  \\ [0.5ex] 
\hline 
Exc. current density & $\text{i}_{0}$ & $30\left[A/m^2\right]$ & $30$ &~\cite{ Monroe_2003}  \\ 
Surface tension & $\gamma$ & $0.556\left[J/m^2\right]$ & $0.22$ &~\cite{ VITOS1998186,tran2016surface}\\
Phase-field interface thickness & $\delta_{PF}$ & $1.5\times10^{-6}\left[m\right]$ & $1.5$ &~\cite{ Arguello2022}\\
Barrier height & $W$ & $W=\frac{12\gamma}{\delta_{PF}}=4.45\times10^{6}\left[J/m^3\right]$ & $1.78$ & computed  \\ 
Gradient energy coefficient & $\kappa_{0}$ & $\kappa_{0}=\frac{3\gamma\delta_{PF}}{2}=1.25\times10^{-6}\left[J/m\right]$ & $0.5$ & computed  \\
Anisotropy strength & $\delta_{aniso}$ & $0.044$ & $0.044$ &~\cite{ tran2016surface,TRAN201948}  \\
Anisotropy mode & $\omega$ & $4$ & $4$ &~\cite{ TRAN201948,ZHENG202040}  \\
Kinetic coefficient  & $L_{\eta}$ & $L_{\eta}=\text{i}_{0}\frac{\gamma}{\text{nF}\text{C}_m^s}=1.81\times10^{-3}\left[1/s\right]$ & $1.81\times10^{-3}$ & computed    \\
Site density electrode & $\text{C}_m^s$ & $7.64\times10^{4}\left[mol/m^3\right]$ & $76.4$ &~\cite{ doi:10.1021/acsenergylett.8b01009} \\
Bulk Li-ion concentration & $\text{C}_{0}$ & $10^{3}\left[mol/m^3\right]$ & $1$ & computed  \\
Conductivity electrode & $\sigma^s$ & $10^{7}\left[S/m\right]$ & $10^{7}$ &~\cite{ CHEN2015376} \\
Conductivity electrolyte & $\sigma^l$ & $1.19\left[S/m\right]$ & $1.19$ &~\cite{ Valo_en_2005}   \\
Diffusivity electrode & $D^s$ & $7.5\times10^{-13}\left[m^2/s\right]$ & $0.75$ &~\cite{ CHEN2015376}   \\
Diffusivity electrolyte & $D^l$ & $3.197\times10^{-10}\left[m^2/s\right]$ & $319.7$ &~\cite{ Valo_en_2005}  \\ [1ex] 
\hline 
\end{tabular}
\label{table:material_parameters} 
\end{table}

\subsection{Lithium dendrite propagation rate: Consistency with experimental data}
\label{subsection:Validation}

Validation of phase-field models with experimental results is a well-known challenge of electrodeposition simulations~\cite{ PhysRevE.92.011301}; partly due to the computational cost of simulating the detailed lithium electrodeposition at the whole-cell scale~\cite{ YURKIV2018609}. One limiting factor is the domain size, defined by the inter-electrode separation distance in experimental cells, which ranges from 1 to 10mm~\cite{ NISHIKAWA201184, Nishikawa_2019, YUFIT2019485}.

We perform 2D simulations to directly compare our simulations against experimental results obtained from thin-cell geometries. We use 2D simulation results to validate the predicted lithium dendrite propagation rates, assuming that the electrodeposition rates are not significantly affected by the problem's size. 

We achieve large simulation sizes through several computational efficiency improvements. We use a sigmoid interpolating function~\cite{ Arguello2022} instead of the popular polynomial interpolating function~\cite{ Zhang_2014, doi:10.1063/1.4905341, CHEN2015376}. We use time-step size adaptivity and mesh mapping in the region of interest (close to the electrode-electrolyte interface). We also use parallel solution strategies and carefully select the phase-field interface thickness ($\delta_{PF}$), given the interface thickness effect on the electrodeposition rate, which ultimately determines the evolution dynamics (motion) of the lithium electrodeposits~\cite{ Arguello2022}.

We compare the simulated lithium dendrite propagation rate against the experimental measurements of Nishikawa et al.~\cite{ NISHIKAWA201184}. Nishikawa et al. studied the dendritic electrodeposition of lithium as a function of total charge (electric current passed over time) using microscope observations. They reported dendrite growth rates of $0.01-0.06\left[\mu m s^{-1}\right]$~\cite{ AKOLKAR201484} using a 1M $\text{LiPF}_\text{6}$ electrolyte, establishing upper and lower bounds for comparison with our simulation results. The wide range of values reported in many studies~\cite{ Crowther_2008, NISHIKAWA201184, Nishikawa_2010} may be caused by the uneven current distribution on the electrode surface~\cite{ NISHIKAWA201184}.

\begin{figure}[h!]
\begin{subfigure}[b]{0.9\linewidth}
    \centering%
	{\includegraphics[height = 6.5cm]{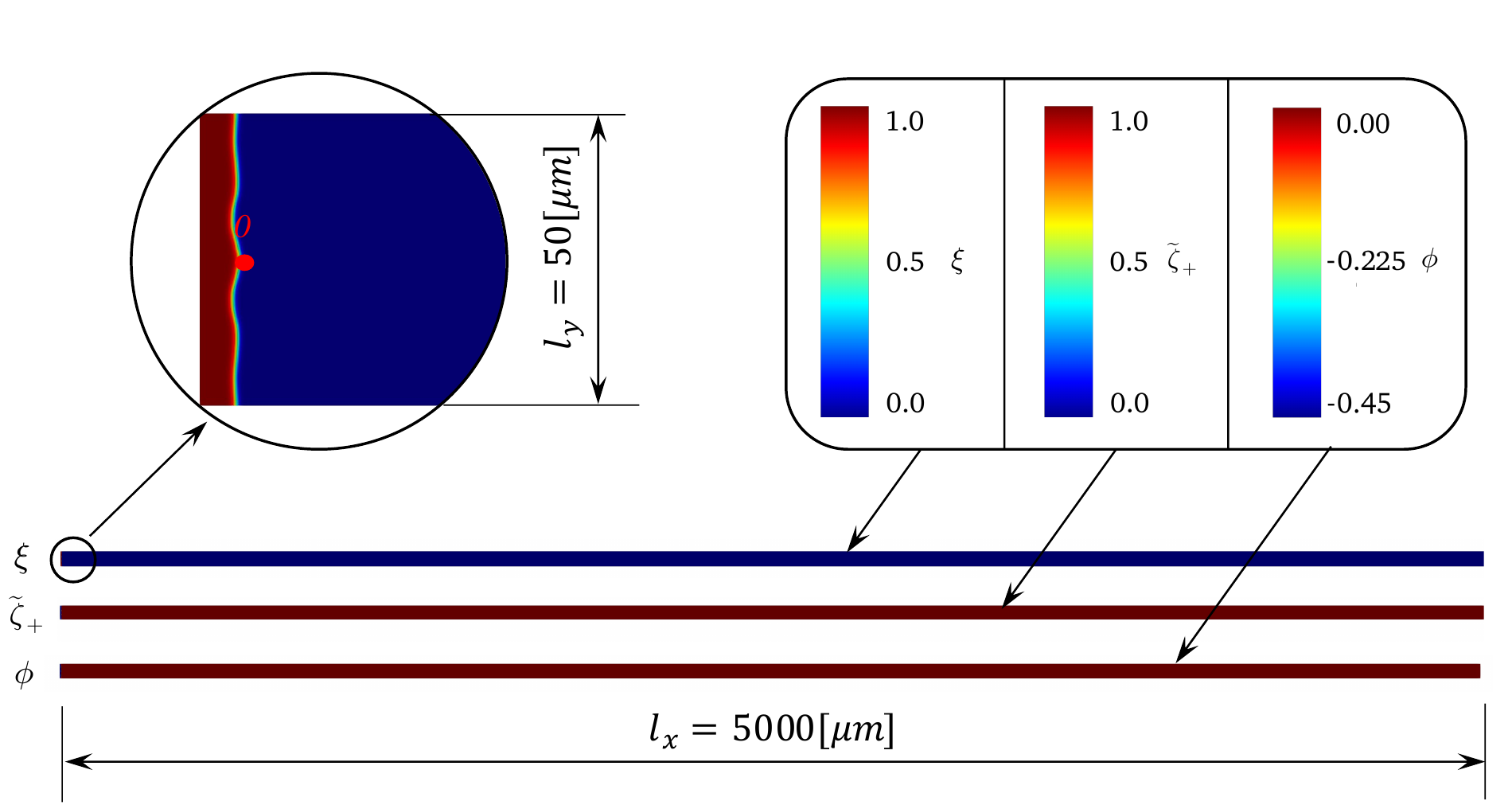}}
	\caption{$t = 0\left[s\right]$.}
\end{subfigure} 
\bigbreak
\begin{subfigure}[b]{0.9\linewidth}
    \centering%
    {\includegraphics[height = 5.8cm]{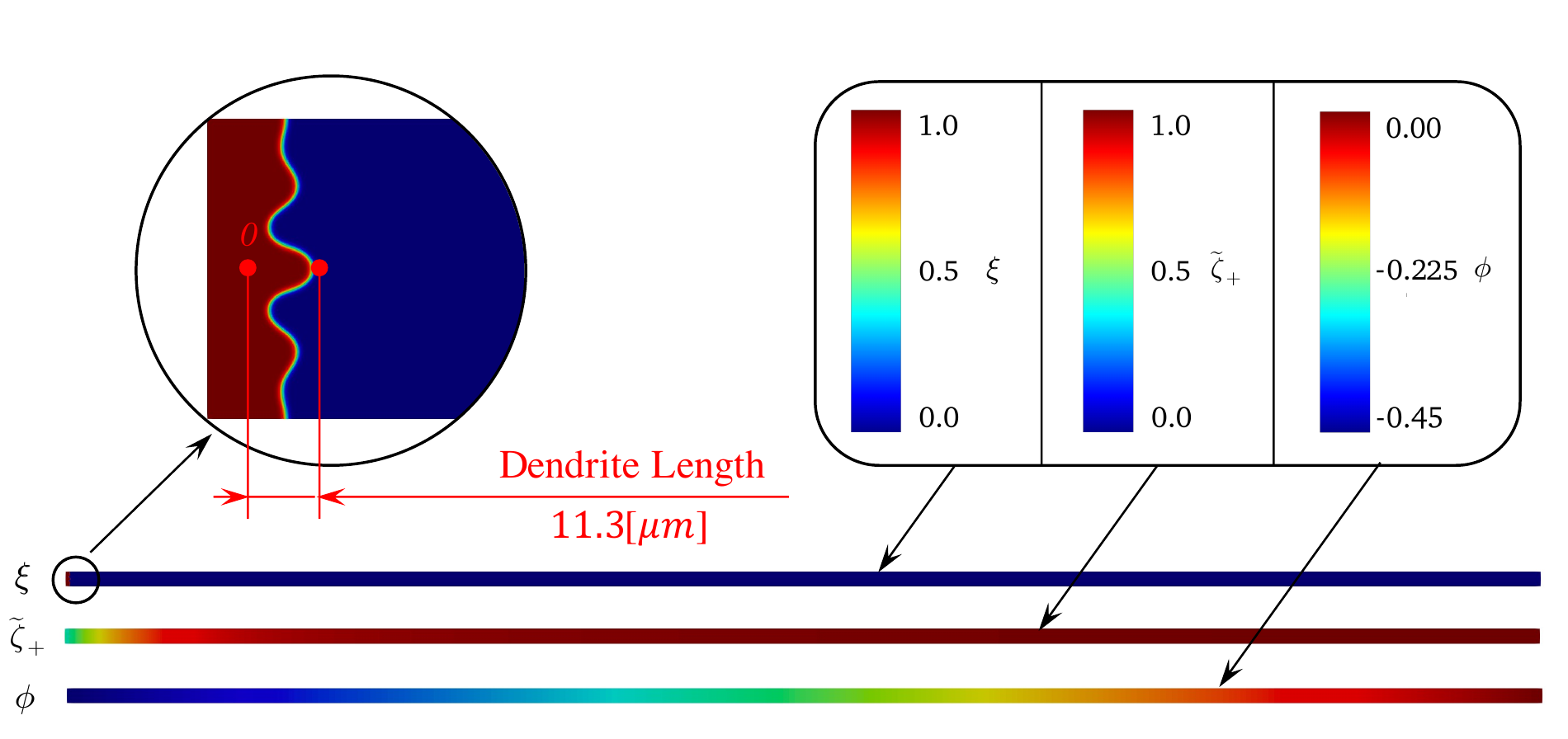}}
    \caption{$t = 400\left[s\right]$.}
\end{subfigure}
\bigbreak
\begin{subfigure}[b]{0.9\linewidth}
    \centering%
    {\includegraphics[height = 5.8cm]{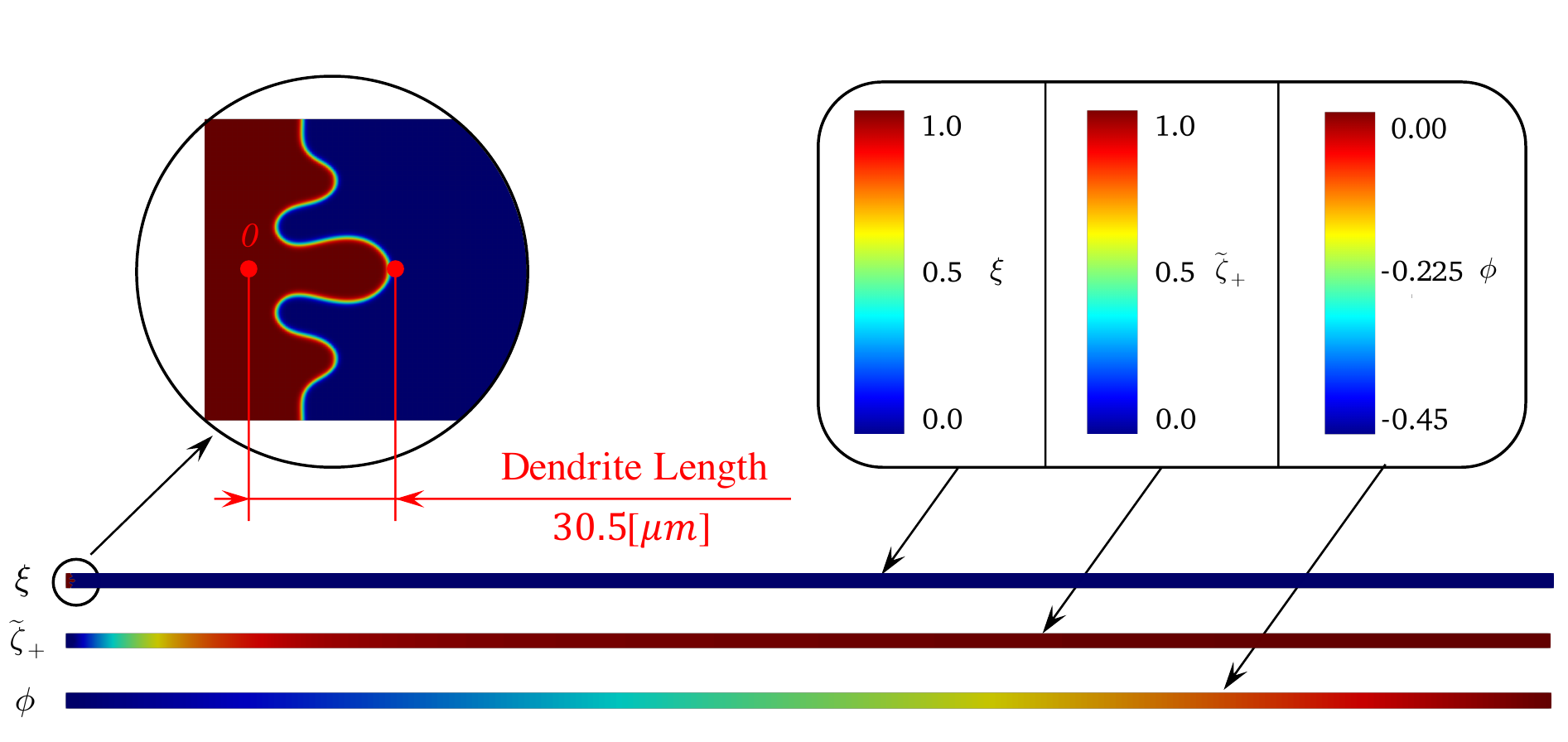}}
    \caption{$t = 800\left[s\right]$.}
\end{subfigure}
\caption{2D phase-field simulation of lithium electrodeposition process at a current density of $i=10\left[mA/cm^2\right]$ with a distance of $l_x=5000\left[\mu m\right]$ between electrodes.}
\label{fig:Expriment_2D_DendriteLength}
\end{figure}

We simulate a lithium electrodeposition process at an applied current density of $i=10\left[mA/cm^2\right]$~\cite{ NISHIKAWA201184}. We use a 2D rectangular domain $l_x=5000\left[\mu m\right]$ long (distance betweenelectrodes), and $l_y=50\left[\mu m\right]$ wide (sufficient width to allow dendrite formation). We use a mesh $8,000\times140$ elements, with mesh distribution that guarantees an $x$-spatial resolution of approximately $0.35\left[\mu m\right]$ ($h<4\delta_{PF}$) in the region of interest of the domain with square elements. We apply a charging electro potential of $\phi_b=-0.45\left[V\right]$ to the cell (a common value in the literature~\cite{ CHEN2015376, doi:10.1021/acsenergylett.8b01009, PhysRevE.92.011301}), well above the threshold overpotential for dendrite growth $\phi_b=-0.37\left[V\right]$~\cite{ doi:10.1021/acsenergylett.8b01009}. We perturb the initial electrode surface by ($\pm 0.5\left[\mu m\right]$) in $x$-direction according to $0.4\sin\left(0.44y+\pi\right)e^{-0.004\left(y-0.5l_y\right)^2}$, consisting of a periodic function combined with a Gaussian distribution to promote dendrite formation at the center of the domain ($\frac{l_y}{2}=25\left[\mu m\right]$).

\begin{figure} [h!]
    \centering%
{\includegraphics[height = 8cm]{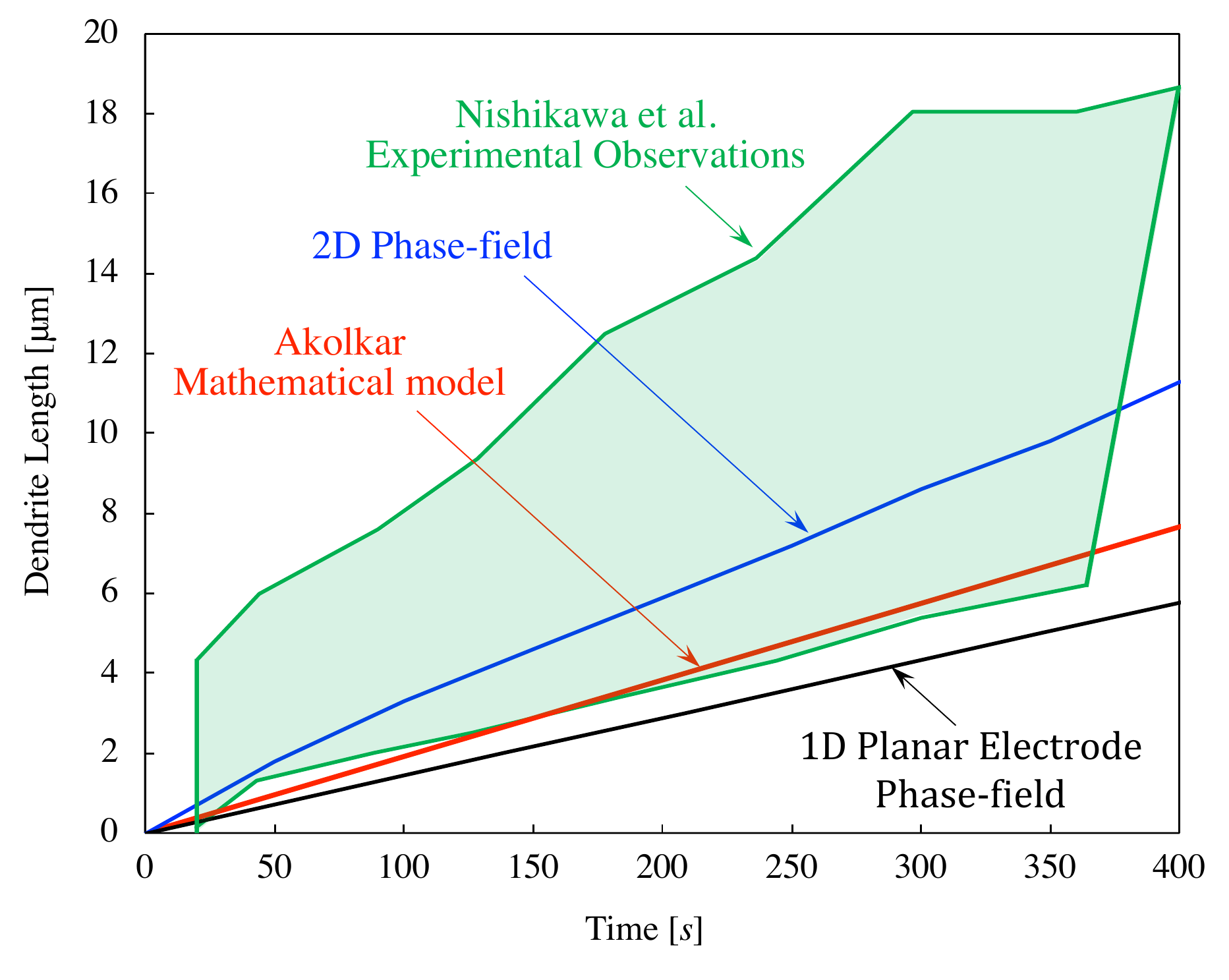}}
\caption{Comparison between 2D phase-field simulations of lithium dendrite growth at a current density of $i=10\left[mA/cm^2\right]$ (blue), experimental data taken from  Nishikawa et al.~\cite{ NISHIKAWA201184} (green), and the Akolkar analytical model predictions~\cite{ AKOLKAR201484} (red). 1D planar electrode phase field simulation for reference (black).}
\label{fig:DendriteLenghtPlot_2D}
\end{figure}

Figure~\ref{fig:Expriment_2D_DendriteLength} shows the initial conditions ($t=0 \left[s\right]$) and the evolution of the system’s variables $\left(\xi, \widetilde{\zeta}_{+}, \phi \right)$. We zoom in the phase-field variable $\xi$ at each instant to emphasize the growth of the lithium dendrite and show the instantaneous dendrite length measurement. Figure~\ref{fig:DendriteLenghtPlot_2D} shows the evolution of the simulated lithium dendrite length over time (blue). While in the range of experimental data (green), the results are sensitive to differences between the transport and kinetic properties and those of the experiments~\cite{ NISHIKAWA201184} and our simulator. Therefore, further well-controlled experimental studies, with detailed characterization of the system parameters, combined with modeling will be necessary to improve the correlation~\cite{ AKOLKAR201484}.

Additionally, Figure~\ref{fig:DendriteLenghtPlot_2D} includes analytical dendrite propagation results for the tip current density developed by Akolkar~\cite{ AKOLKAR201484} for comparison. Akolkar's model calculates the dendrite propagation rate by analyzing various overpotentials that develop at the dendrite tip and the flat electrode surface, assuming a current density of $i=10\left[mA/cm^2\right]$ applied and a constant radius of $r=1\left[\mu m\right]$ at the dendrite's tip. This simplification constitutes departs from our phase-field model, where the dendrite's tip radius is variable throughout the simulation. This difference may partly explain the higher growth rate obtained in our simulations. Furthermore, we also include 1D planar electrode phase-field results using an identical set-up as to the 2D simulation~\cite{ Arguello2022}. 

\subsection{2D Simulation of lithium dendrite growth}
\label{sc:2Dsim}

The following 2D numerical experiment compares our model's features to other phase-field simulations reported in the literature. This comparison verifies our work as a preliminary step to continue with a series of larger and more complex 3D simulations of lithium dendrite growth in metal anode batteries latter in this paper.

\begin{figure}[h!]
\begin{subfigure}[b]{0.2\linewidth}
    \centering%
	{\includegraphics[height = 3.2cm]{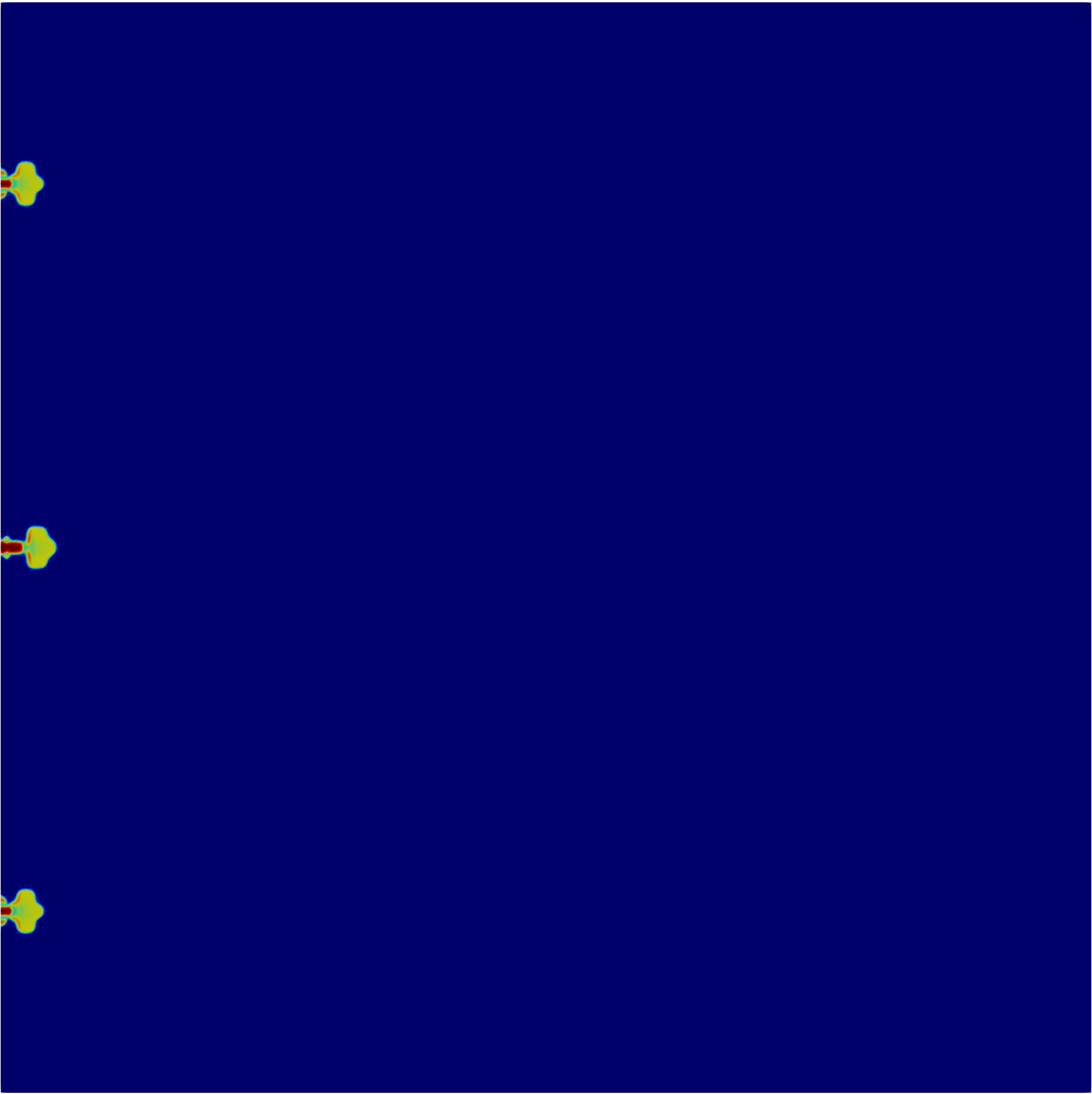}}
\end{subfigure} 
\begin{subfigure}[b]{0.2\linewidth}
    \centering%
    {\includegraphics[height = 3.2cm]{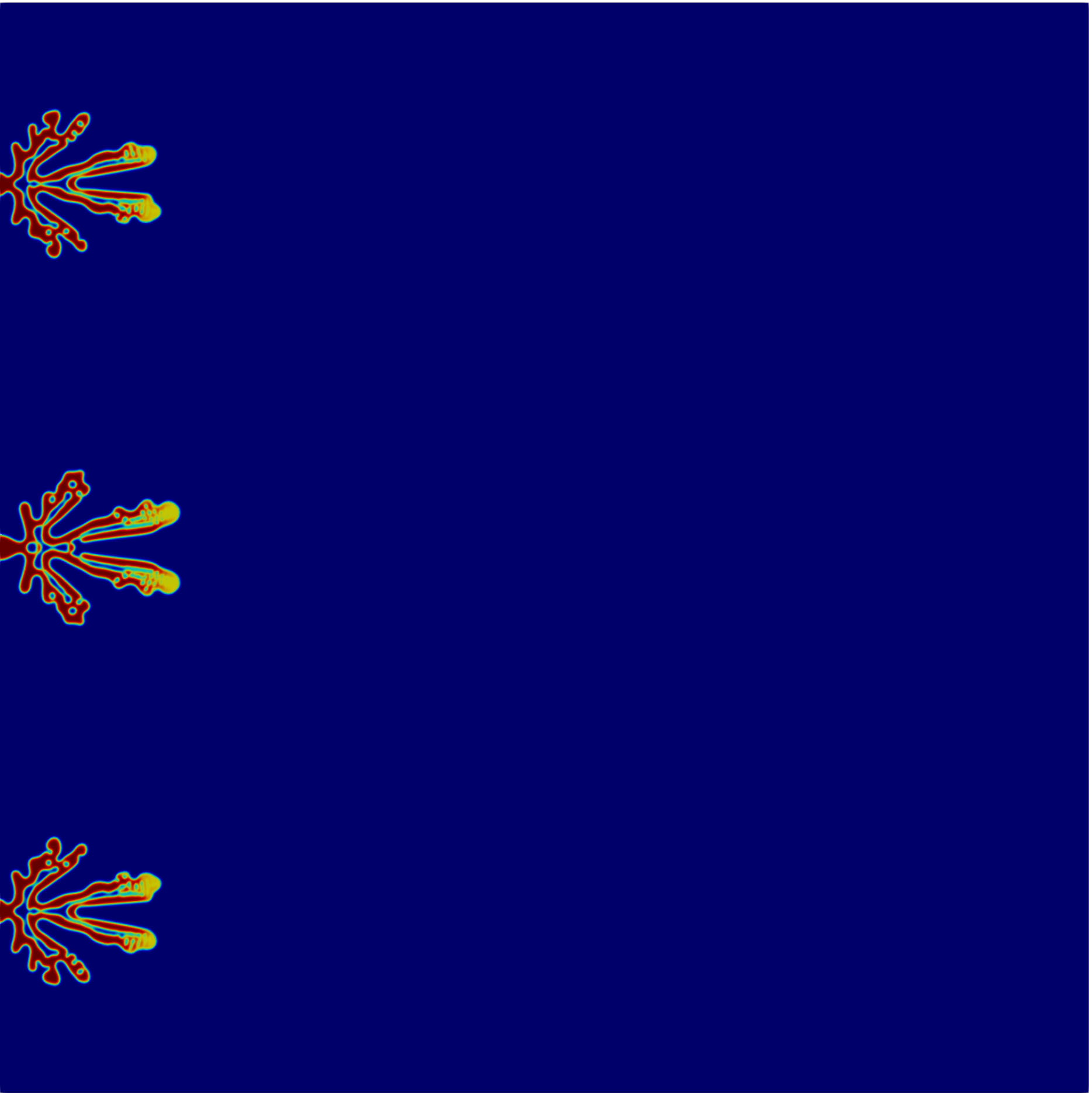}}
\end{subfigure}
\begin{subfigure}[b]{0.2\linewidth}
    \centering%
    {\includegraphics[height = 3.2cm]{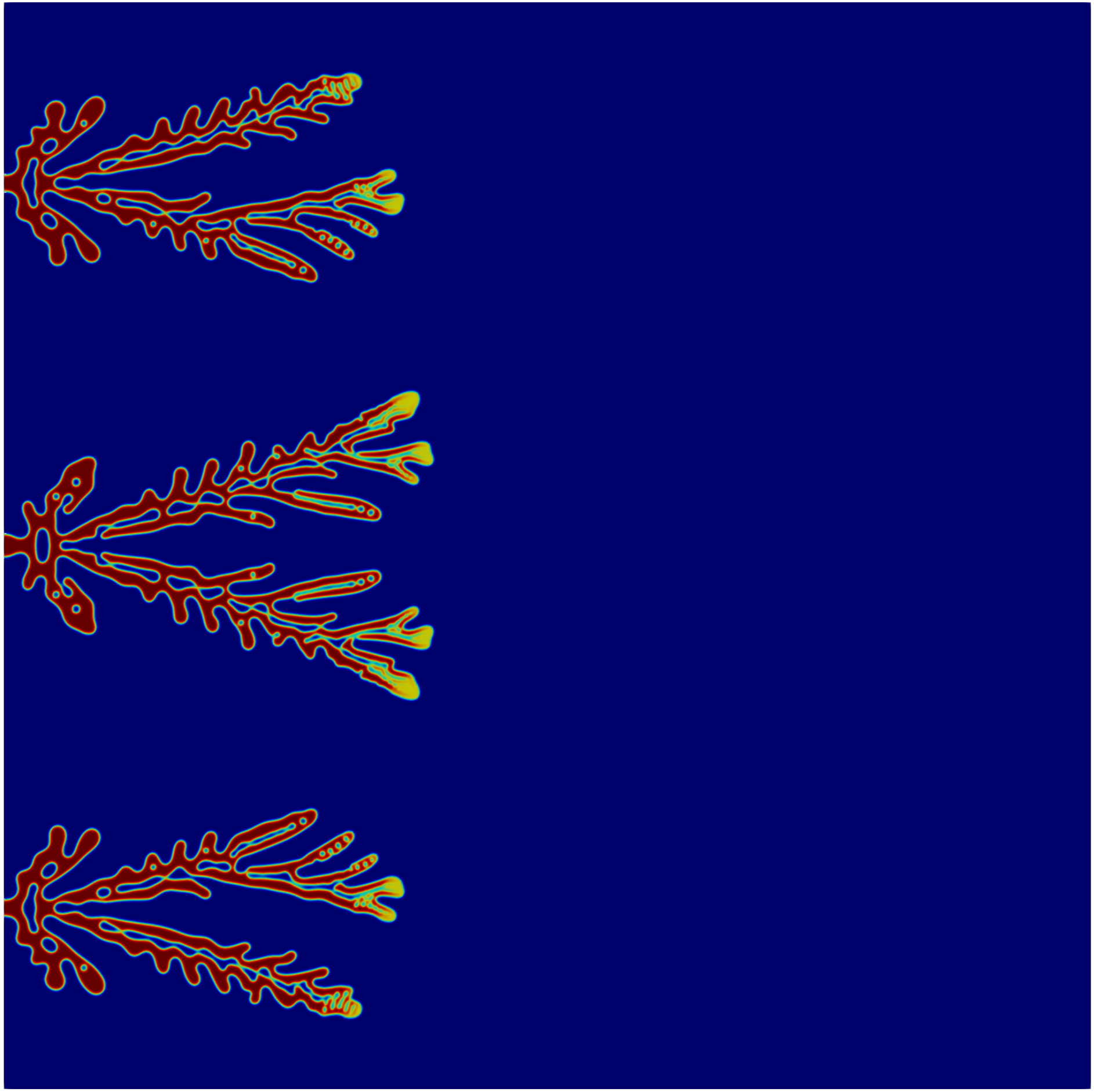}}
\end{subfigure}
\begin{subfigure}[b]{0.2\linewidth}
    \centering%
    {\includegraphics[height = 3.2cm]{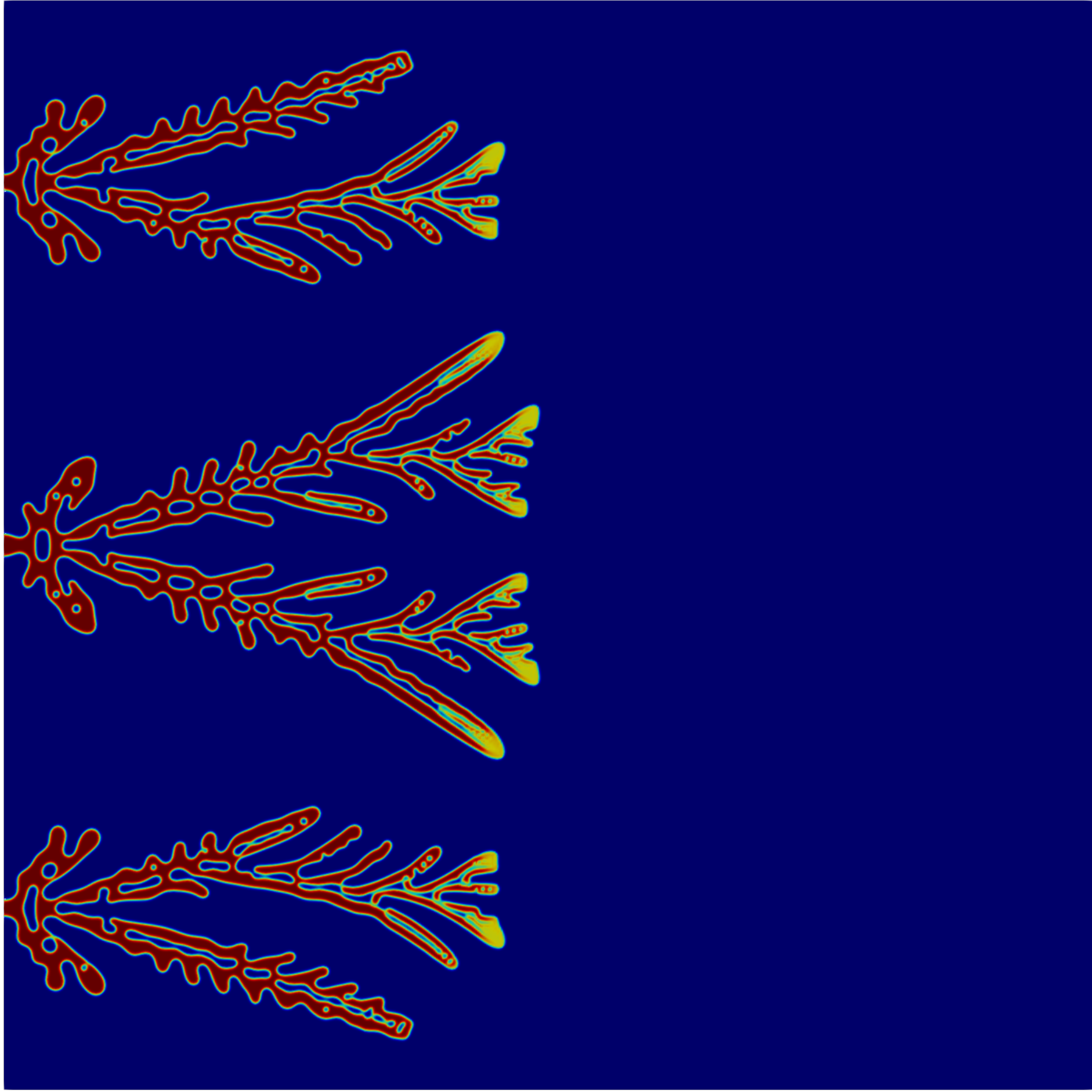}}
\end{subfigure}
\begin{subfigure}[b]{0.1\linewidth}
    \centering%
    {\includegraphics[height = 3.2cm]{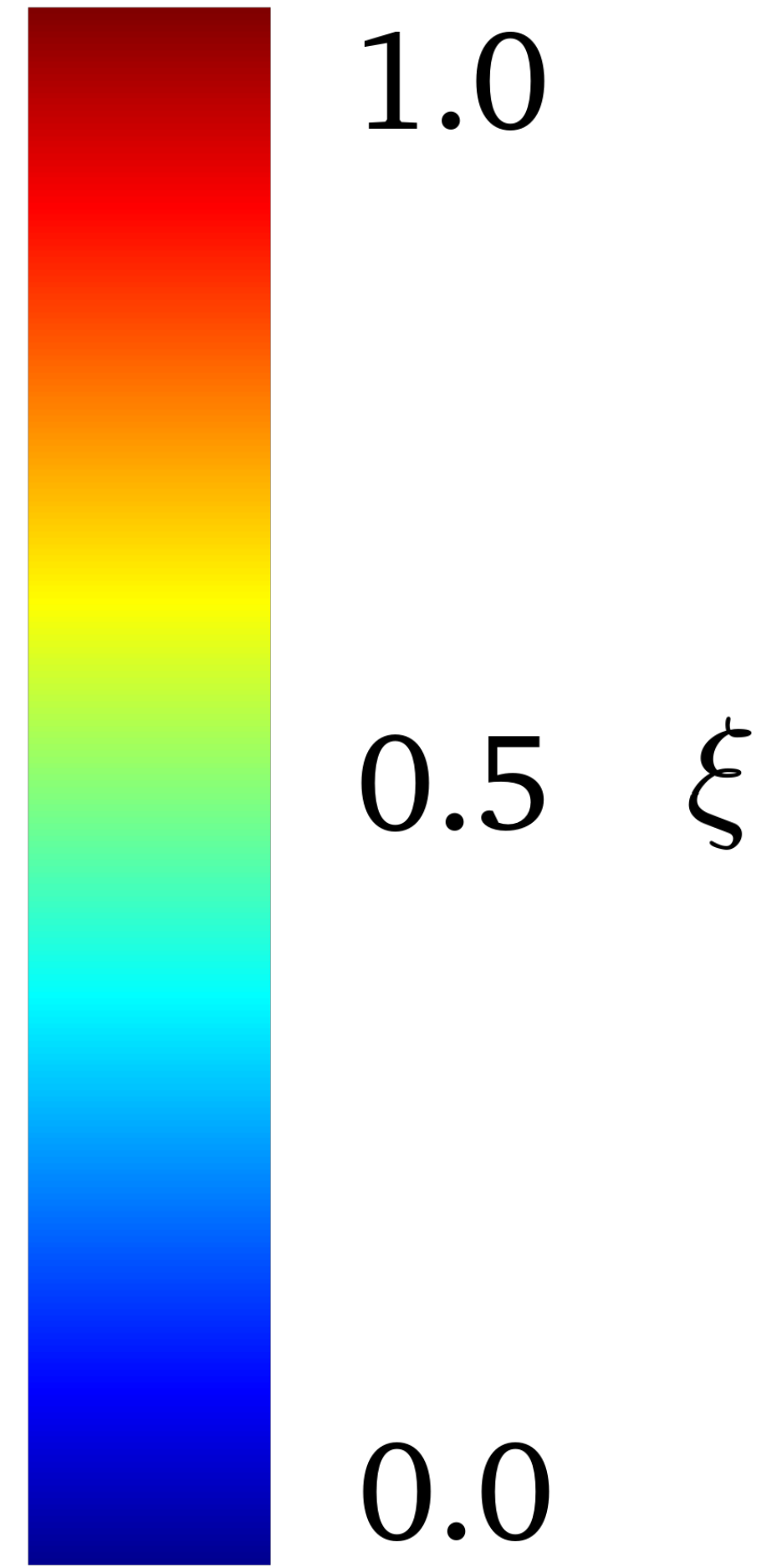}}
\end{subfigure}
\bigbreak
\begin{subfigure}[b]{0.2\linewidth}
    \centering%
	{\includegraphics[height = 3.2cm]{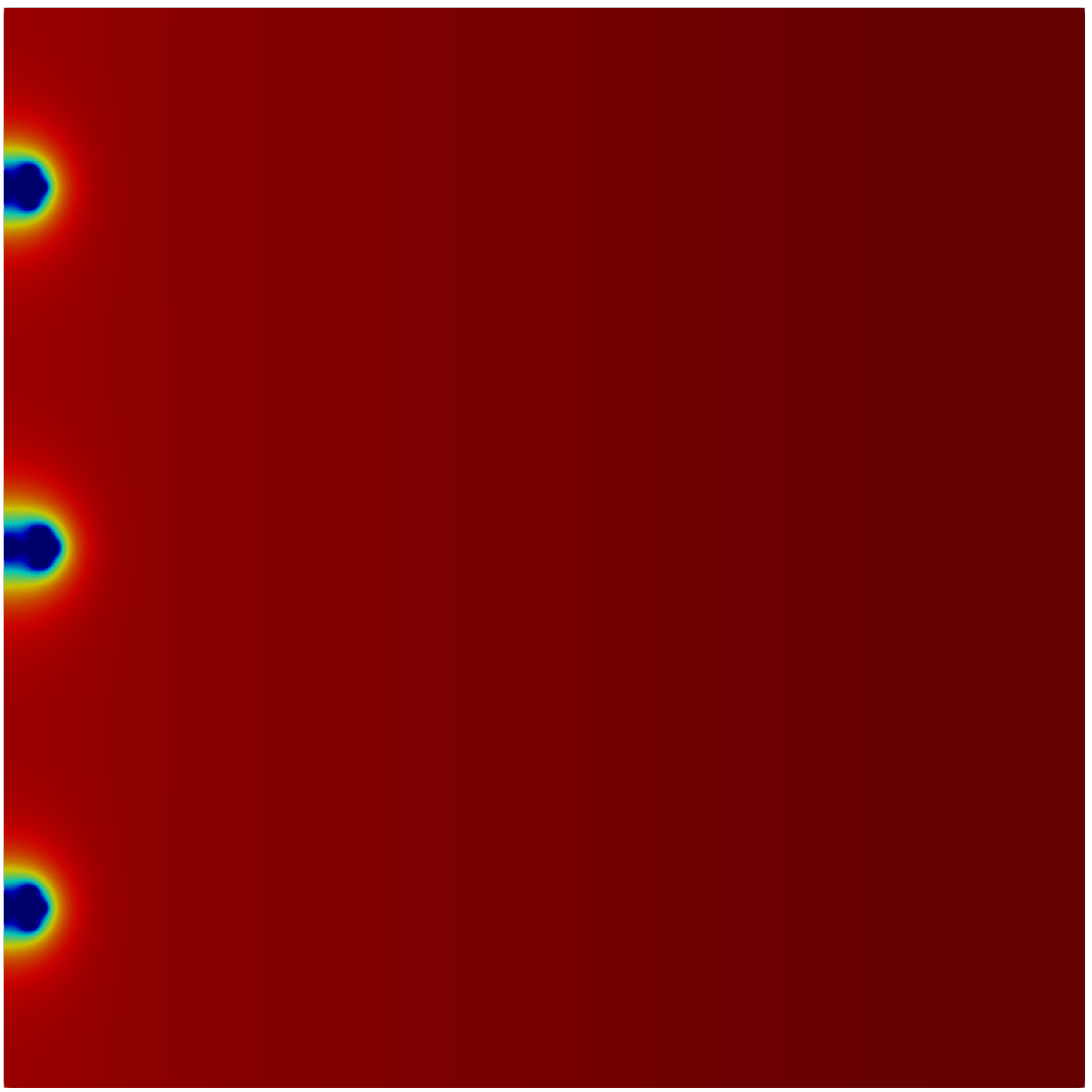}}
\end{subfigure} 
\begin{subfigure}[b]{0.2\linewidth}
    \centering%
    {\includegraphics[height = 3.2cm]{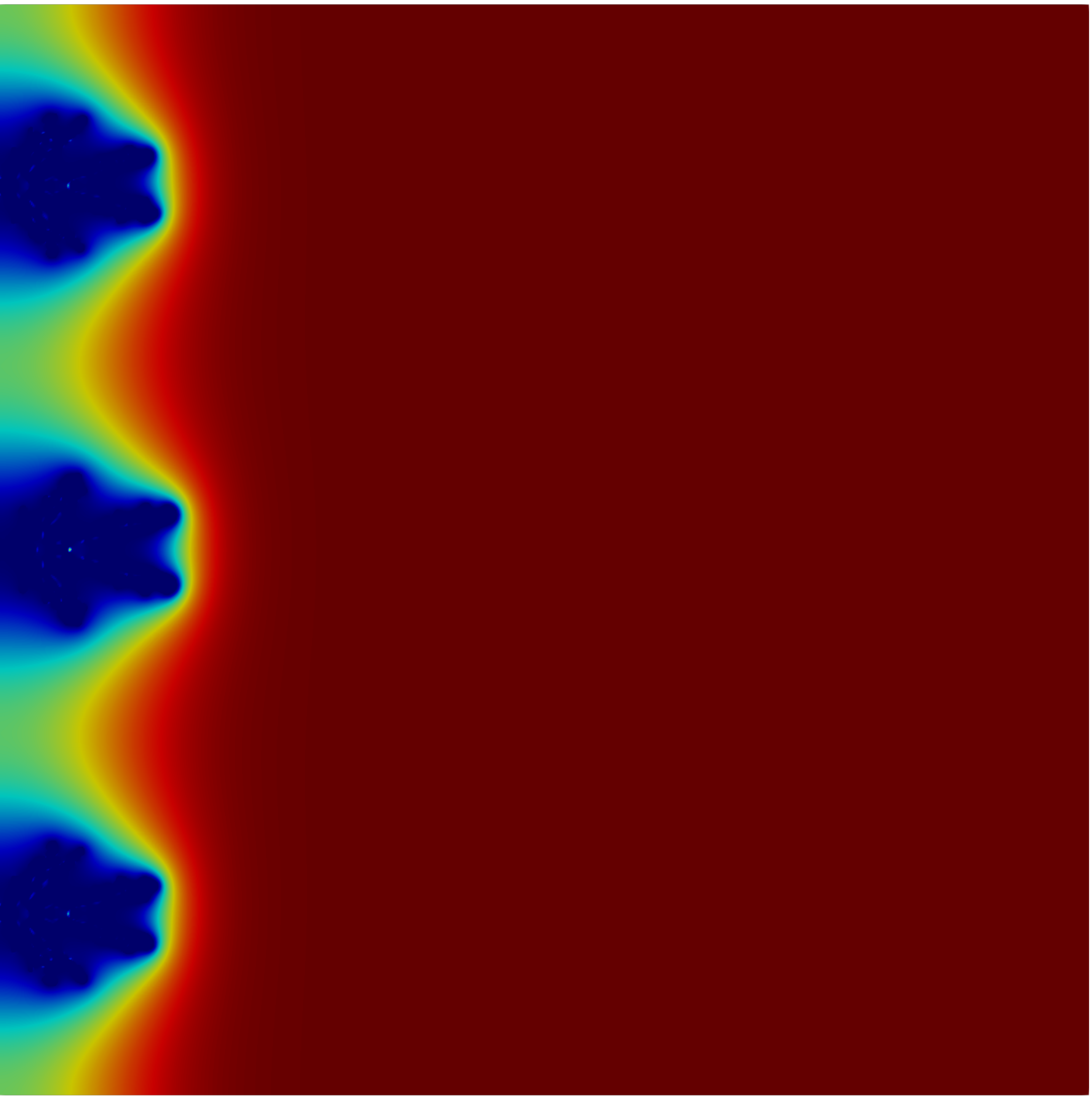}}
\end{subfigure}
\begin{subfigure}[b]{0.2\linewidth}
    \centering%
    {\includegraphics[height = 3.2cm]{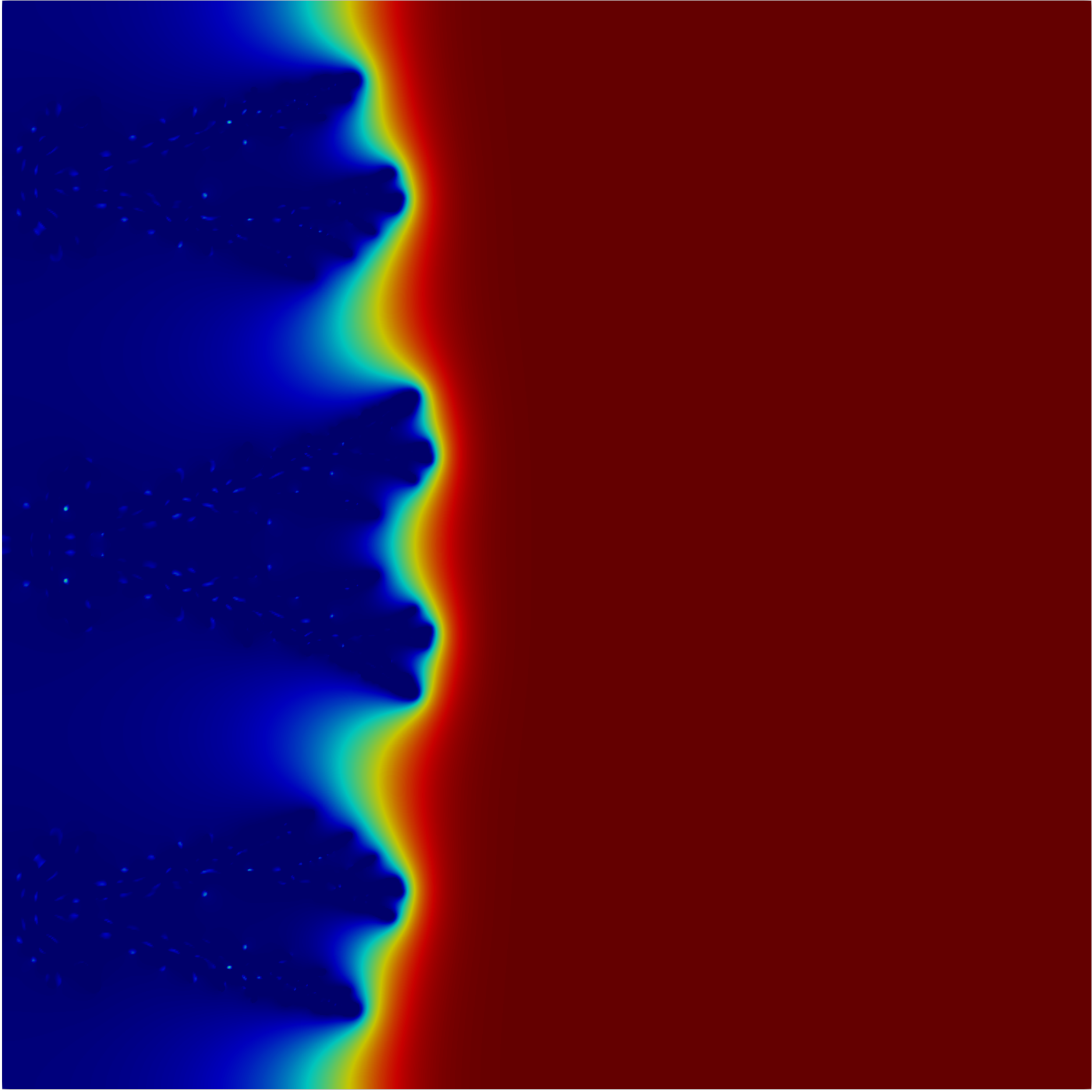}}
\end{subfigure}
\begin{subfigure}[b]{0.2\linewidth}
    \centering%
    {\includegraphics[height = 3.2cm]{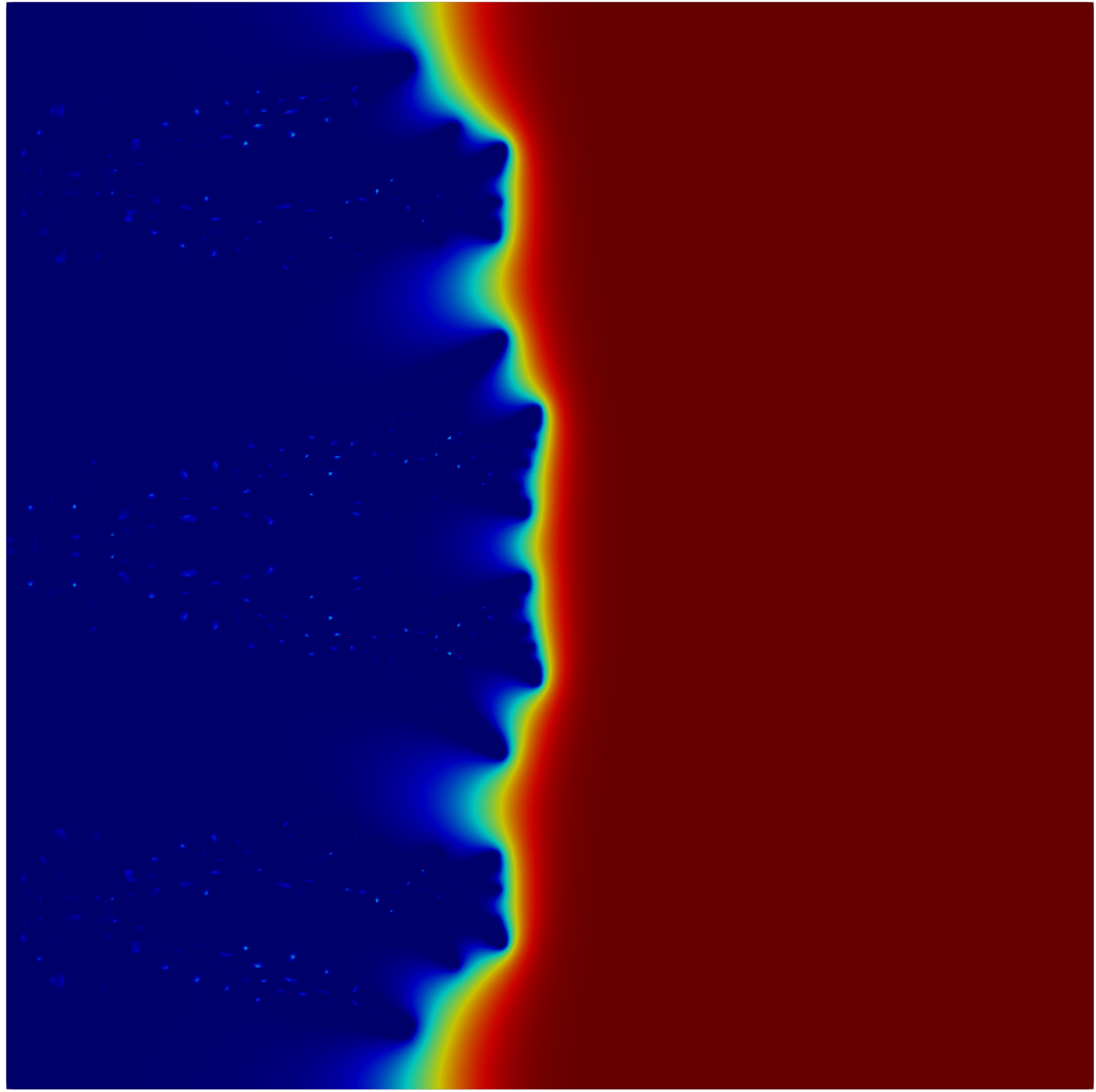}}
\end{subfigure}
\begin{subfigure}[b]{0.1\linewidth}
    \centering%
    {\includegraphics[height = 3.2cm]{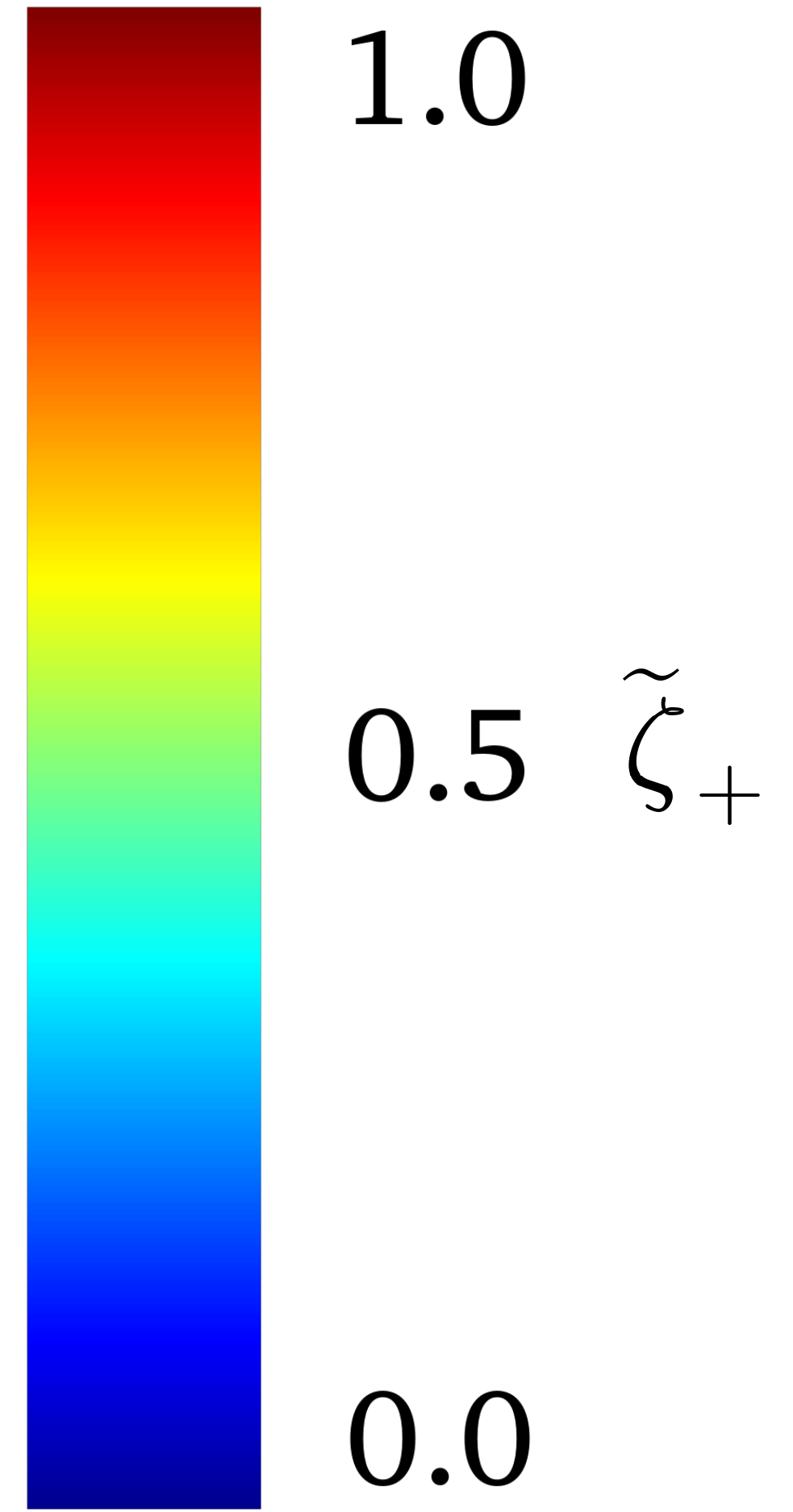}}
\end{subfigure}
\bigbreak
\begin{subfigure}[b]{0.2\linewidth}
    \centering%
	{\includegraphics[height = 3.2cm]{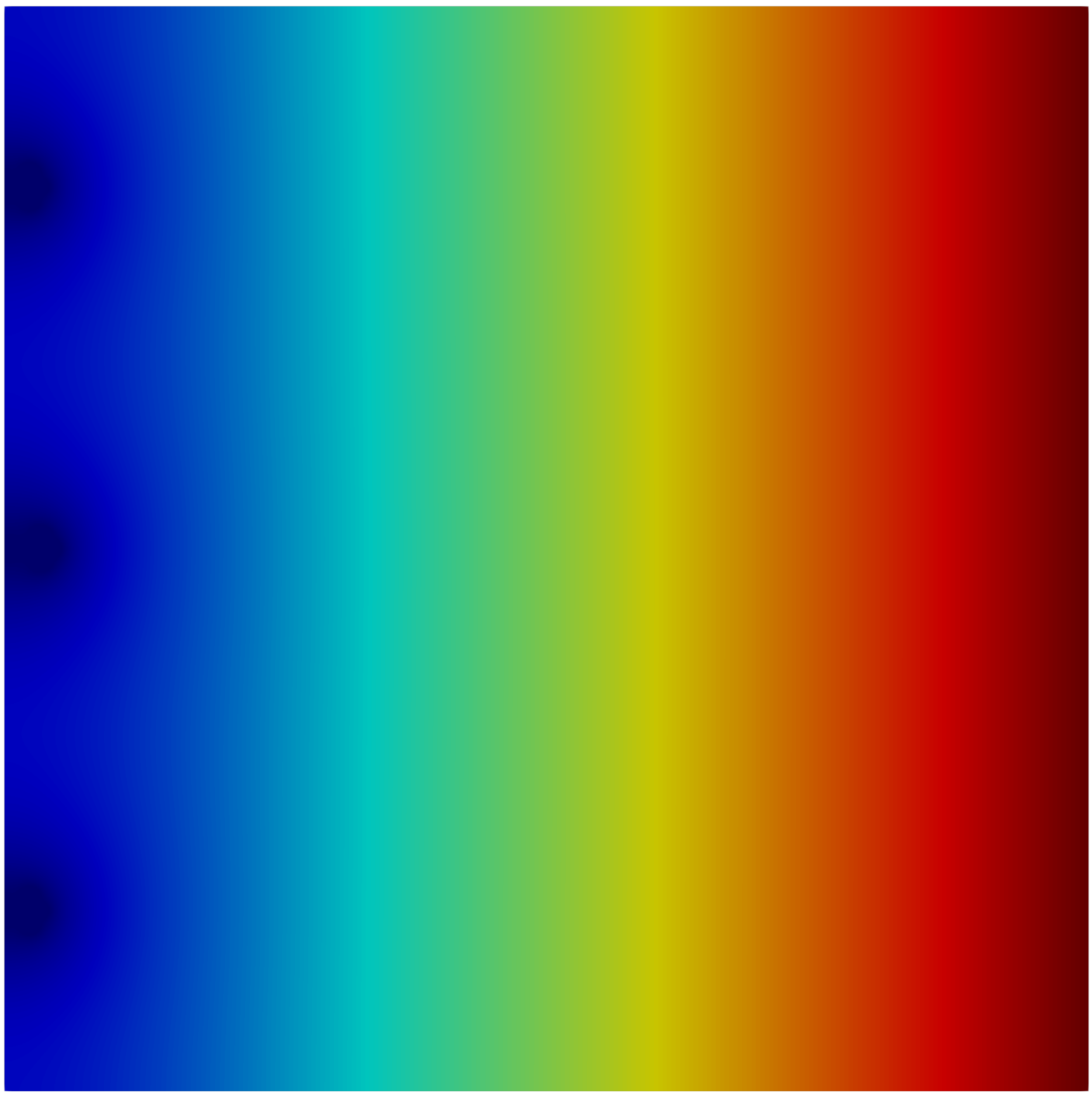}}
	\caption{$t = 5 \left[s\right]$.}
\end{subfigure} 
\begin{subfigure}[b]{0.2\linewidth}
    \centering%
    {\includegraphics[height = 3.2cm]{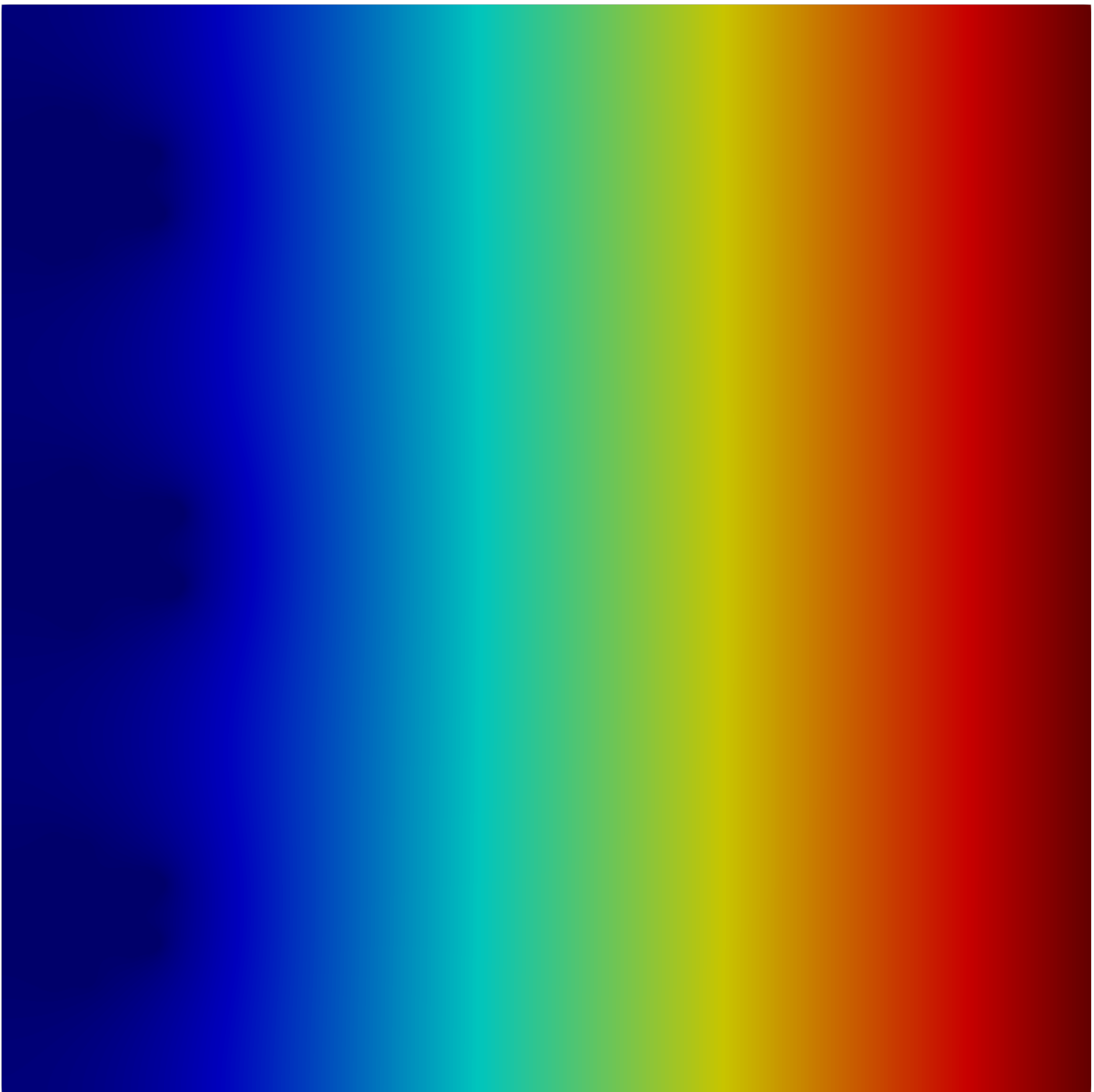}}
    \caption{$t = 20 \left[s\right]$.}
\end{subfigure}
\begin{subfigure}[b]{0.2\linewidth}
    \centering%
    {\includegraphics[height = 3.2cm]{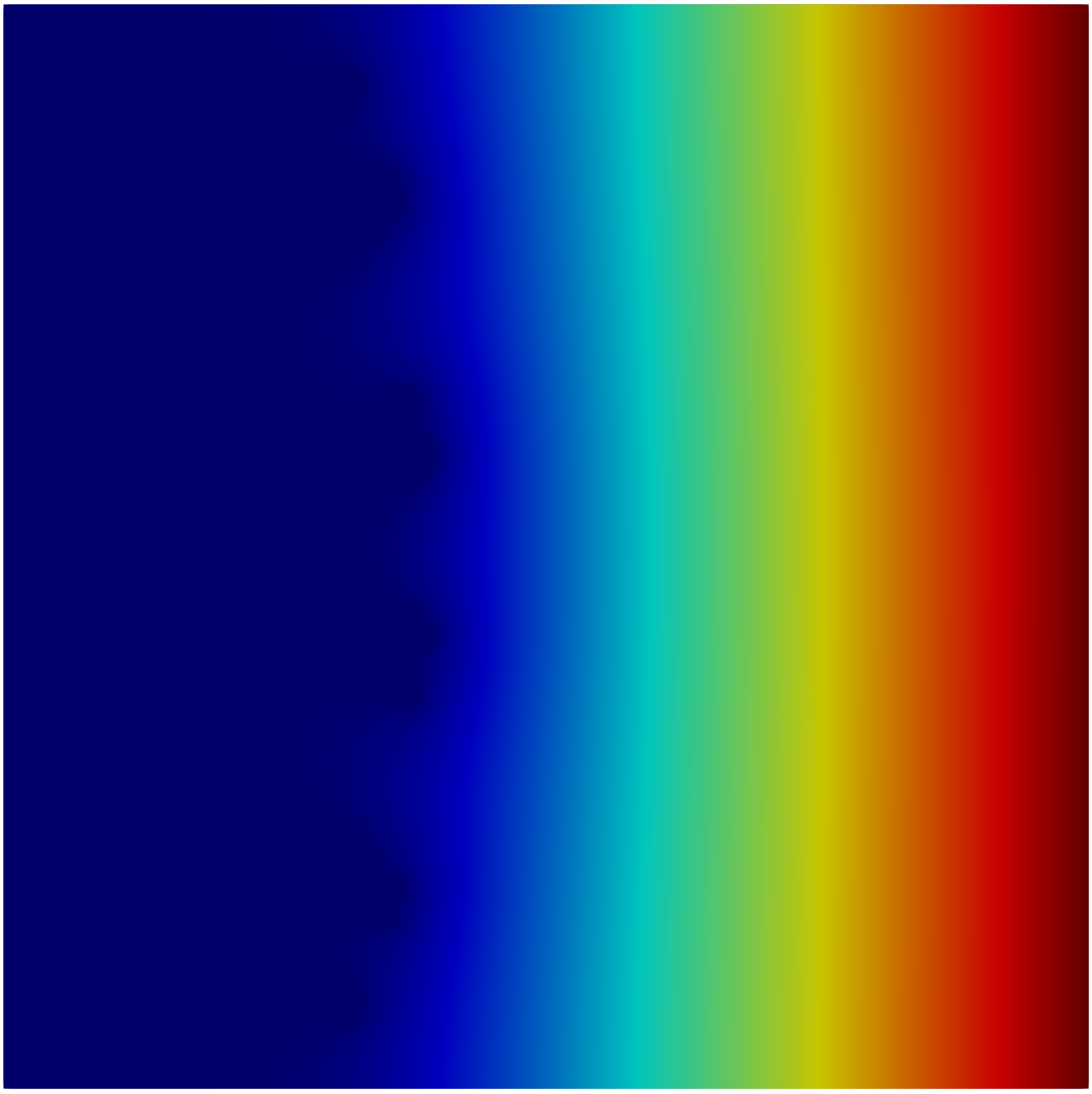}}
    \caption{$t = 50 \left[s\right]$.}
\end{subfigure}
\begin{subfigure}[b]{0.2\linewidth}
    \centering%
    {\includegraphics[height = 3.2cm]{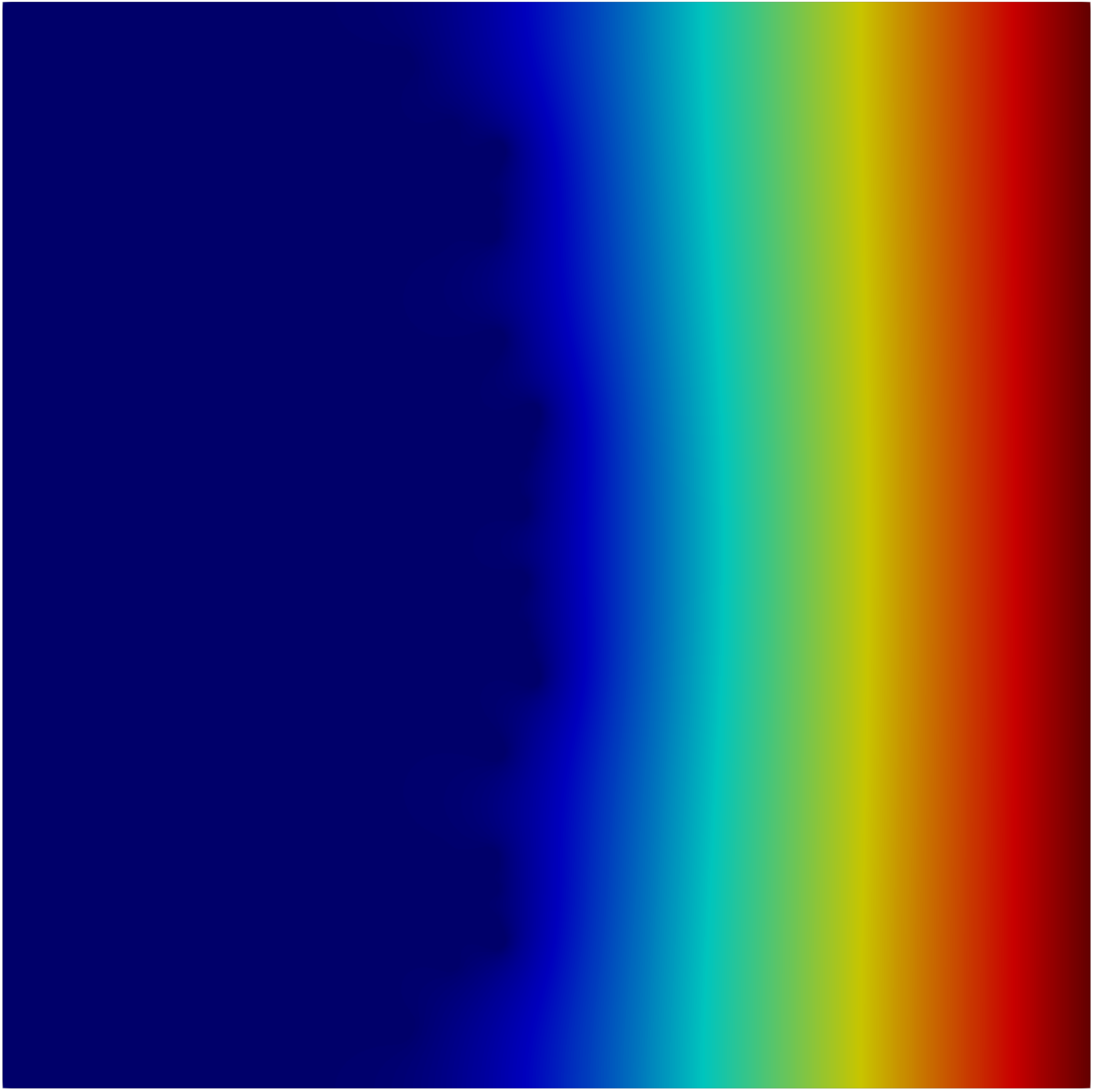}}
    \caption{$t = 70 \left[s\right]$.}
\end{subfigure}
\begin{subfigure}[b]{0.1\linewidth}
    \centering%
    {\includegraphics[height = 3.2cm]{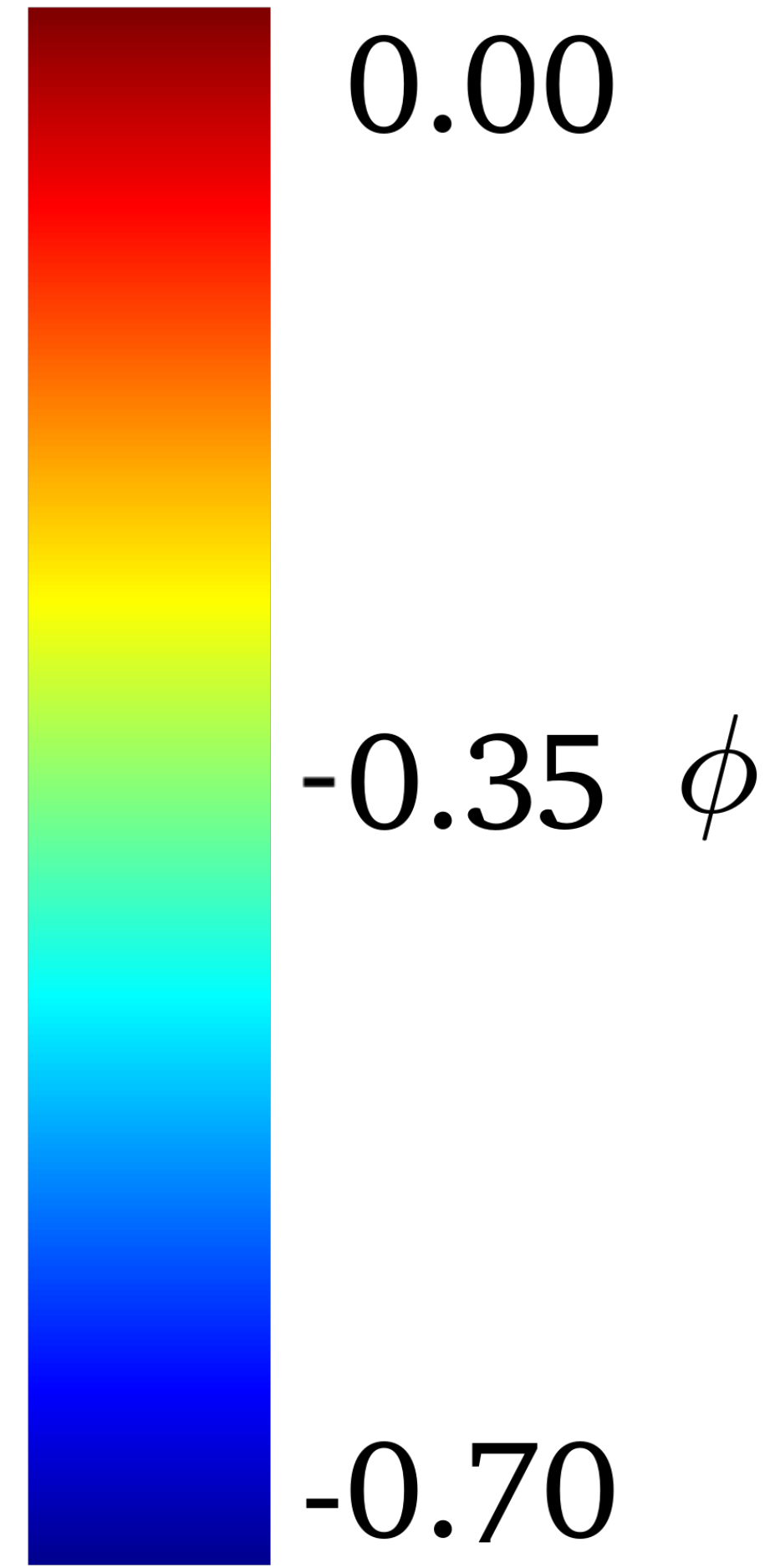}}
    \caption{$.$}
\end{subfigure}
\caption{2D phase-field simulation of bush-like lithium dendrite formation under $\phi_b=-0.7\left[V\right]$ charging potential. The upper row shows the lithium solid-phase evolution $\xi$; the second row shows the lithium-ion distribution $\widetilde{\zeta}_{+}$; and the third row shows the electric potential field $\phi$. Square domain set as $360\times360\left[\mu m^2\right]$.}
\label{fig:2D_MultiSeed_evolut}
\end{figure}

In this case, we use a 2D square domain set as $360\times360\left[\mu m^2\right]$, within the range of typical distances between electrodes in metal anode batteries~\cite{ BAI20182434}.  We use a $500\times720$ structured mesh with a mapping in the $x$-direction to obtain square elements of size $0.5\times0.5\left[\mu m^2\right]$ in the region of interest (see Section~\ref{subsection:3D_Single_nuclei} for details). We apply a charging electro potential of $\phi_b=-0.7\left[V\right]$ to the cell, with an artificial nucleation region with 3 protrusions (ellipsoidal seeds), equally spaced, growing from the left boundary, with semi-axes $4\left[\mu m\right]\times1\left[\mu m\right]$. Figure~\ref{fig:2D_MultiSeed_evolut} shows the evolution of our system’s variables $\left(\xi, \widetilde{\zeta}_{+}, \phi \right)$ at different time steps. The corresponding temporal-spatial distribution of the phase-field variable $\xi$ depicts the dynamic morphological evolution of the simulated lithium dendrite. 

Figure~\ref{fig:2D_MultiSeed_evolut} shows the evolution of bush-like lithium dendrite formation, having morphological resemblance to the dendritic patterns reported in previous phase-field studies of dendrite growth~\cite{ YURKIV2018609, CHEN2021229203}. The lithium dendrites grow from each nucleation site and move towards the opposite electrode, following the electric field ($\vec{E}=-\nabla\phi$). As the dendrite grows, side branches appear, growing across the electrolyte, in agreement with lithium dendrite experiments~\cite{ C6EE01674J}. The computed width range of lithium branches is between 5 and 13 $\left[\mu m\right]$. These bush-like morphologies grow fast across the electrolyte region and penetrate through porous separators, becoming potentially dangerous as they can produce battery short circuits~\cite{ BAI20182434}. The spatial distribution of the lithium-ion concentration $\widetilde{\zeta}_{+}$, and electric potential $\phi$, corresponding to the second and third rows of Figure~\ref{fig:2D_MultiSeed_evolut} respectively, show that they follow the same trend reported in other phase-field models of lithium dendrite growth~\cite{ YURKIV2018609, doi:10.1063/1.4905341}. The lithium-ion concentration remains equal to the bulk concentration, $\widetilde{\zeta}_{+}=1$, through the electrolyte phase and decreases near the dendrite front (solid phase) due to the electrodeposition process~\cite{ CHEN2015376}. The electric potential has a large gradient that spreads over the electrolyte phase and remains constant and equal to $\phi_b$ at the electrode phase. As the lithium dendrite moves across the electrolyte region, the gradient of the electric potential (electric field) increases, in agreement with simulation results reported by Hong et al.~\cite{ doi:10.1063/1.4905341}.

\begin{figure}[h!]
    \centering%
{\includegraphics[height = 6cm]{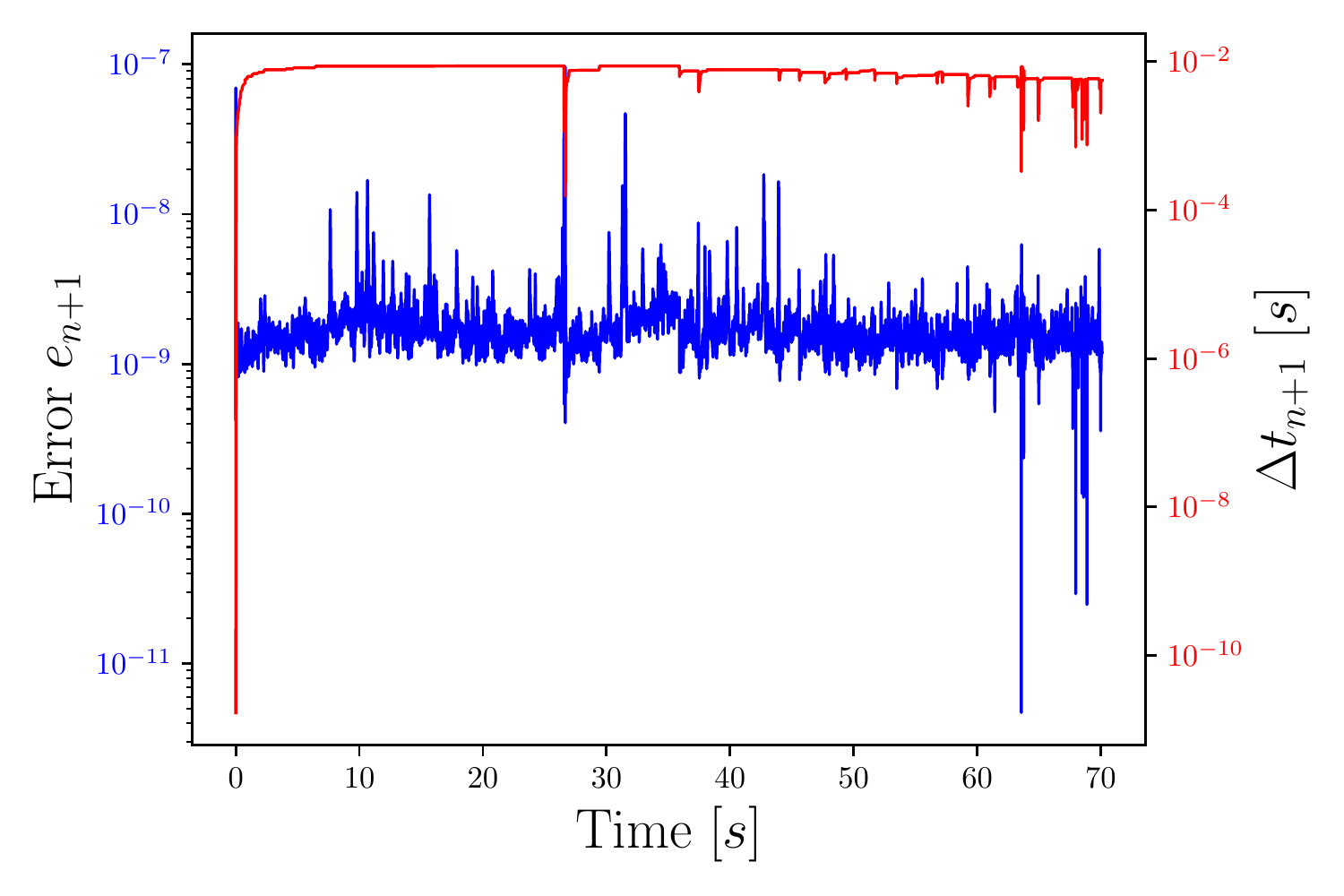}}
\caption{Performance of time-adaptive strategy in 2D lithium dendrite growth simulation. Weighted local truncation error (blue) and time-step size (red) vs time.}
\label{fig:DeltaTvsT_3seed_2D}
\end{figure}

Figure~\ref{fig:DeltaTvsT_3seed_2D} illustrates the interplay between the weighted truncation error estimate $e_{n+1}$~\eqref{eq:WLE2} (blue) and the time-step size evolution $\Delta t_{n+1}$~\eqref{eq:DT} (red), throughout the $70 \left[s\right]$ of simulation. The figure shows that the simulation initially requires a small time-step size of $\Delta t_0 = 10^{-10}\left[s\right]$ to converge as the phase field develops its interface thickness and the ion concentration and electro potential distributions achieve equilibrium. The time-step size increases sharply, achieving a stationary size of $\Delta t_{n+1}=10^{-2}\left[s\right]$, after $5\left[s\right]$ of simulation. In addition, the time-step size starts to decrease slightly after $40\left[s\right]$, which corresponds with the acceleration of lithium dendrite propagation rate as it approaches the opposite electrode~\cite{ doi:10.1063/1.4905341}. The maximum and minimum tolerances for the time-adaptive scheme are $10^{-7}$ and $10^{-9}$ in this case, allowing the time-step size to decrease or increase accordingly when error tolerances bounds are reached.

\subsection{3D simulation of lithium dendrite growth: Single nucleus experiment}
\label{subsection:3D_Single_nuclei}

\begin{figure} [h]
\begin{subfigure}[b]{0.4\linewidth}
    \centering%
{\includegraphics[height = 6.0cm]{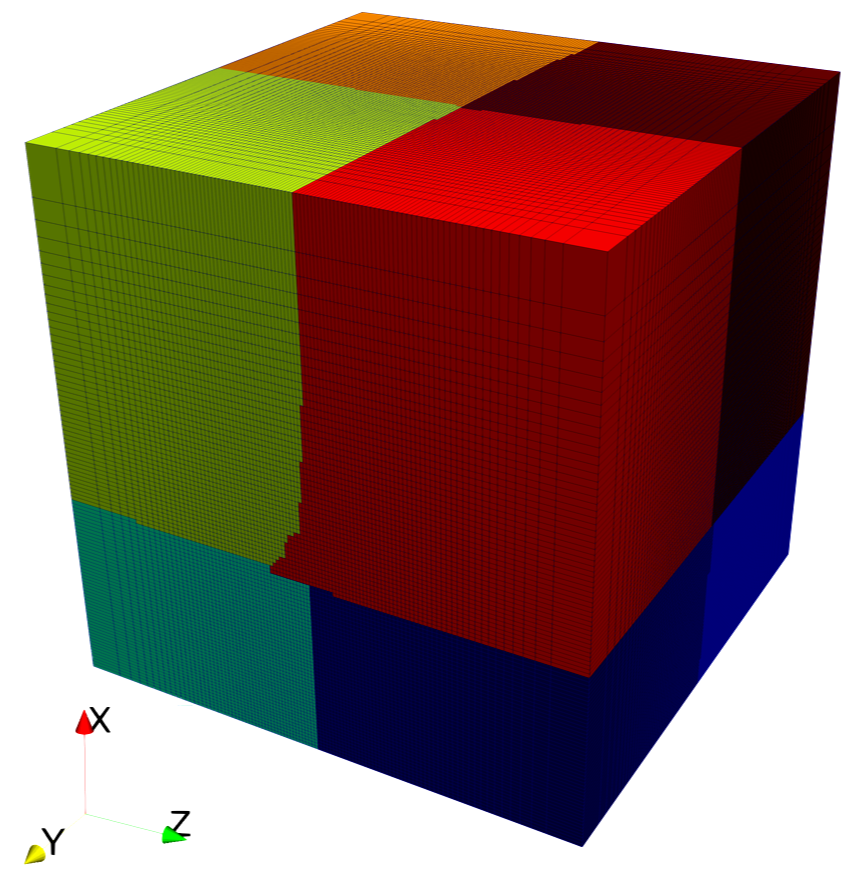}}
\caption{3D mesh partition in 8 processors.}
\label{fig:3Dmesh_split}
\end{subfigure}
\begin{subfigure}[b]{0.4\linewidth}
    \centering%
{\includegraphics[height = 6.0cm]{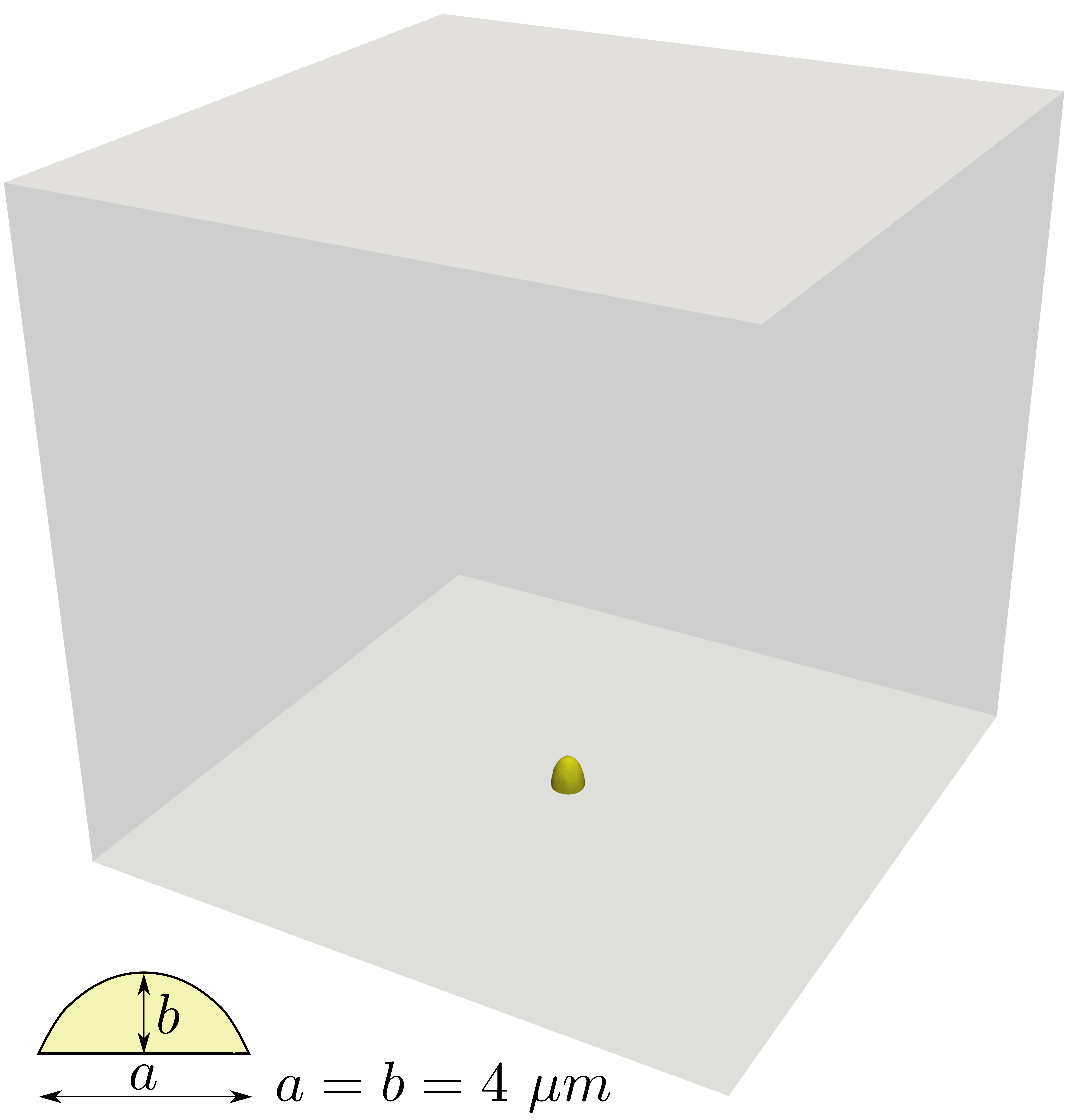}}
\caption{Simple nucleation.}
\label{fig:Simple_nucleation}
\end{subfigure}
\caption{3D mesh partition in 8 processors, each one represented by a different color (\subref{fig:3Dmesh_split}), and geometry of the initial protrusion for the simple nucleation experiment (\subref{fig:Simple_nucleation}).}
\label{fig:3D_CompuDomain}
\end{figure}

This section describes 3D phase-field simulations of lithium dendrite formation to study three-dimensional highly branched "spike-like" dendritic patterns, commonly observed experimentally. These patterns form under high current density loads, which correspond to fast battery charge~\cite{ jana2019electrochemomechanics, ding2016situ, TATSUMA20011201}. We select a geometrical unit that characterizes a real cell structure~\cite{ YURKIV2018609, Trembacki_2019}. We choose a computational domain of $80 \times 80 \times 80 \left[\mu m^3\right]$. Consequently, given the domain size, we expect growth rates up-to two orders of magnitude faster-than-normal, due to the short separation between electrodes $l_x = 80 \left[\mu m\right]$.

We use a 3D structured mesh with eight-node hexahedral elements. We distribute the mesh to focus the node's mapping on the area of interest (see Figure~\ref{fig:3Dmesh_split}). In particular, in the $x$-direction $x_r=\left(2/\pi\right)arcsin{\left(x_u\right)}$; where $x_u$ is the node's $x$ coordinate normalized by $l_x$, before mapping (uniform distribution), and $x_r$ is the node's mapped coordinate.  The $\arcsin$ function transitions smoothly, inducing a relatively small variation in the mesh size within $50\%$ of the physical domain where the lithium electrodeposit process occurs~\cite{ YURKIV2018609, MU2019100921, Zhang_2021}. The mapping produces a finer mesh to properly capture the phase-field interface thickness (4 elements in the interface~\cite{ Arguello2022}) and the steepest gradients of $\widetilde{\zeta}_{+}$ and $\phi$.  We use a $120^3$ tensor-product mesh with a mesh size of $0.4\left[\mu m\right]$ in the region of interest (bottom half of the domain). We partition the mesh into eight processors. Figure~\ref{fig:3Dmesh_split} identifies each core with a different color, showing that the tensor-product mesh can efficiently allocate resources in the region of interest (more resources/colors allocated to the bottom half of the domain).  As in the previous 2D simulation, we apply a charging electro-potential value of $\phi_b=-0.7\left[V\right]$ to the cell, with an artificial nucleation region formed by a single protrusion (ellipsoidal seed), with its center at $\left(0;l_y/2;l_z/2\right)$, and semi-axes $4\left[\mu m\right]\times2\left[\mu m\right]\times2\left[\mu m\right]$ as Figure~\ref{fig:Simple_nucleation} shows. For visualization purpose, here we rotate the 3D figures to depict a vertical $x$-axis, showing an upright lithium dendrite growth, in agreement with the convention used for lithium dendrite experiments~\cite{ NISHIKAWA201184, ding2016situ, TATSUMA20011201}. 
 
\begin{figure}[h!]
\begin{subfigure}[b]{0.32\linewidth}
    \centering%
{\includegraphics[height = 5.15cm]{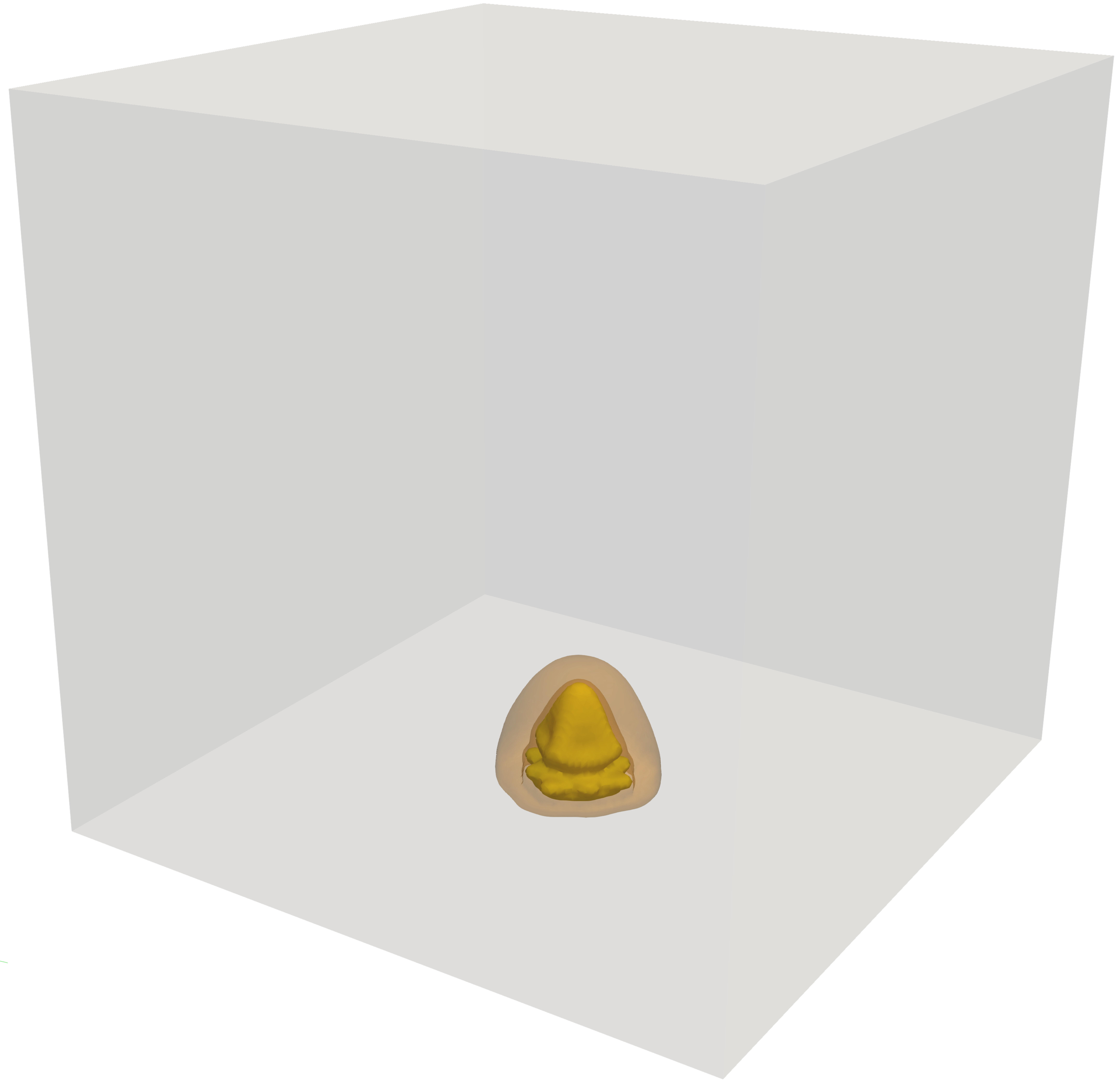}}
\caption{$t = 0.1\left[s\right]$.}
\end{subfigure}
\begin{subfigure}[b]{0.32\linewidth}
    \centering%
    {\includegraphics[height = 5.15cm]{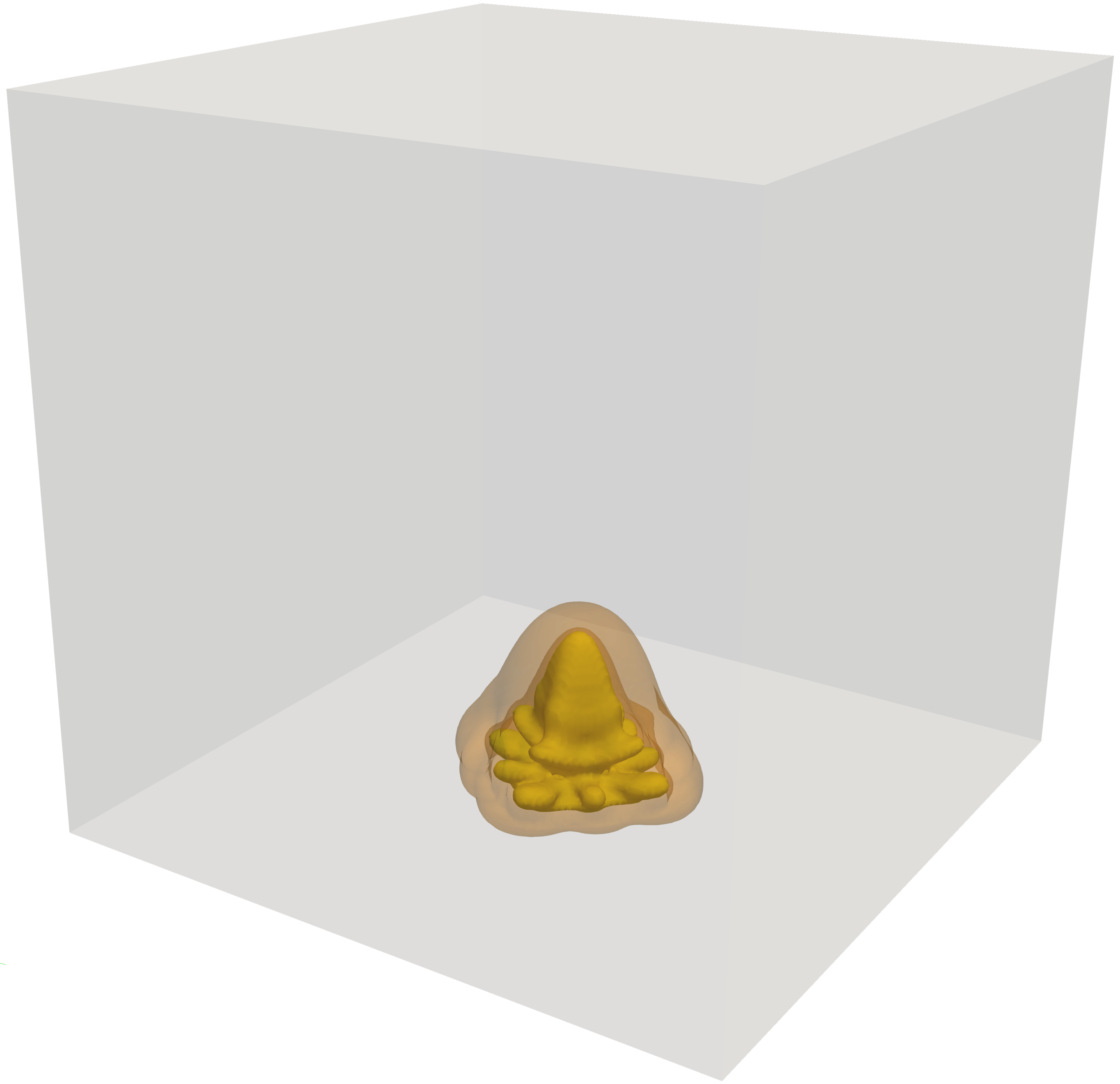}}
    \caption{$t = 0.2\left[s\right]$.}
\end{subfigure}
\begin{subfigure}[b]{0.32\linewidth}
    \centering%
    {\includegraphics[height = 5.15cm]{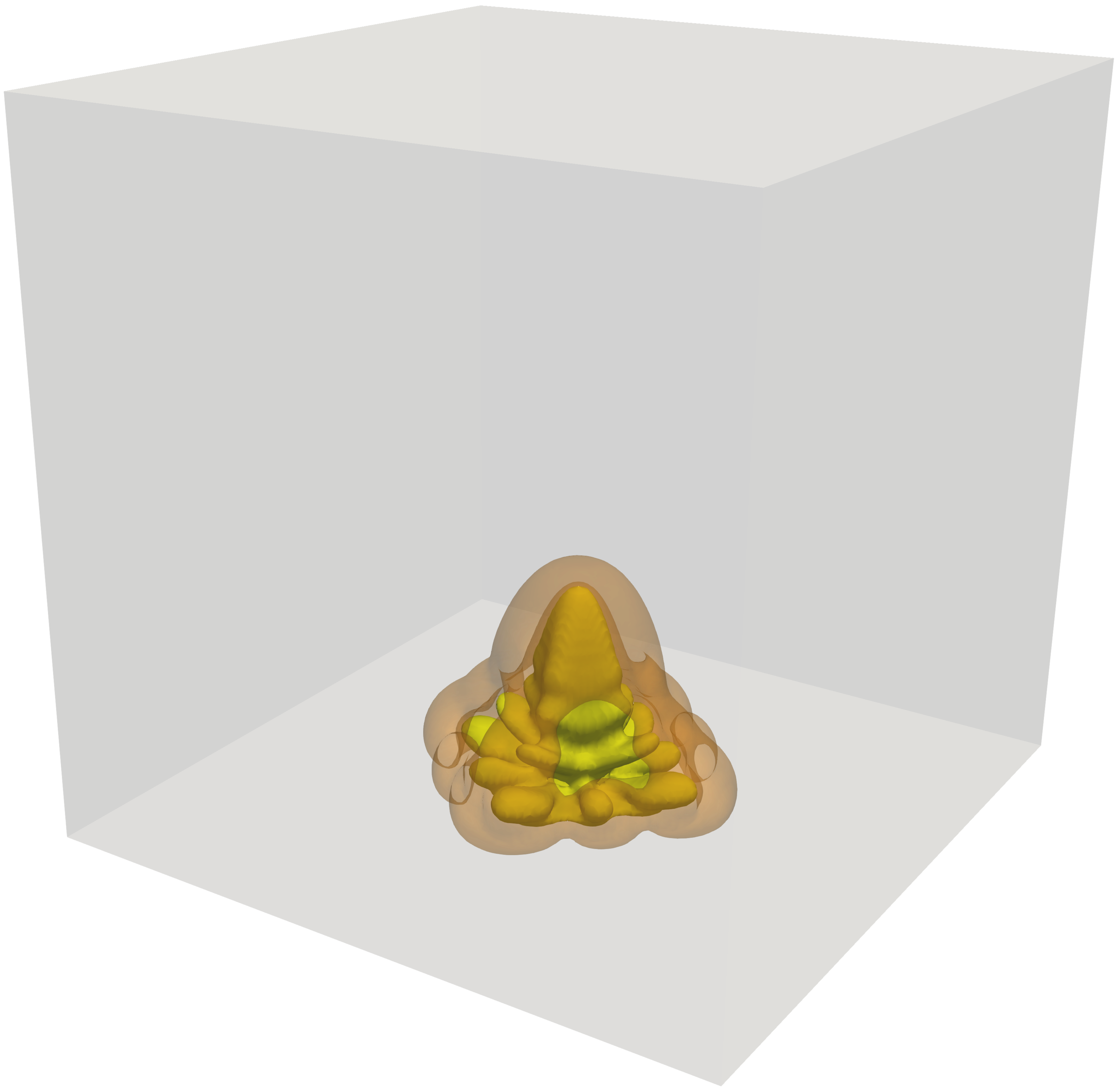}}
    \caption{$t = 0.3\left[s\right]$.}
\end{subfigure}
\begin{subfigure}[b]{0.32\linewidth}
    \centering%
    {\includegraphics[height = 5.15cm]{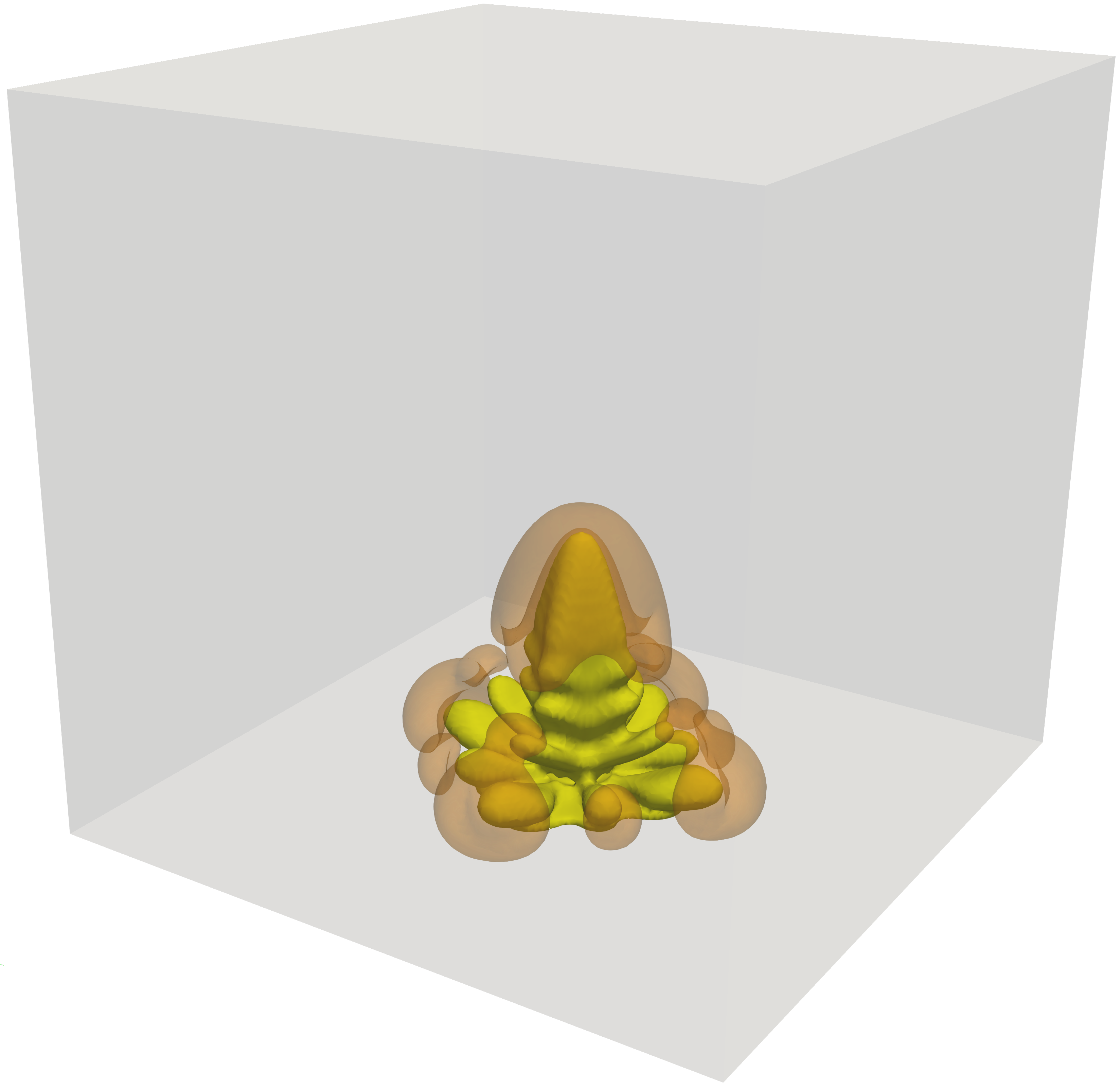}}
    \caption{$t = 0.4\left[s\right]$.}
\end{subfigure}
\begin{subfigure}[b]{0.32\linewidth}
    \centering%
    {\includegraphics[height = 5.15cm]{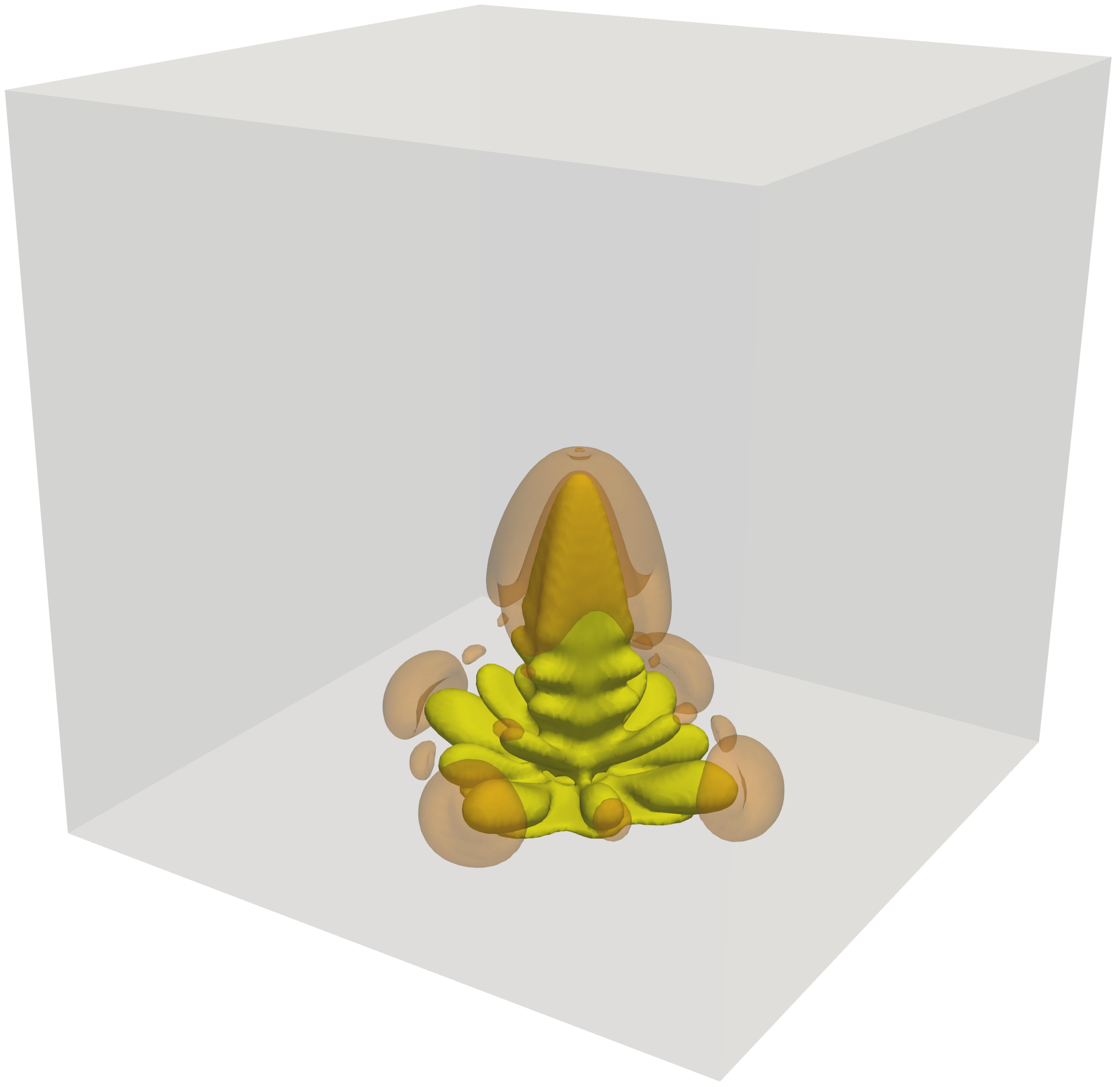}}
    \caption{$t = 0.5\left[s\right]$.}
\end{subfigure}
\begin{subfigure}[b]{0.32\linewidth}
    \centering%
    {\includegraphics[height = 5.15cm]{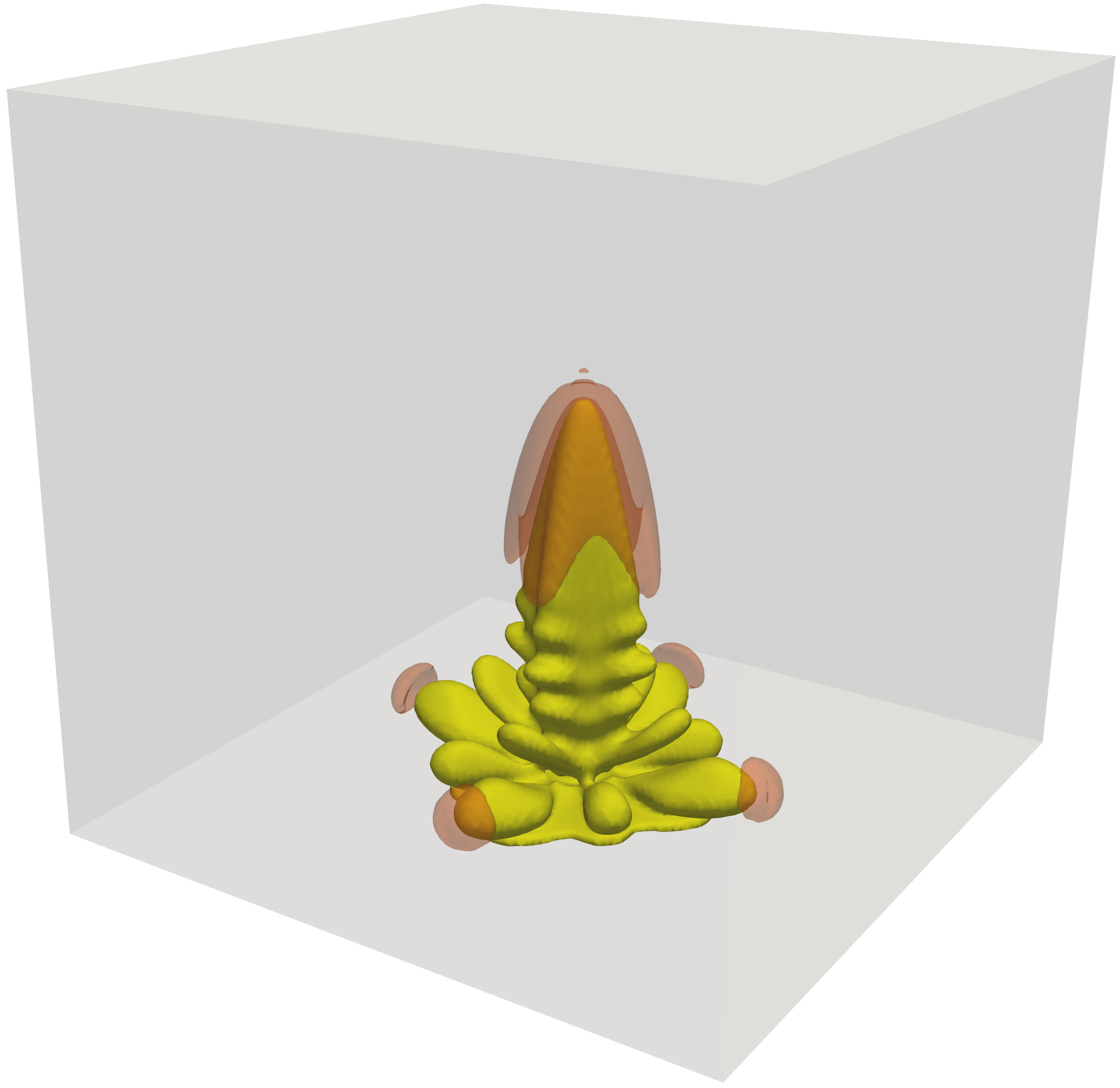}}
    \caption{$t = 0.6\left[s\right]$.}
\end{subfigure}
\caption{Spike-like lithium dendrite formation under $\phi_b=-0.7\left[V\right]$ charging potential. The electrodeposited lithium is represented with a yellow isosurface plot of the phase-field variable $\xi$. Electrolyte regions with enriched concentration of lithium-ion ($\widetilde{\zeta}_{+}>1$) represented with orange volumes. Cube domain set as $80 \times 80 \times 80 \left[\mu m^3\right]$.}
\label{fig:SingleSeed_evolut}
\end{figure}

Figure~\ref{fig:SingleSeed_evolut} shows the morphological evolution of the simulated lithium dendrite (isosurface plot of the phase-field variable $\xi$). The simulation forms a spike-like, symmetric, and highly branched pattern, consisting of the main trunk and sets of four equal side branches growing in each horizontal direction. A result consistent with the body-centered cubic (bcc) crystallographic arrangement of lithium metal~\cite{ YURKIV2018609}. Our model includes the four folded surface anisotropy~\eqref{eq:KAPPA3D}. In-situ optical microscopic investigations~\cite{ ding2016situ,TATSUMA20011201} report spike-like lithium dendrite formation. These dendritic patterns grow when high (over-limiting) current densities are applied to the cell ($>20\left[mA/cm^2\right]$)~\cite{ BAI20182434}. In such cases,  the rate of lithium deposition overcomes the rate of solid-electrolyte interface  formation, allowing the lithium deposit to grow almost free from the influence of the interface~\cite{ BAI20182434}. The enrichment of lithium-ion concentration appears in the vicinity of dendrite tips, reaching peak values of up to $\widetilde{\zeta}_{+}=2.1$, and triggering tip-growing dendritic lithium. Electrolyte regions with higher lithium-ion concentration ($\widetilde{\zeta}_{+}>1$) are represented by orange volumes in Figure~\ref{fig:SingleSeed_evolut}. It is worth mentioning that this phenomenon was previously reported by Hong et al.~\cite{ doi:10.1063/1.4905341} in 2D simulations of lithium dendrite formation, proposing a compositionally graded electrolyte, which could potentially suppress the dendrite initiation.

\begin{figure}[h!]
    \centering%
{\includegraphics[height = 6.5cm]{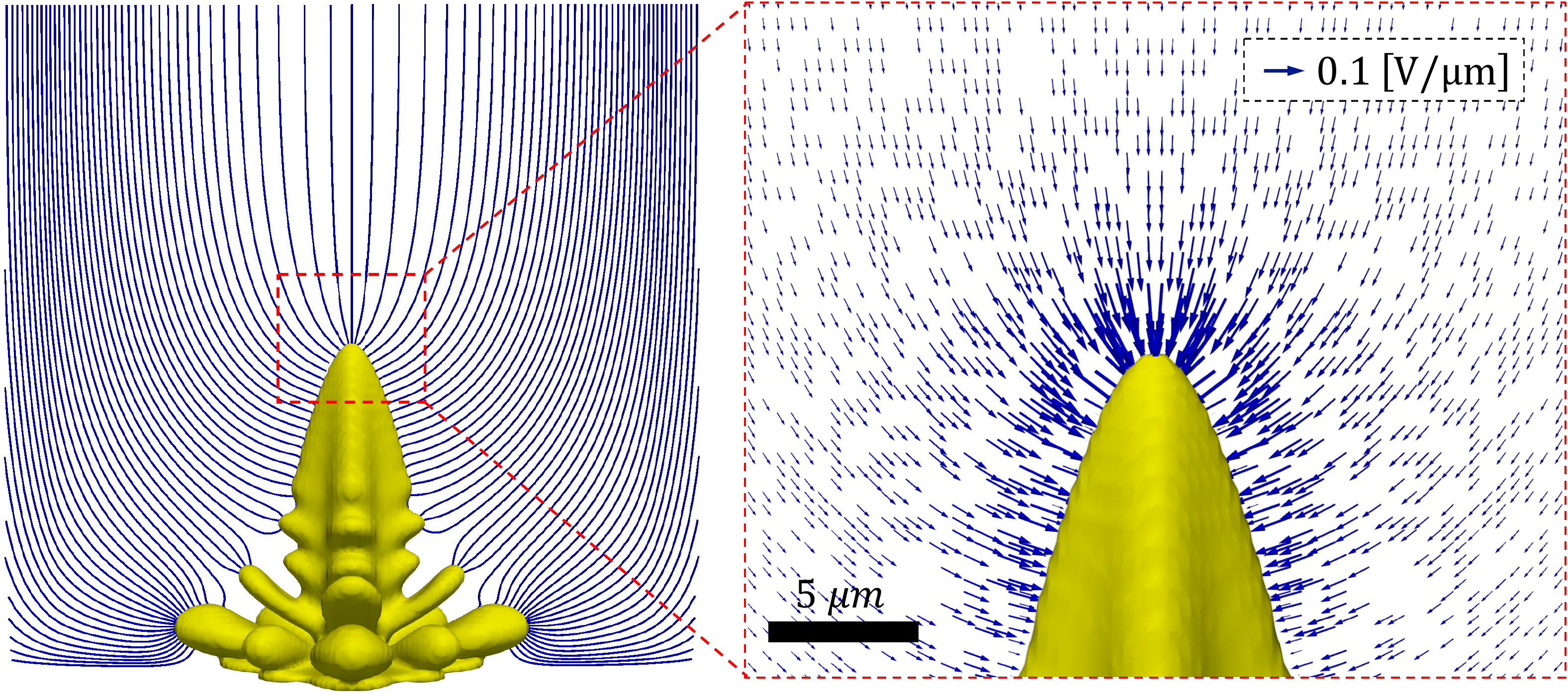}}
\caption{Overlay of electric field distribution (blue streamlines and vectors) with dendrite morphology at time $t = 0.5\left[s\right]$. Streamline plane set at $y=40\left[\mu m\right]$.}
\label{fig:ElectField_SingleSeed3D}
\end{figure}

We calculate the electric field distribution by differentiation of the resolved electric potential $\vec{E}=-\nabla\phi^h$. The magnified view in Figure~\ref{fig:ElectField_SingleSeed3D} shows how the electric field localizes in the vicinity of the dendrite tip, leading to an enriched concentration of the lithium-ion it induces due to the strong migration from the surrounding regions (see~\eqref{eq:Diffusion_Eq} and~\cite{ doi:10.1063/1.4905341}).

\begin{figure}[h!]
    \centering%
{\includegraphics[height = 6.5cm]{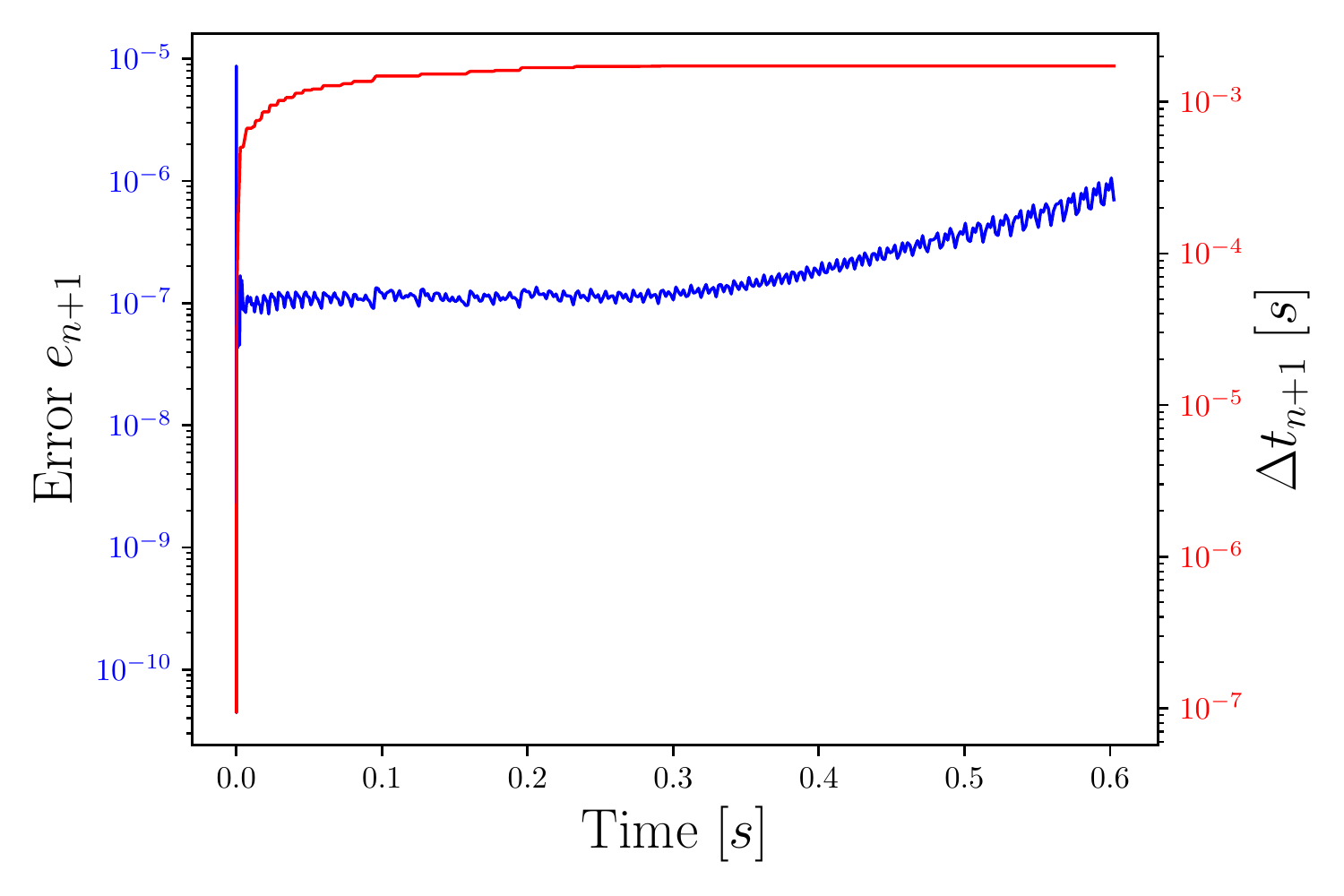}}
\caption{Time adaptivity for 3D spike-like dendrite growth simulation.}
\label{fig:DeltaTvsT_SingleSeed3D}
\end{figure}

Figure~\ref{fig:DeltaTvsT_SingleSeed3D} shows the performance of the time-adaptive scheme, throughout the $0.6\left[s\right]$ of simulation. Starting with a small time-step of $\Delta t_0 = 10^{-7}\left[s\right]$ to achieve convergence, followed by a rapid increase in size, until reaching a stationary value of about $\Delta t_{n+1}=10^{-3}\left[s\right]$. The evolution of the weighted truncation error $e_{n+1}$ (blue) stays close to the minimum tolerance limit ($10^{-7}$) during $0.3\left[s\right]$ of simulation; beyond this point, the error estimate increases due to the acceleration of lithium dendrite propagation rate as the dendrites approach the positive electrode. The time-step size remains unchanged since the error estimate remains within the error bounds. 

\begin{figure} [h]
    \centering%
{\includegraphics[height = 6.5cm]{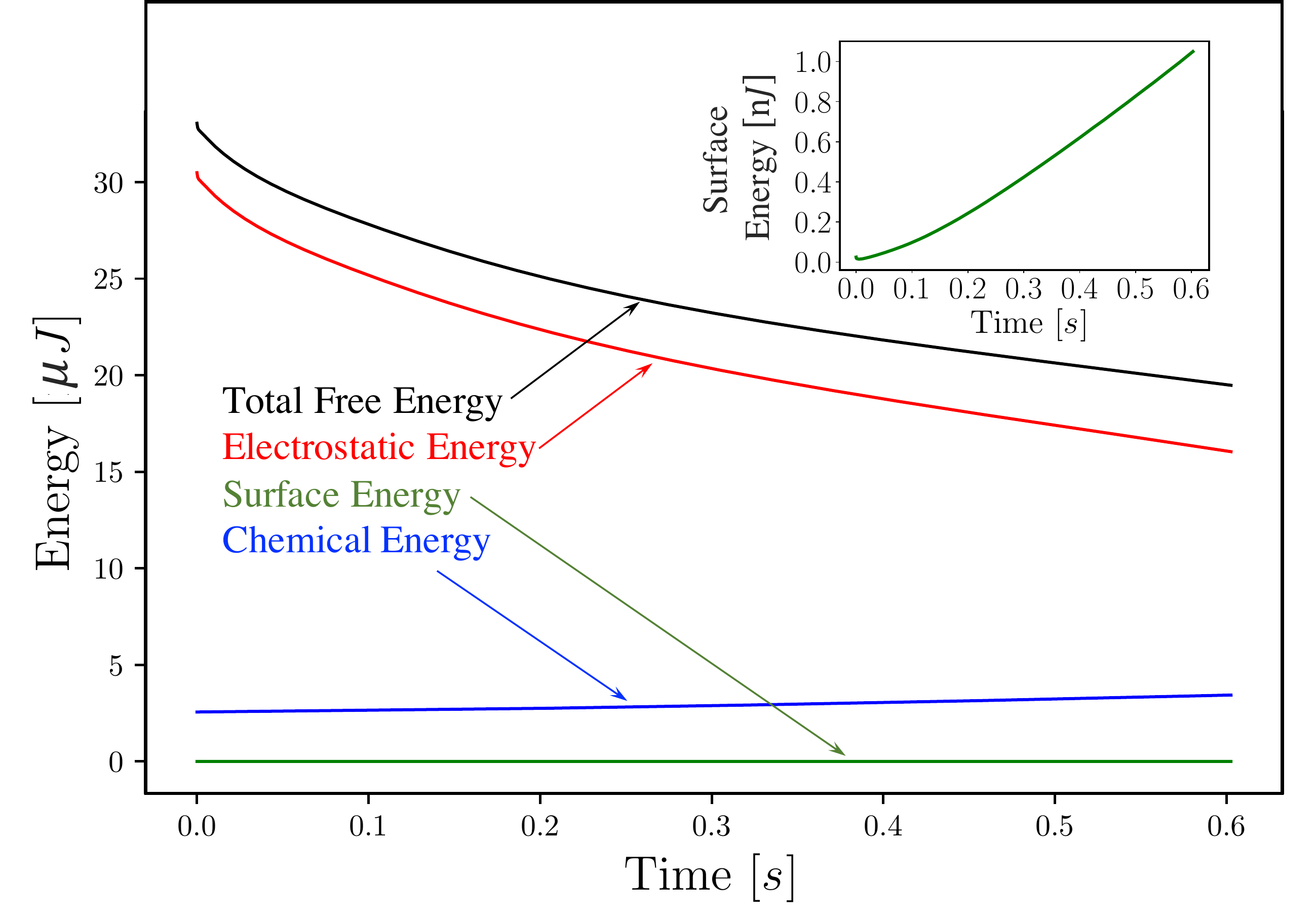}}
\caption{Energy time series for 3D spike-like dendrite growth simulation. The inset plots the increasing surface energy in smaller scale for better appreciation.}
\label{fig:EnergyvsTime_SingleSeed3D}
\end{figure}

The construction of phase-field models satisfies an a priori nonlinear stability relationship, expressed as a time-decreasing free-energy functional; nevertheless, standard discrete approximations do not inherit this stability property. Thus, Figure~\ref{fig:EnergyvsTime_SingleSeed3D} shows the evolution of the Gibbs free energy of the system $\Psi$, see~\eqref{eq:gibbsfreeen}, using our adaptive time integration scheme. We plot the total energy curve (black), as well as three additional energy curves that correspond to each one of its terms, namely, the Helmholtz (chemical) free energy $\int_{V}\text{f}_{\text{ch}}dV$ (blue), surface energy $\int_{V}\text{f}_{\text{grad}}dV$ (green), and electrostatic energy $\int_{V}\text{f}_{\text{elec}}dV$ (red), as the figure indicates.  Figure~\ref{fig:EnergyvsTime_SingleSeed3D} shows that the total systems' discrete free energy does not increase with time. Therefore, we obtain discrete energy stable results using our second-order backward-difference (BDF2) time-adaptive marching scheme. Alternative, provably unconditionally stable second-order time accurate methods may deliver larger time-step sizes for phase-field models~\cite{ GOMEZ20115310, Sarmiento:2017, Wu:2014, hawkins2012numerical, Vignal:2017}, but these are beyond the scope of this work. Additionally, the system's chemical and surface energies increase as the lithium surface area grows as time progresses. In parallel, the electrostatic energy decreases in time. This interaction is consistent with the electrodeposition process, where the system stores the applied electrostatic energy as electrochemical energy as the battery charges.

\subsection{3D Simulation of lithium dendrite growth: Multi-nuclei experiment}

\begin{figure}[h!]
\begin{subfigure}[b]{0.32\linewidth}
    \centering%
{\includegraphics[height = 5.25cm]{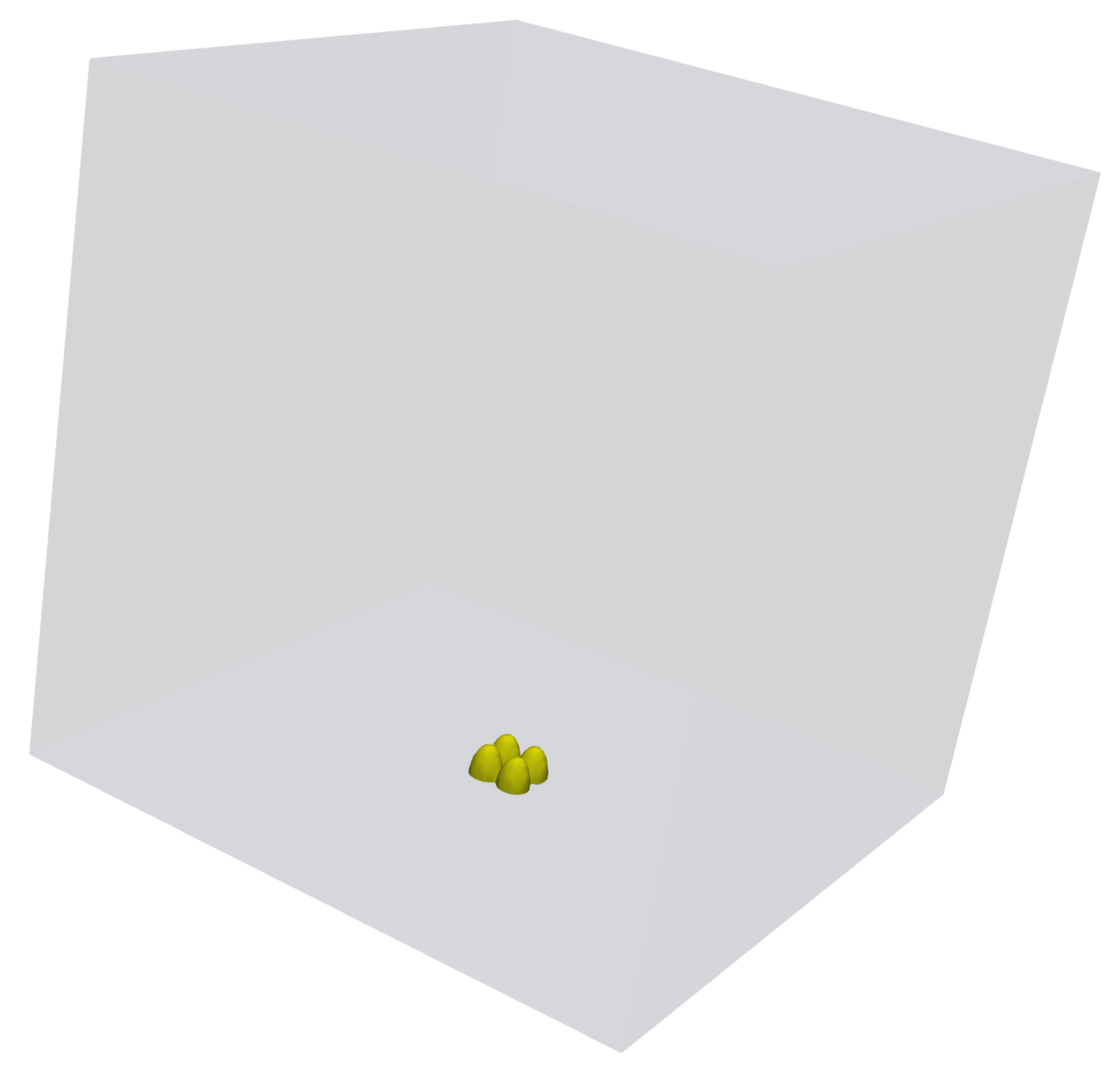}}
\caption{$t = 0.0\left[s\right]$.}
\end{subfigure}
\begin{subfigure}[b]{0.32\linewidth}
    \centering%
{\includegraphics[height = 5.25cm]{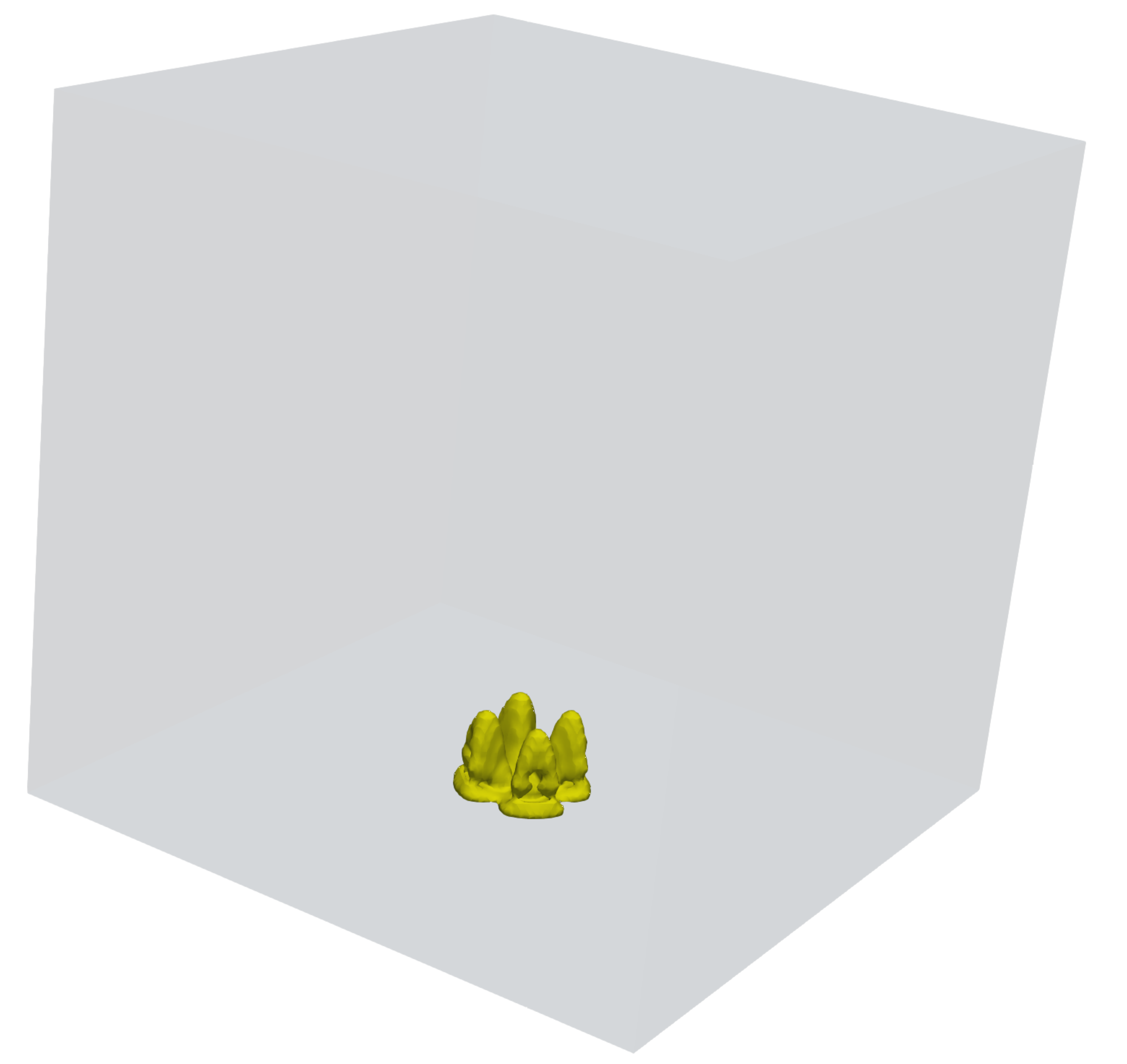}}
\caption{$t = 0.1\left[s\right]$.}
\end{subfigure}
\begin{subfigure}[b]{0.32\linewidth}
    \centering%
{\includegraphics[height = 5.25cm]{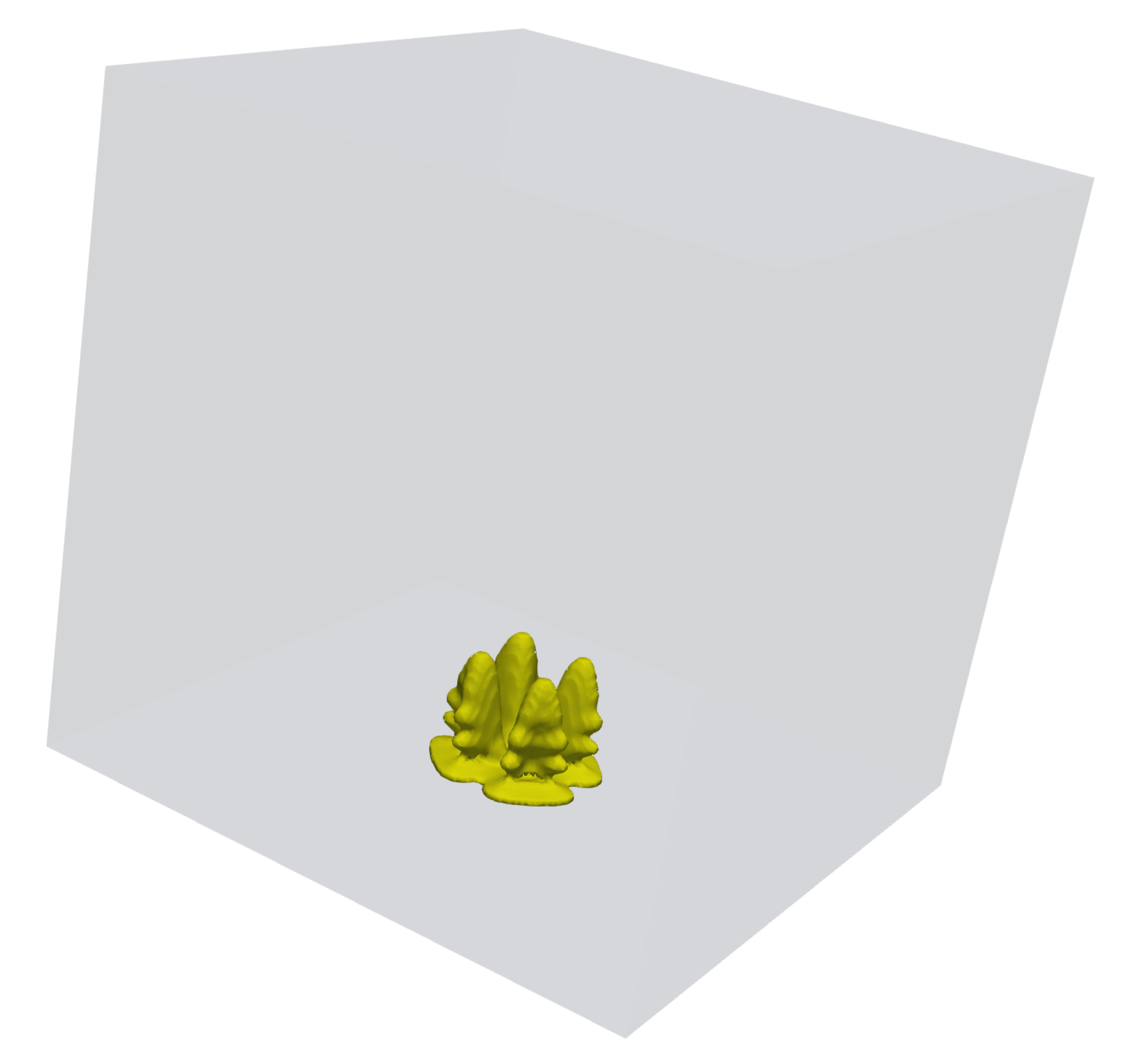}}
\caption{$t = 0.2\left[s\right]$.}
\end{subfigure}
\begin{subfigure}[b]{0.32\linewidth}
    \centering%
    {\includegraphics[height = 5.25cm]{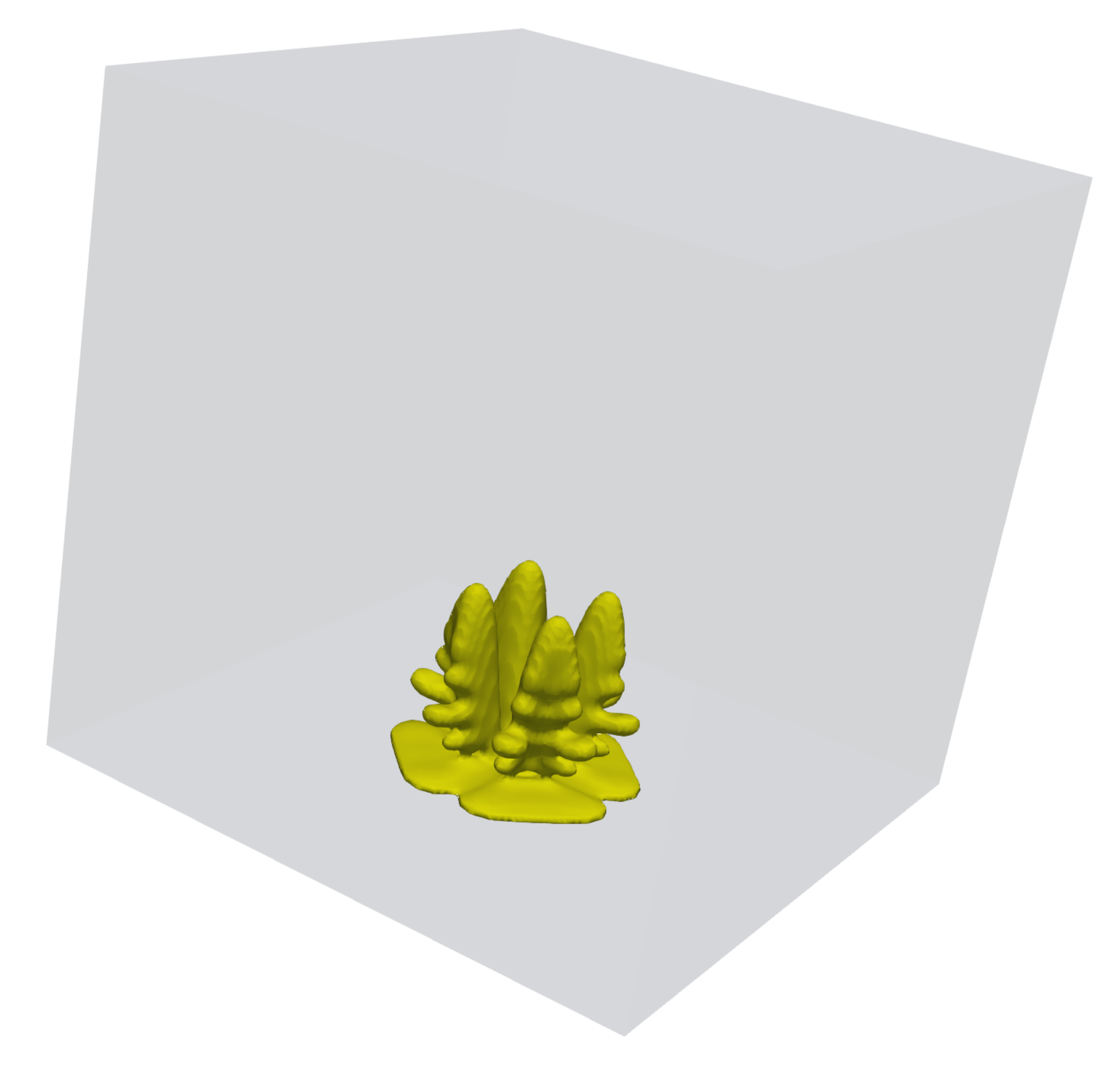}}
    \caption{$t = 0.4\left[s\right]$.}
\end{subfigure}
\begin{subfigure}[b]{0.32\linewidth}
    \centering%
    {\includegraphics[height = 5.25cm]{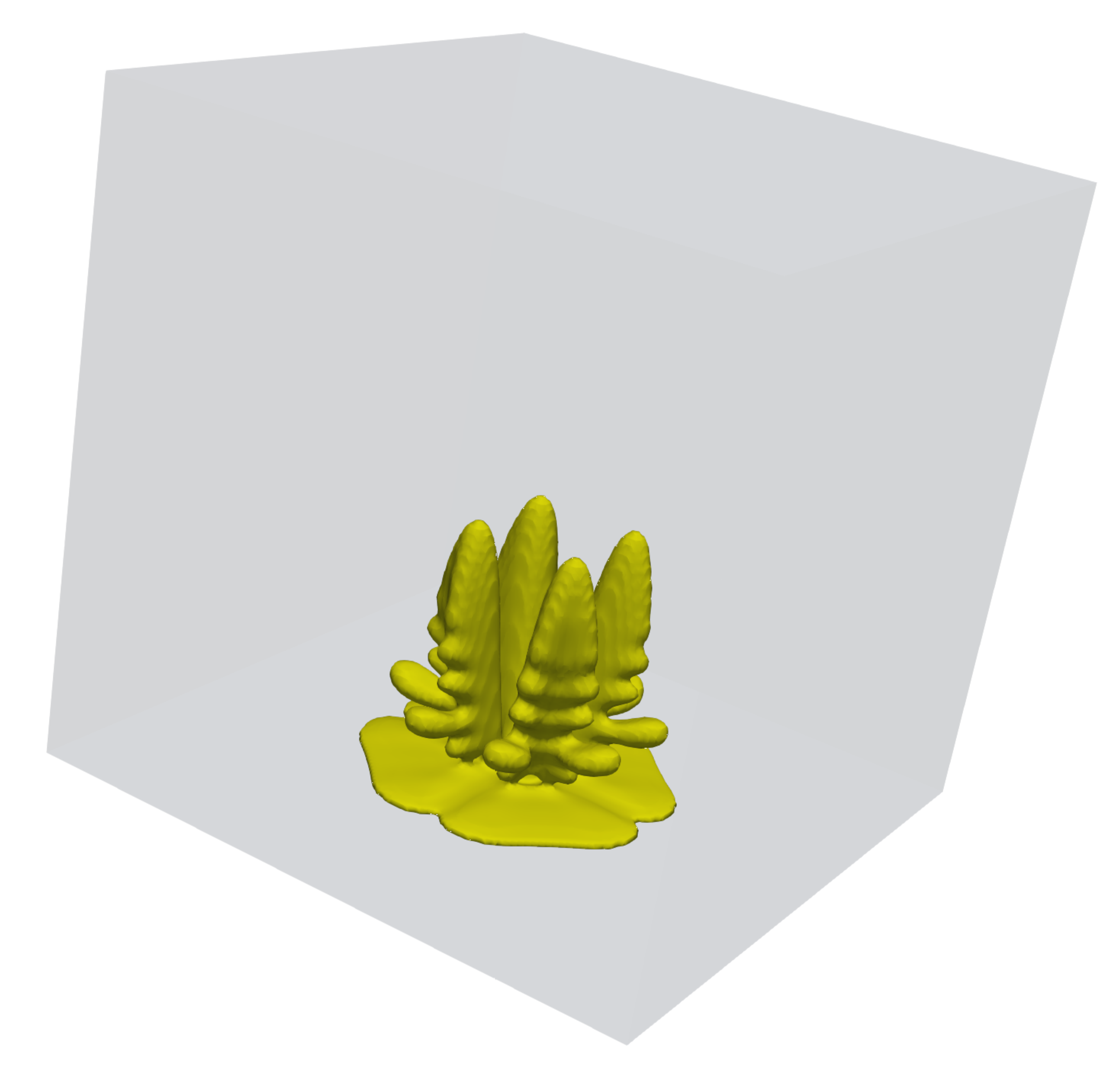}}
    \caption{$t = 0.6\left[s\right]$.}
\end{subfigure}
\begin{subfigure}[b]{0.32\linewidth}
    \centering%
    {\includegraphics[height = 5.25cm]{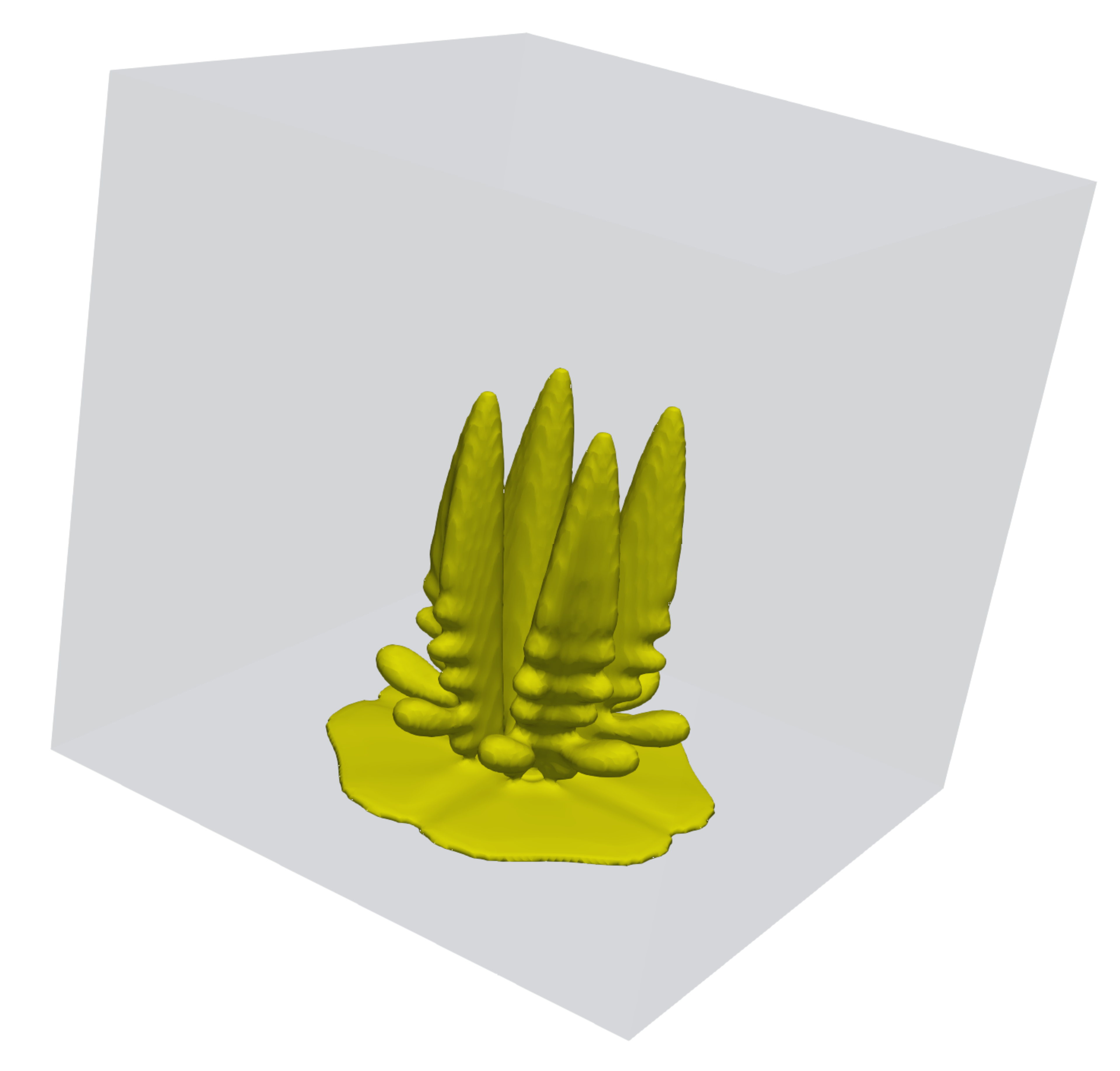}}
    \caption{$t = 0.8\left[s\right]$.}
\end{subfigure}
\caption{Spike-like lithium dendrite formation from multiple nucleation sites under $\phi_b=-0.7\left[V\right]$ charging potential. The electrodeposited lithium is represented with a yellow isosurface plot of the phase-field variable $\xi$. Cube domain set as $80 \times 80 \times 80 \left[\mu m^3\right]$.}
\label{fig:MultipleSeed_evolut}
\end{figure}

The lithium dendrite nucleation depends on local inhomogeneities that may arise from different causes, such as defects and impurities in the metal anode, imperfect contact between the electrode and electrolyte caused by the development of a solid-electrolyte interface, and variations in the local concentration or temperature in the electrolyte~\cite{ doi:10.1021/acsenergylett.8b01009}. Given the random nature of the nucleation phenomenon, we need to deal with some degree of randomness and uncertainty when defining the artificial nuclei in the simulation. We study the simulations' sensitivity to the artificial nuclei size, shape, and proximity following 2D studies that show this dependance~\cite{ CHEN2015376, ELY2014581, YURKIV2018609}. Thus, the following 3D numerical experiment tests the simulation's sensitivity to the nuclei distribution and proximity by comparing four-nuclei morphological results with those previously obtained with a single artificial protrusion. The simulation setup is similar to the previous 3D experiment with a different nucleation arrangement. Therefore, we form  the artificial nucleation region with four protrusions (ellipsoidal seeds), with semi-axes $4\left[\mu m\right]\times2\left[\mu m\right]\times2\left[\mu m\right]$, and centres located at $\left(y,x,z\right)=\left(0,38,38\right)$, $\left(0,42,38\right)$, $\left(0,38,42\right)$ and $\left(0,42,42\right)$ .

\begin{figure}
\begin{subfigure}[b]{0.35\linewidth}
    \centering%
{\includegraphics[height = 4.5cm]{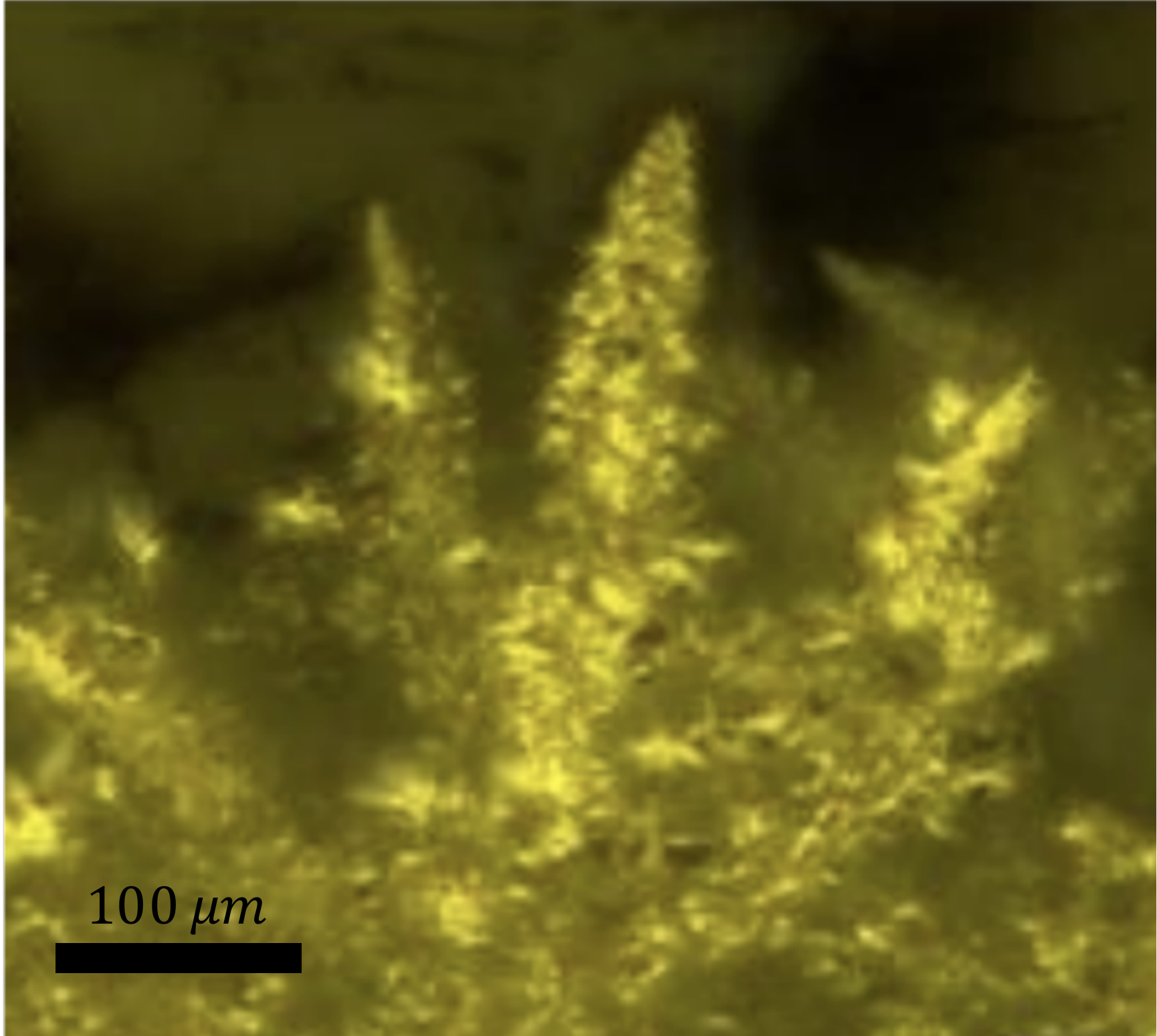}}
\caption{$10 \left[mA/cm^2\right]$.}
\label{fig:Experiment_1}
\end{subfigure}
\begin{subfigure}[b]{0.35\linewidth}
    \centering%
{\includegraphics[height = 4.5cm]{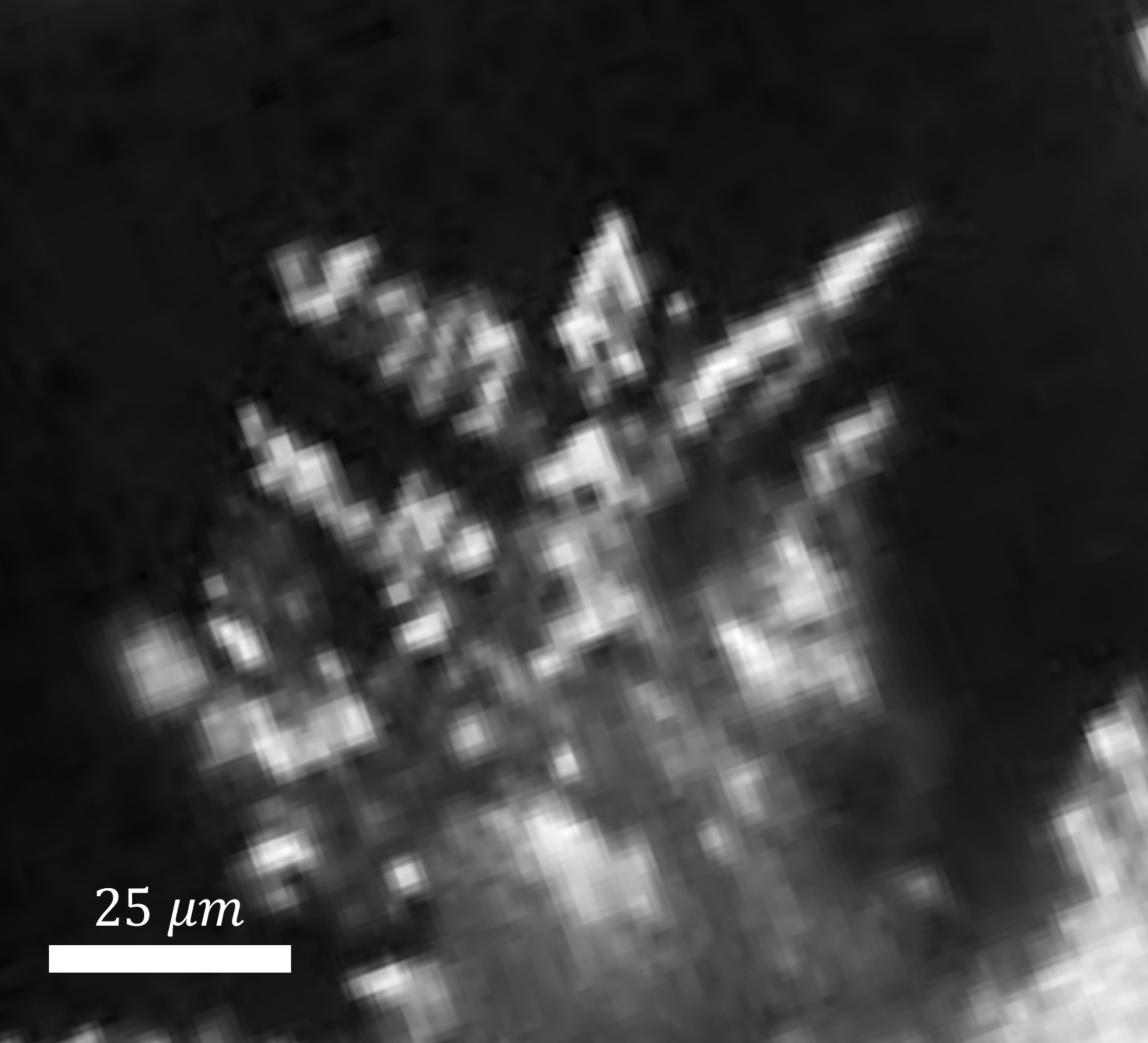}}
\caption{$3 \left[mA/cm^2\right]$.}
\label{fig:Experiment_2}
\end{subfigure}
\caption{Photographs of lithium electrodes with electrochemical deposition of lithium. The deposition condition for (\subref{fig:Experiment_1}), as performed by Ding~\cite{ ding2016situ}, consist of $10 \left[mA/cm^2\right]$ current density applied for 1 hour in 1M LiTFS/DME/DOL electrolyte, with working distance between electrodes set about $2 \left[mm\right]$. The deposition condition for (\subref{fig:Experiment_2}), as performed by Tatsuma~\cite{ TATSUMA20011201}, consist of $3 \left[mA/cm^2\right]$ current density applied for 1 hour in 1M LiCl$\text{O}_4$ electrolyte, with working distance between electrodes set about $3 \left[mm\right]$ (reproduced with Journal's permission).}
\label{fig:Exp_SpikeLikeLiDendrite}
\end{figure}

Figure~\ref{fig:MultipleSeed_evolut} shows the morphological evolution of the simulated lithium dendrite (isosurface plot of the phase-field variable $\xi$). In agreement with previous numerical experiments, the simulation forms a spike-like, symmetric, and highly branched pattern. The dendrite morphology consists of four main trunks growing from each nucleus, with pairs of orthogonal branches developing to the sides. 

The dendrite growth does not occur perpendicular to the stack but at an angle of about $20^\circ$ between main trunks. The stacks seem to repel each other, showing morphological similarity with dendritic patterns observed in lithium experiments performed by Ding~\cite{ ding2016situ}, see Figure~\ref{fig:Exp_SpikeLikeLiDendrite}. As a consequence of the reduced distance between electrodes (limited by the computational cost), the time scales we simulate are significantly shorter ($\sim$ two orders of magnitude faster) than the experimentatl ones~\cite{ NISHIKAWA201184, ding2016situ, TATSUMA20011201}.

\begin{figure}
\begin{subfigure}[b]{0.42\linewidth}
    \centering%
{\includegraphics[height = 5cm]{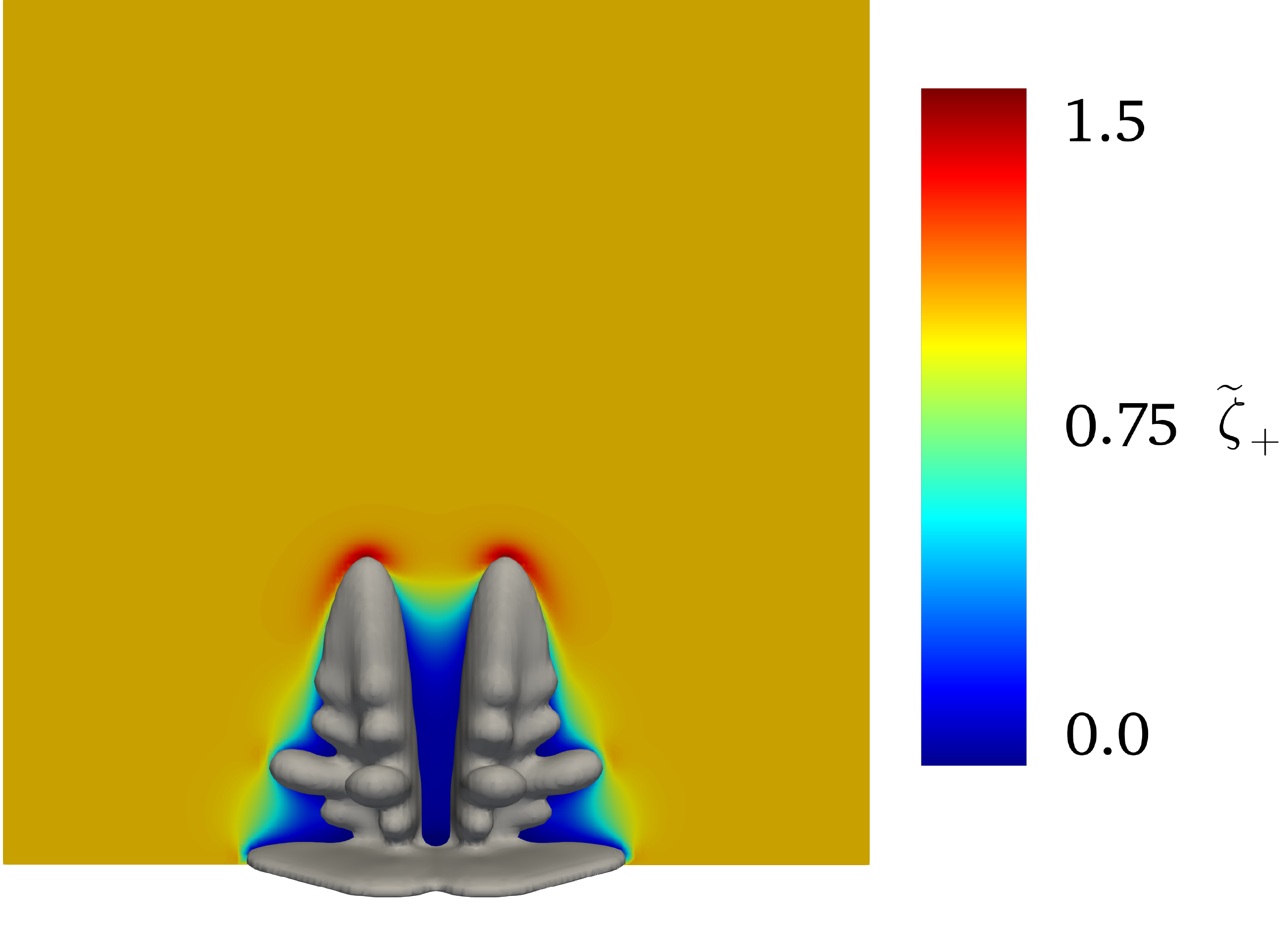}}
\caption{$\widetilde{\zeta}_{+}$ at $t = 0.5\left[s\right]$	.}
\label{fig:C_Dist_MultSeed3D}
\end{subfigure}
\begin{subfigure}[b]{0.42\linewidth}
    \centering%
{\includegraphics[height = 5cm]{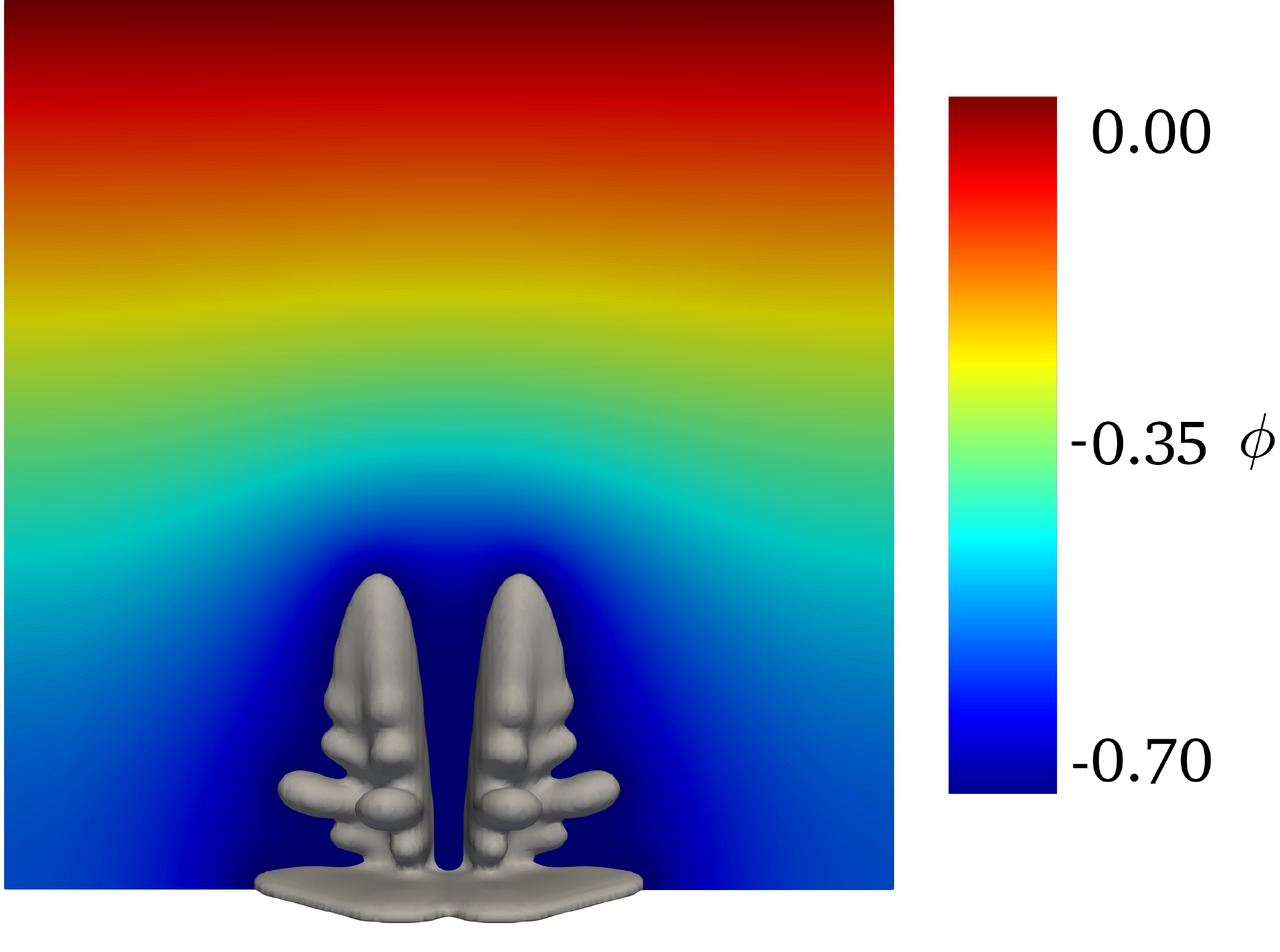}}
\caption{$\phi$ at $t = 0.5\left[s\right]$.}
\label{fig:Phi_Dist_3D_MultSeed3D}
\end{subfigure}
\caption{ Spatial distribution overlay of lithium-ion concentration (\subref{fig:C_Dist_MultSeed3D}), and electric potential (\subref{fig:Phi_Dist_3D_MultSeed3D}), with dendrite morphology at $t = 0.5\left[s\right]$. Contour plane set at $y=35\left[\mu m\right]$.}
\label{fig:C_Phi_3D_MultSeed_Dist}
\end{figure}

\begin{figure}
\begin{subfigure}[b]{0.32\linewidth}
    \centering%
{\includegraphics[height = 4.9cm]{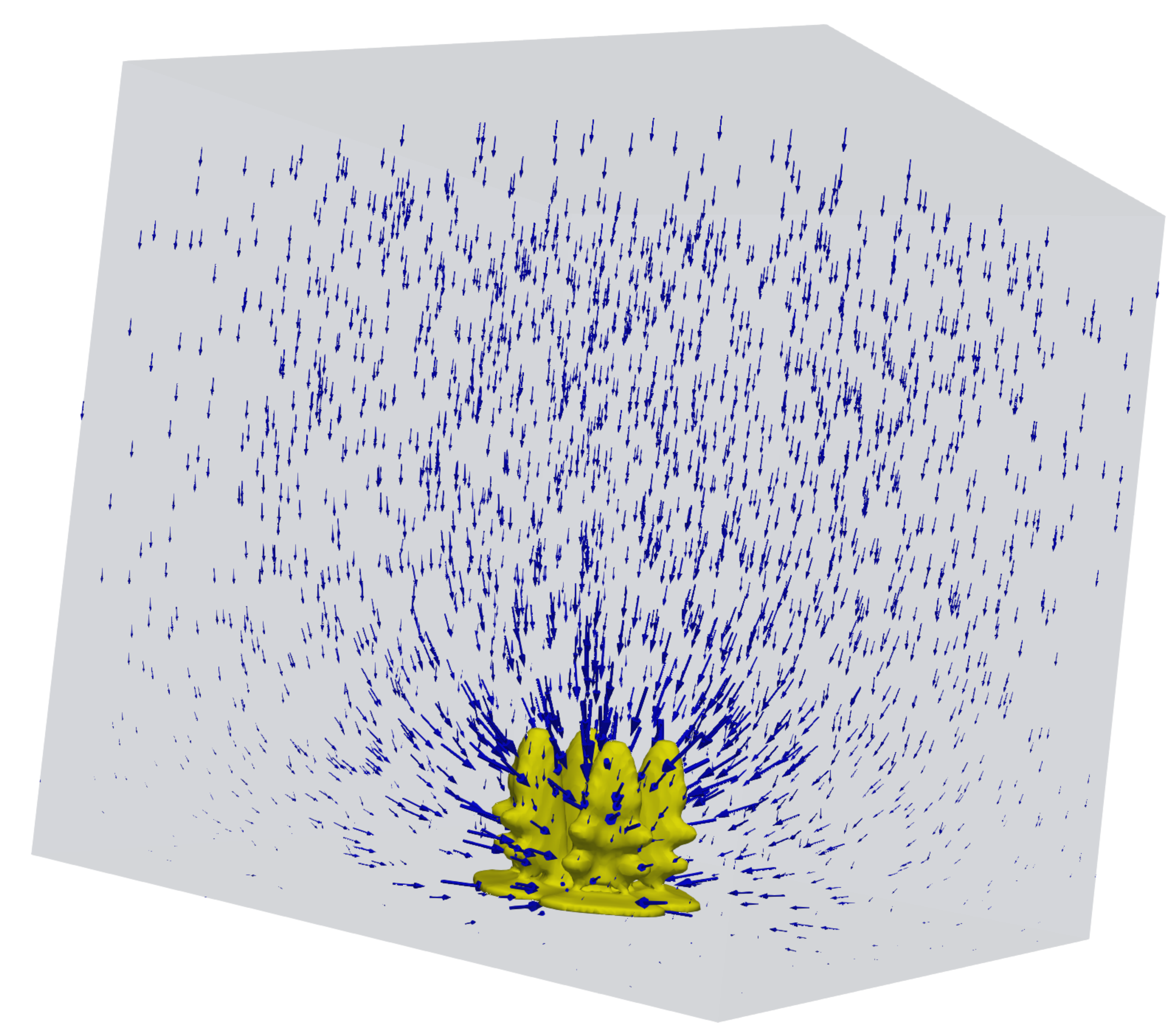}}
\caption{$t = 0.25\left[s\right]$.}
\end{subfigure}
\begin{subfigure}[b]{0.32\linewidth}
    \centering%
{\includegraphics[height = 4.9cm]{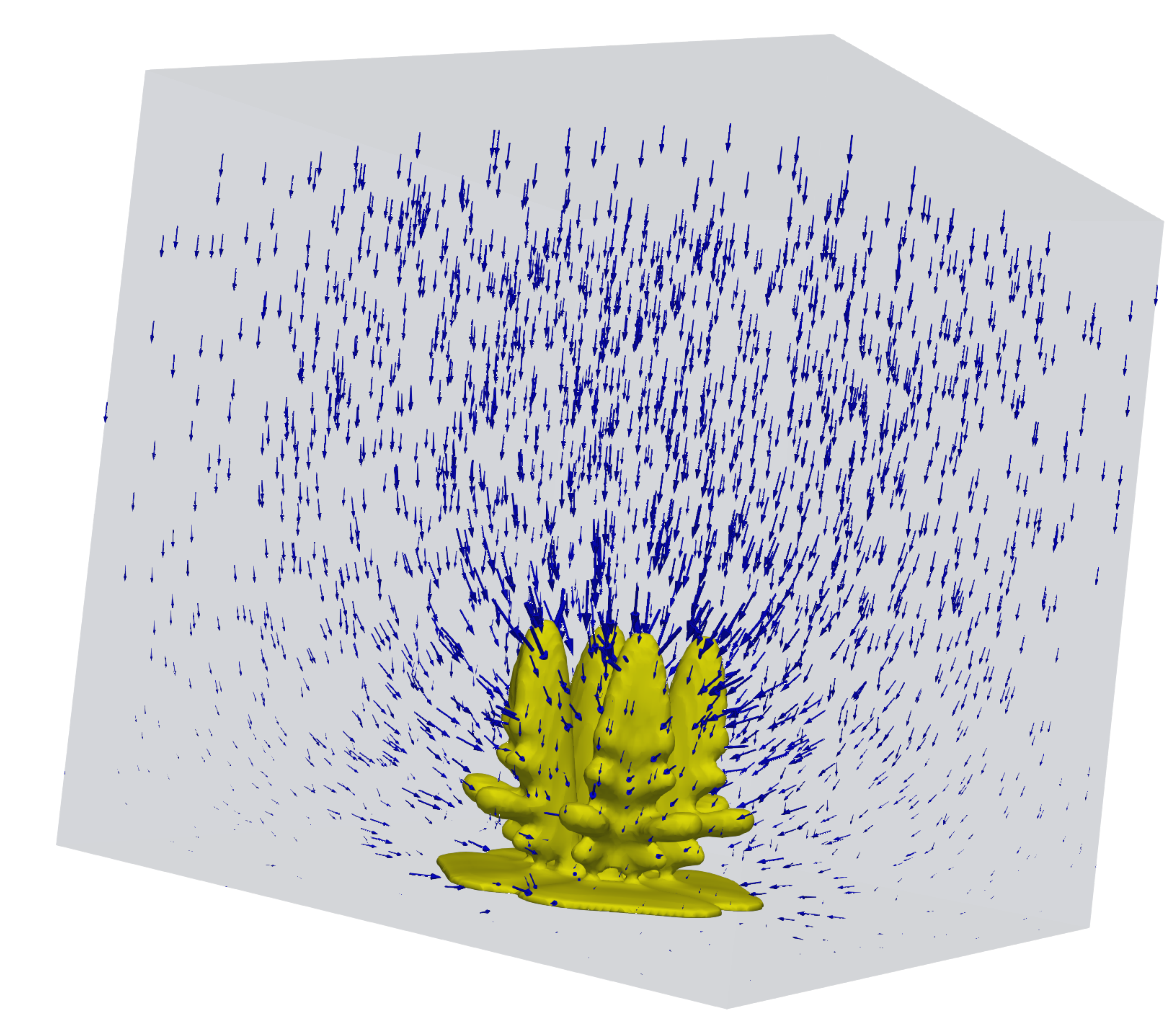}}
\caption{$t = 0.50\left[s\right]$.}
\end{subfigure}
\begin{subfigure}[b]{0.32\linewidth}
    \centering%
    {\includegraphics[height = 4.9cm]{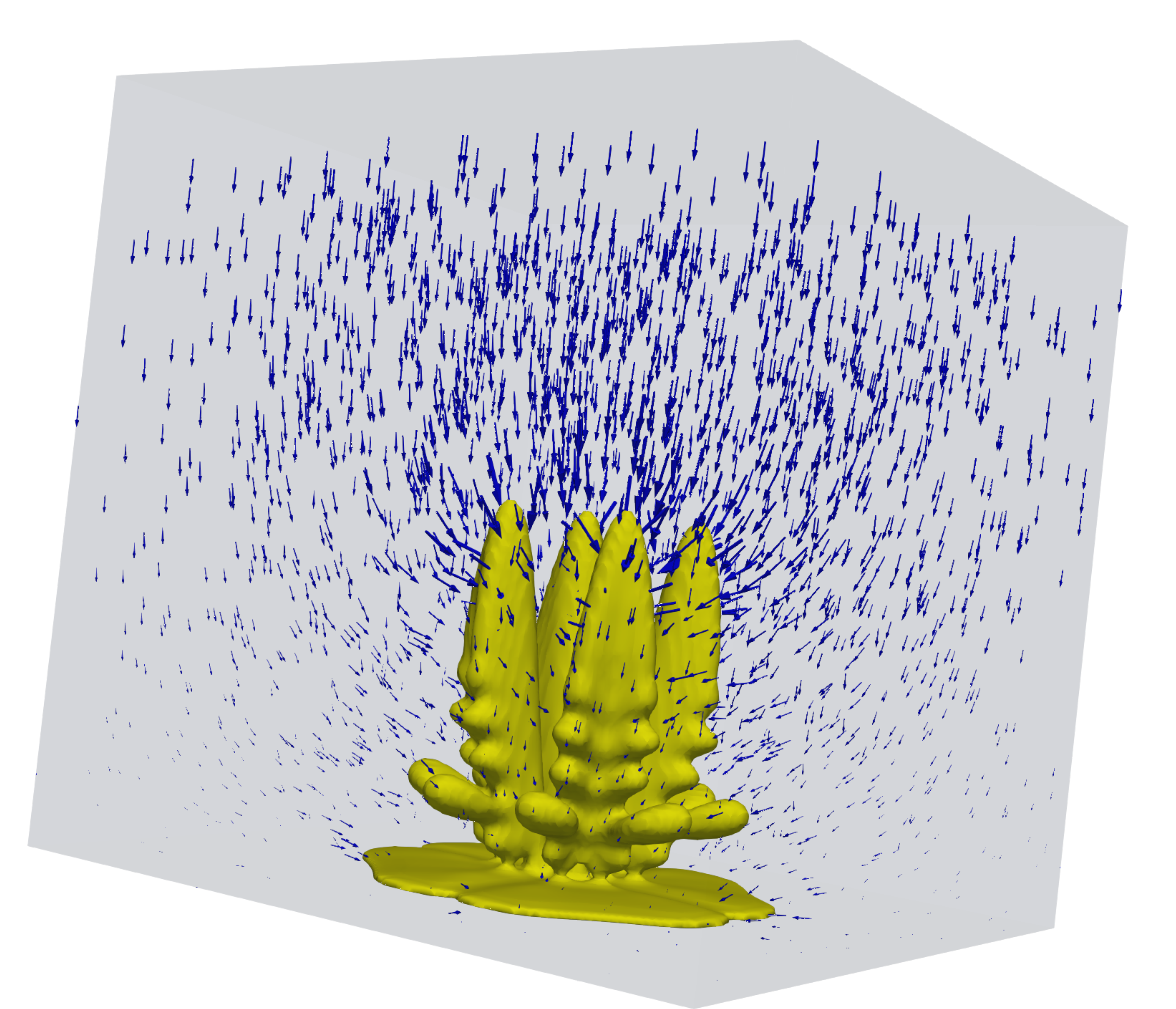}}
    \caption{$t = 0.75\left[s\right]$.}
\end{subfigure}
\bigbreak
\begin{subfigure}[b]{0.32\linewidth}
    \centering%
{\includegraphics[height = 4.9cm]{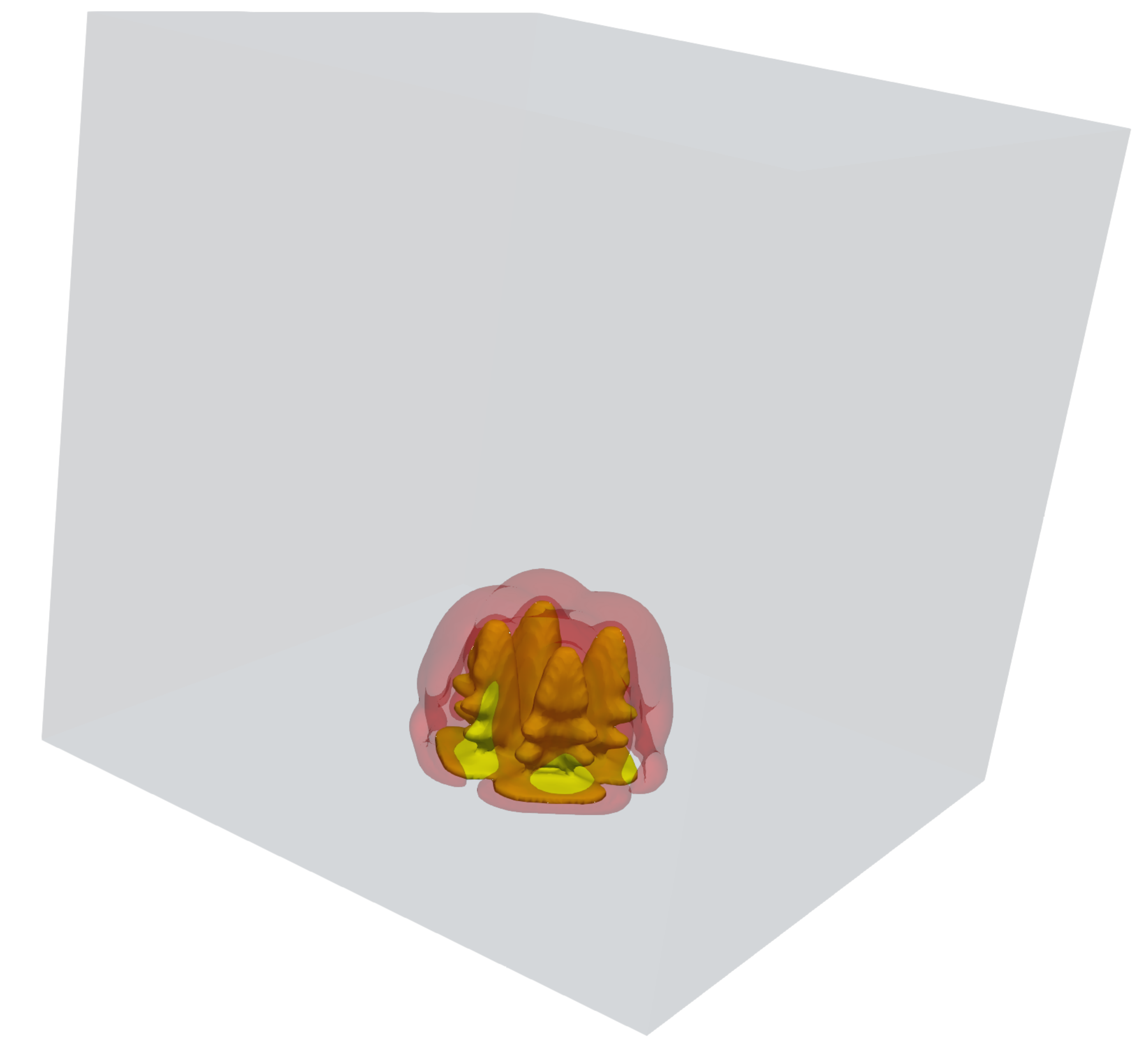}}
\caption{$t = 0.25\left[s\right]$.}
\end{subfigure}
\begin{subfigure}[b]{0.32\linewidth}
    \centering%
{\includegraphics[height = 4.9cm]{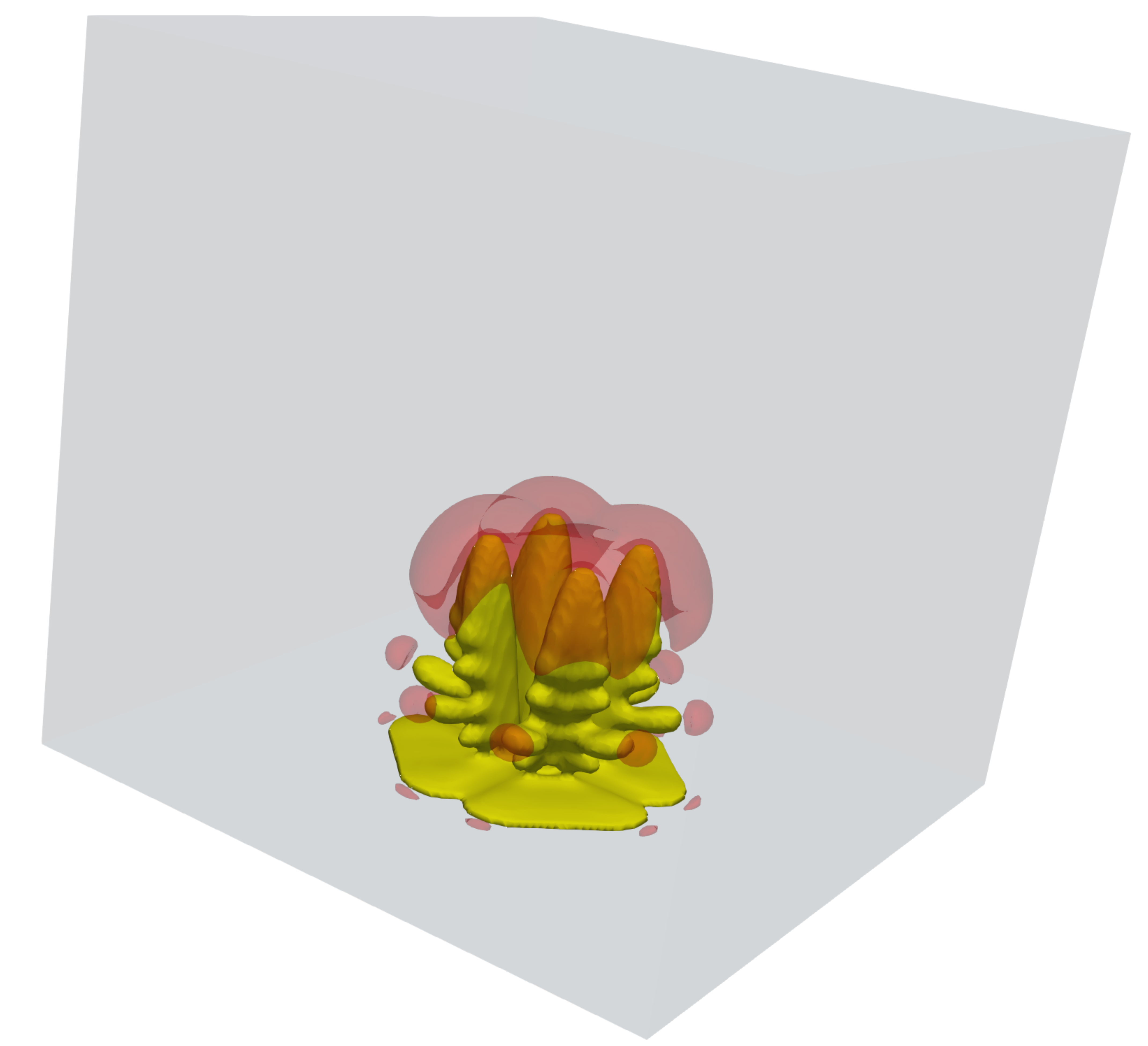}}
\caption{$t = 0.50\left[s\right]$.}
\end{subfigure}
\begin{subfigure}[b]{0.32\linewidth}
    \centering%
    {\includegraphics[height = 4.9cm]{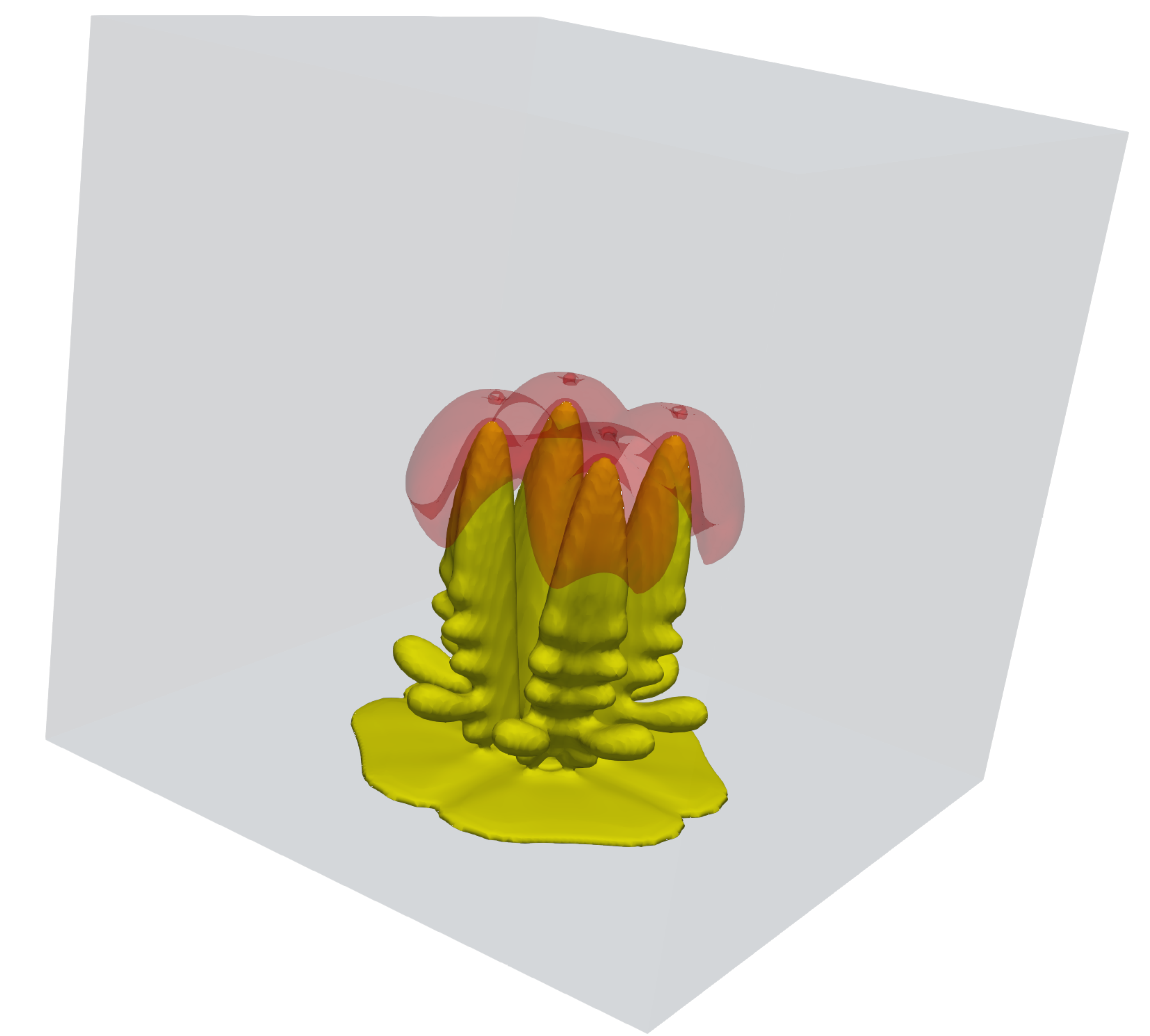}}
    \caption{$t = 0.75\left[s\right]$.}
\end{subfigure}
\caption{Simulation of spike-like lithium dendrite formation from multiple nucleation sites. The upper row shows the overlay of the electric field evolution $\vec{E}=-\nabla\phi$ (blue vectors) with dendrite morphology; the second row exhibit the lithium ion enrichment effect $\widetilde{\zeta}_{+}>1$ (red volume) surrounding the dendrite tips.}
\label{fig:E_Li-ion_MultiSeed_evolut}
\end{figure}

Figure~\ref{fig:C_Dist_MultSeed3D} details the simulated dendritic morphology at time $t = 0.5\left[s\right]$, together with the spatial distribution of the lithium-ion concentration $\widetilde{\zeta}_{+}$ in the electrolyte region. No side branches form facing the center of the nucleation arrangement (between the dendrites), where the deposition process depletes the lithium-ion concentration (shown in blue). Figure~\ref{fig:C_Dist_MultSeed3D} also shows how dendrite tips with higher lithium-ion concentration consume lithium ions from the dendrite valley~\cite{ doi:10.1063/1.4905341}, creating a shadow that inhibits branching growth in the spatial proximity of more developed adjacent dendrites. As a consequence, the side branches do not grow dendrites facing the center of the nucleation arrangement (unfavorably oriented dendrites) due to the 3D interactions with other dendrites (adjacent to the main trunks)~\cite{ TAKAKI201321}. Therefore, 2D phase-field models of dendrite electrodeposition cannot capture this 3D phenomenon.

Figure~\ref{fig:Phi_Dist_3D_MultSeed3D} shows the spatial distribution of the electric potential $\phi$ overlaid with the lithium dendrite at the same instant $t = 0.5\left[s\right]$. In agreement with the 2D simulations of Section~\ref{sc:2Dsim}, the electric potential has a steep gradient over the electrolyte phase and its steepness increases as the lithium dendrite grows.

Figure~\ref{fig:E_Li-ion_MultiSeed_evolut} shows the evolution of the electric field $\vec{E}=-\nabla\phi$ and the enriched lithium-ion concentration ($\widetilde{\zeta}_{+}>1$) in the vicinity of the dendrite tips (red volumes represent enriched concentration). As before, a larger electric field localizes in the vicinity of the dendrite tips, leading to a higher lithium-ion concentration with peak values of $\widetilde{\zeta}_{+}=1.8$. The electric migration forces overcome the concentration diffusion gradient and drive lithium cations from surrounding regions with lower concentrations (i.e., Li-ion depletion of valley regions) to accumulate around dendrite tips, triggering tip-growth with highly branched dendritic lithium.

\begin{figure}
\begin{subfigure}[b]{0.47\linewidth}
    \centering%
{\includegraphics[height = 5cm]{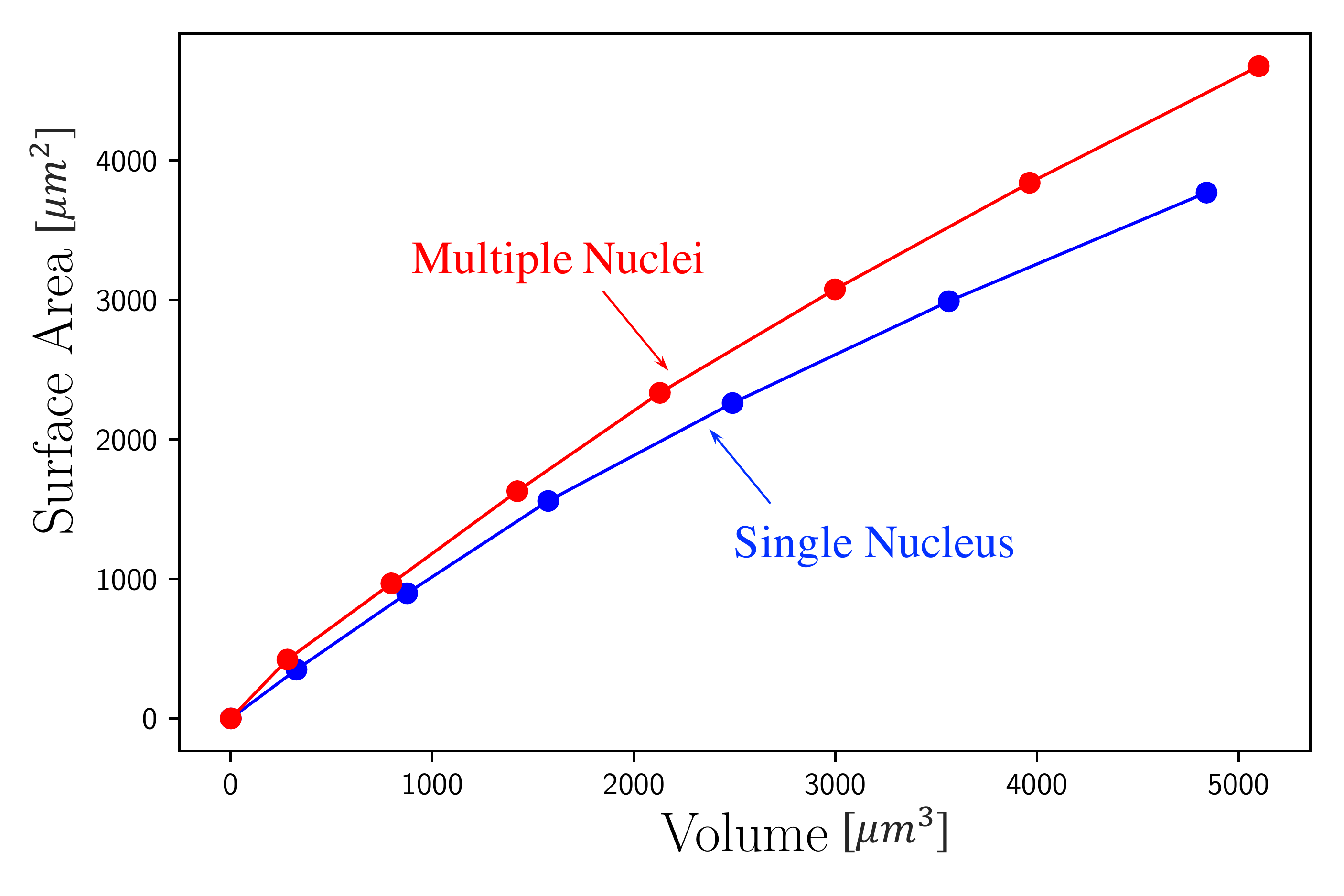}}
\caption{Volume vs Surface Area.}
\label{fig:VolvsSurf_3DSeed}
\end{subfigure}
\begin{subfigure}[b]{0.42\linewidth}
    \centering%
{\includegraphics[height = 5cm]{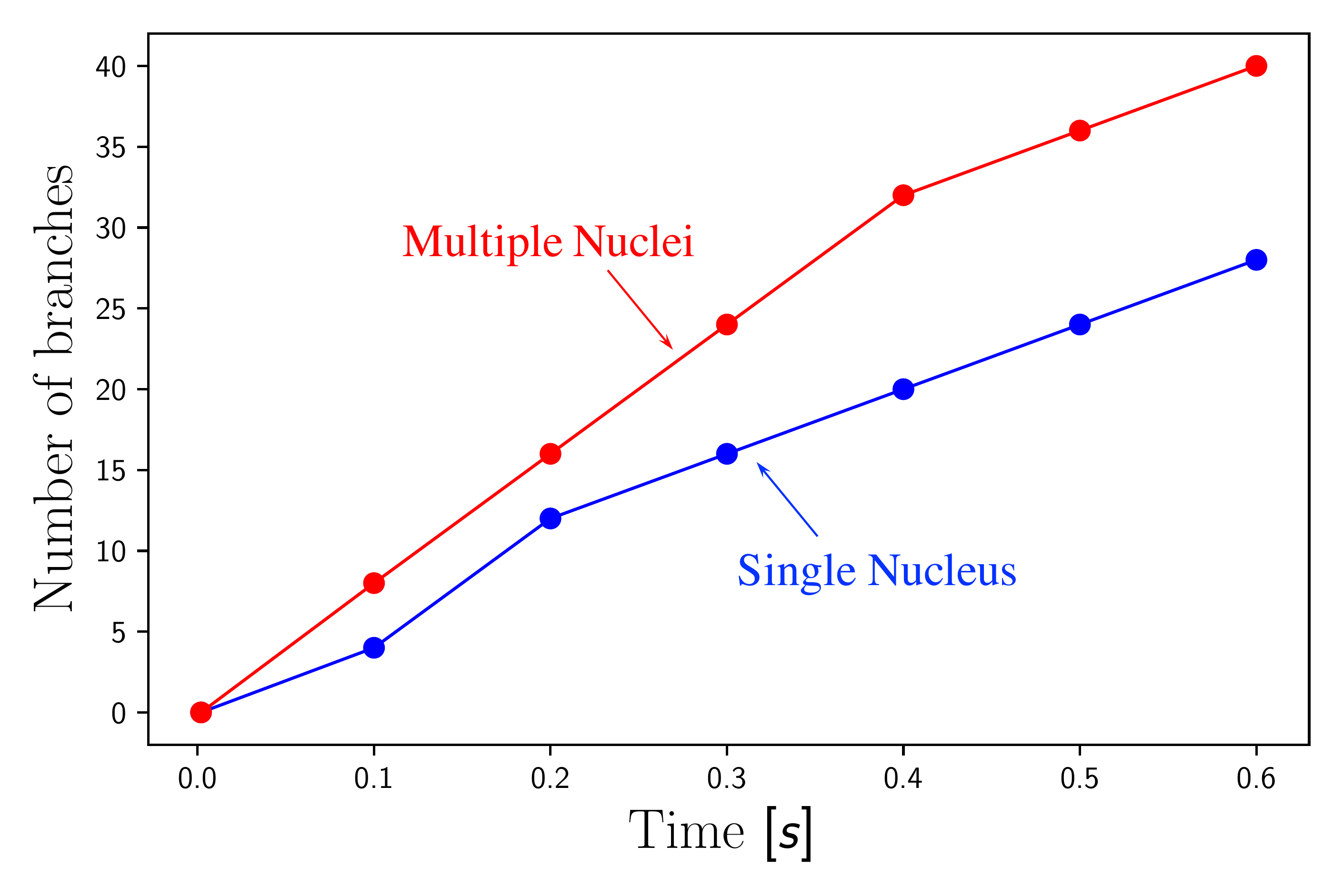}}
\caption{Side branches vs t.}
\label{fig:BranchesvsTime_3DSeed}
\end{subfigure}
\caption{Morphological comparison between 3D simulations of lithium dendrite growth (single and multiple-nuclei), in terms of the evolution of volume vs surface area ratio \subref{fig:VolvsSurf_3DSeed}, and number of side branches developed over time \subref{fig:BranchesvsTime_3DSeed}.}
\label{fig:3D_Morpholoy_Compare}
\end{figure}

We analyze the dendritic patterns' morphology evolution using single and multiple nuclei. Following Yufit et al.~\cite{ YUFIT2019485}, we characterize the morphology by tracking the dendrites' volume-specific area ratio ($\mu m^2 / \mu m^3$) and the branch number evolution in time. Figure~\ref{fig:VolvsSurf_3DSeed} plots the growth of the volume versus the surface area for the 3D lithium patterns we simulate. The volume-specific area average ratios of 0.83 and 0.91 $\left[\mu m^2 / \mu m^3\right]$ of the single and multiple nuclei simulations, respectively; where a higher $\text{surface area/volume}$ ratio is indicative of a more branched shape. Given the lack of experimental data available in the literature for quantitative characterization of the spike-like lithium morphologies, we rely on experimental results available for zinc dendrites. Yufit et al.~\cite{ YUFIT2019485} report values between 0.86 and 1.04 $\left[\mu m^2 / \mu m^3\right]$ for experimental formation of dendrites in zinc batteries. These values are slightly higher ratios than the ones we obtain for lithium. Nevertheless, these values are reasonable since zinc has a hexagonal crystal structure and tends to grow more branching (fractal) dendrites than lithium, which has a cubic crystal structure (less inherent anisotropy)and, thus, produces needle (spiky) dendrites~\cite{ PhysRevE.92.011301}. Figure~\ref{fig:BranchesvsTime_3DSeed} compares the number of branches developed over time in each case. The simulatilons produce ratios of 48 and 72 branches per second [$1/\left[s\right]$] for the single and multiple nuclei simulations, respectively. We compute the number of branches by visual inspection of the simulated morphologies, where we consider new protuberances as incipient branches under development. Direct comparison with dendrite experiments was impractical in this case, due to the faster dynamics of the simulation (small distance between electrodes).

\begin{figure}
    \centering%
{\includegraphics[height = 6.5cm]{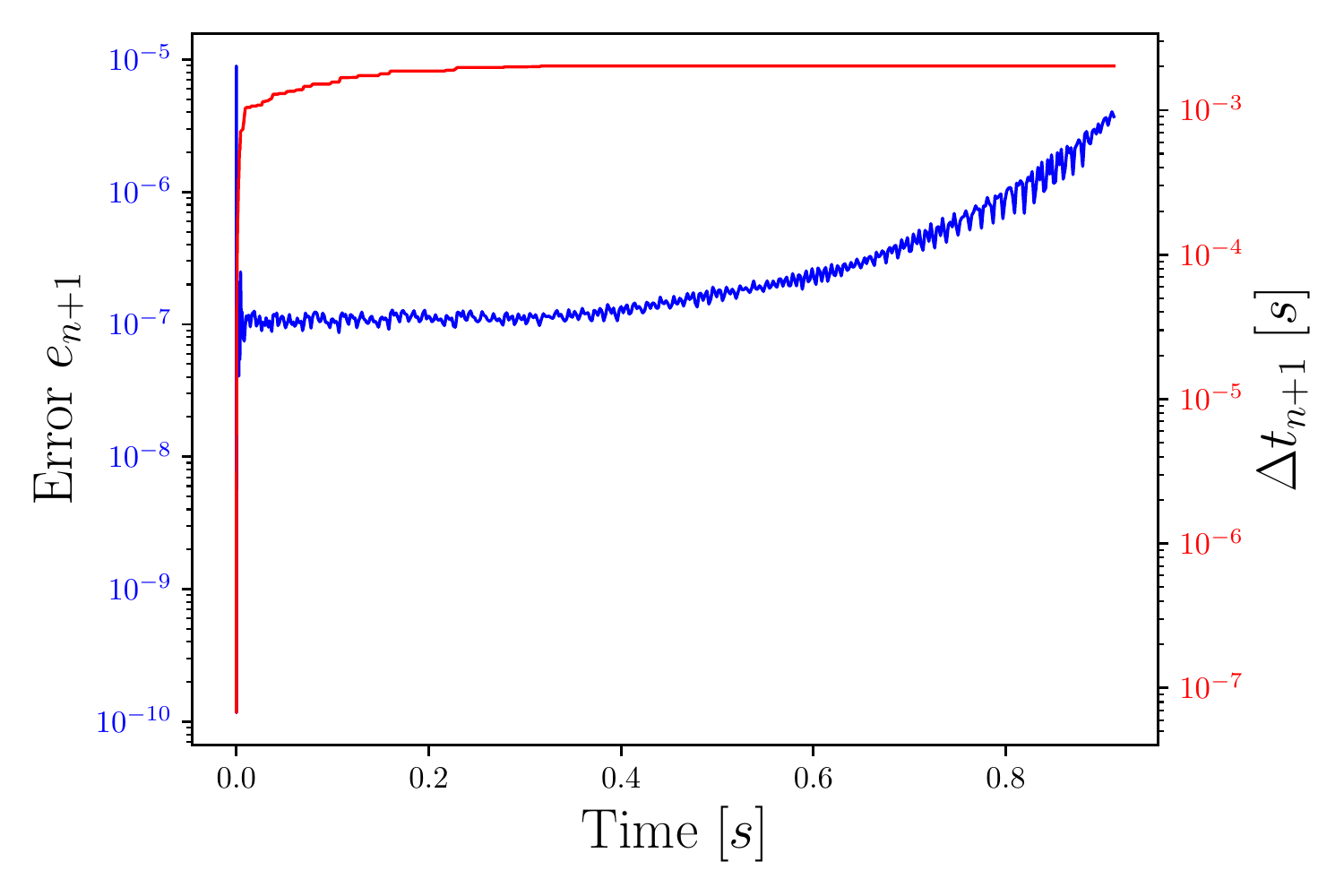}}
\caption{Time adaptivity plot for 3D spike-like four-nuclei dendrite growth simulation.}
\label{fig:DeltaTvsT_MultipleSeed3D}
\end{figure}

Figure~\ref{fig:DeltaTvsT_MultipleSeed3D} shows the performance of the time-adaptive scheme, throughout the $0.8\left[s\right]$ of the simulation. Starting with a small time-step of $\Delta t_0 = 10^{-7}\left[s\right]$ to initially achieve convergence, followed by a rapid increase in size, until reaching a stationary value of about $\Delta t_{n+1}=10^{-3}\left[s\right]$. The weighted truncation error $e_{n+1}$ (blue) stays close to the minimum tolerance limit ($10^{-7}$) during $0.4\left[s\right]$ of the simulation, after which the estimated error grows due to the acceleration of lithium dendrite propagation rate as it approaches the positive electrode. The time-step size remains unchanged since the error estimate remains within the tolerances

\begin{figure}
    \centering%
{\includegraphics[height = 6.5cm]{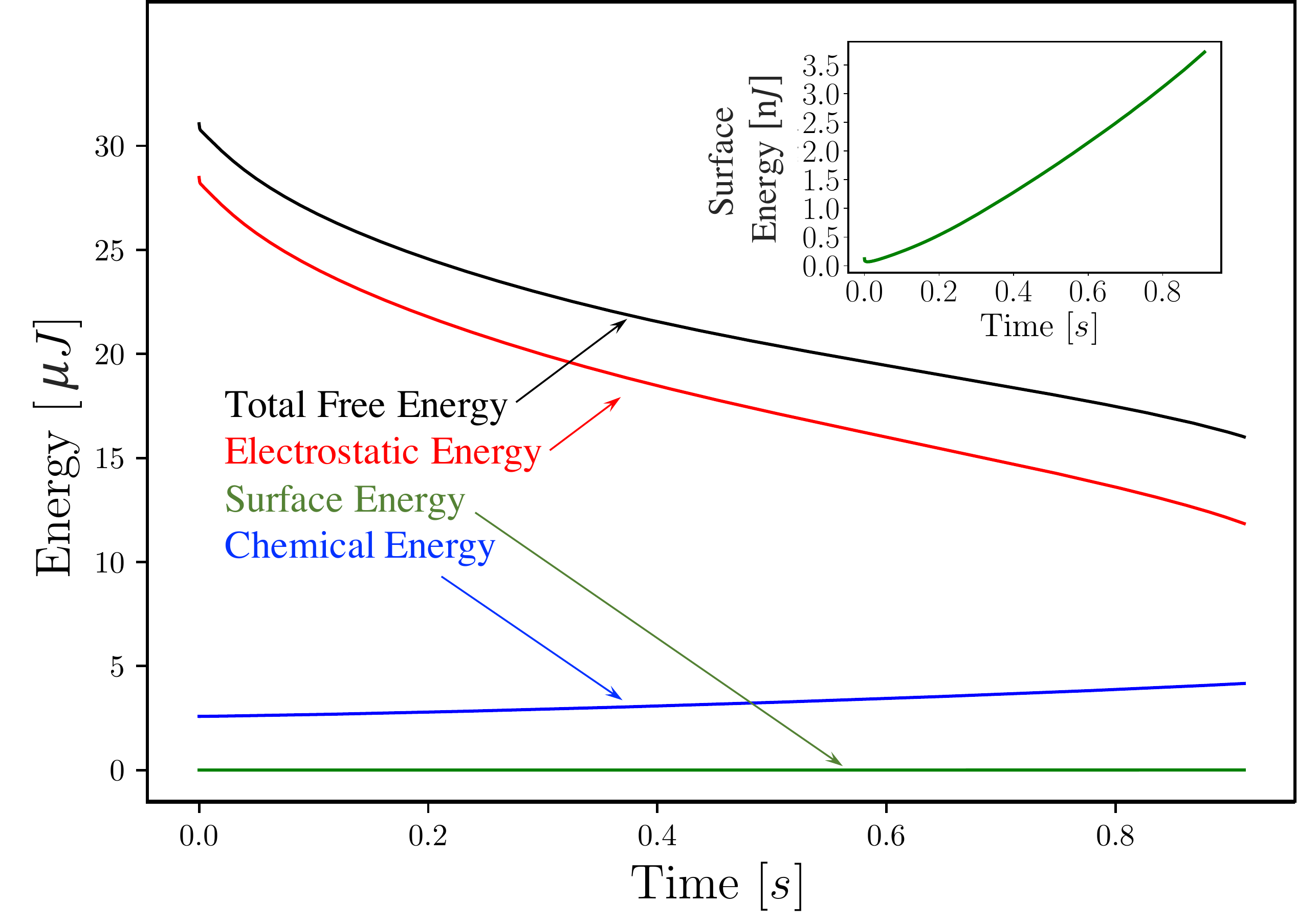}}
\caption{Energy time series for 3D spike-like four-nuclei dendrite growth simulation. The inset plots the increasing surface energy in smaller scale for better appreciation.}
\label{fig:EnergyvsTime_MultipleSeed3D}
\end{figure}

Figure~\ref{fig:EnergyvsTime_MultipleSeed3D} shows the systems' energy evolution.  The total energy curve (black) does not increase with time, showing energy stable results. As before, the system's chemical (blue) and surface energies (green) increase with time as the lithium surface area increases, while the electrostatic energy (red) decreases. This energy reduction happens as the process transforms electrostatic energy and stores it as electrochemical energy (battery charge). Figure~\ref{fig:EnergyvsTime_MultipleSeed3D} shows that the surface energy for the four-nuclei simulation is almost four times larger than the single nucleus surface energy of Figure~\ref{fig:EnergyvsTime_SingleSeed3D}. The proportionately larger surface area due to the four seeds explains this scaling.

\section{Conclusions}
\label{section:concl}

Using a phase-field model, we describe the dendrite formation process during metal (lithium) electrodeposition. We present the weak variational form of the coupled equations describing electrochemical interactions during the battery charge cycle and its finite element implementation using an open-source library. We use a time-adaptive strategy and detail its implementation and parametrization. The numerical experiments demonstrate numerically that the adaptive strategy produces energy stable results when we use a minimum tolerance for the time-adaptive scheme of $\text{tol}_{\text{min}}=10^{-7}$. We reduce the computational cost of simulating the detailed lithium electrodeposition at the scale of the whole cell (distance between electrodes of about $l_x=5000\left[\mu m\right]$) by using time-step size adaptivity, mesh mapping to concentrate the mesh in the region of interest, parallel computation, and a careful selection of the phase-field interface thickness ($\delta_{PF}$) considering the applied electro potential value of $\phi_b=-0.45\left[V\right]$.

We compare our 2D simulation results (lithium dendrite propagation rate) with experimental measurements by Nishikawa et al. \cite{NISHIKAWA201184} at an applied current density of $i=10\left[mA/cm^2\right]$. The growth rates we predict ($\sim 0.04\left[\mu m s^{-1}\right]$) are within the range of experimental results ($0.01-0.06\left[\mu m s^{-1}\right]$ \cite{NISHIKAWA201184}). Further, we compare the two-dimensional behavior of our model with previous phase-field simulations from the literature. The 2D evolution of the phase-field order parameter $\xi$ under $\phi_b=-0.70\left[V\right]$ applied voltage depicts the dynamic growth of “bush-like” lithium dendrites. The spatial distribution of the lithium-ion concentration $\widetilde{\zeta}_{+}$, and electric potential $\phi$, follow the same trend reported in~\cite{ doi:10.1063/1.4905341, YURKIV2018609, CHEN2021229203}. 

We also perform 3D simulations of lithium dendrite formation to explain 3D highly branched "spike-like" dendritic patterns formed under high current density status (fast battery charge). We study the 3D morphological evolution of dendritic lithium under $\phi_b=-0.70\left[V\right]$ applied voltage. We mimic the cubic crystal structure and surface anisotropy of lithium by using a 3D four-fold anisotropy model~\eqref{eq:KAPPA3D} of William et al.~\cite{ GEORGE2002264} to simulate crystal growth. 

We analyze the sensitivity of the model to initial conditions by comparing results using different setups (single nucleus versus a four-nuclei arrangement) for accounting for the random nature of the nucleation process. We obtain spike-like, symmetric, and highly-branched patterns in both cases. We obtain volume to specific-area average ratios of 0.83 and 0.91 $\left[\mu m^2 / \mu m^3\right]$  for the single and multiple nuclei simulations, respectively; where a higher $\text{surface area/volume}$ ratio indicates a more branched dendritic shape.

We study the 3D distribution of the electric field and lithium-ion concentration to better understand the mechanism behind spike-like dendrite growth. In both simulations,  the electric field increases in the vicinity of the dendrite tips, increasing the lithium-ion concentration that peaks at $\widetilde{\zeta}_{+}=2.1$ for a single nucleus and $1.8$ for multiple nuclei. Thus, electric migration forces overcome the concentration diffusion gradient, causing lithium cations to move from less concentrated surrounding regions (i.e., lithium-ion depletion of valley regions) and accumulate around dendrite tips, triggering tip-growing and highly branched dendritic lithium. In multiple-nuclei simulation we observe that the deposition process depletes the lithium-ion concentration at the center of the nucleation arrangement. Thus, 3D interactions with other dendrites creates a shadow that inhibits branching growth in the spatial proximity of more developed adjacent dendrites~\cite{ TAKAKI201321}. Therefore, 2D phase-field models of dendrite electrodeposition cannot capture this 3D phenomenon.

Finally, unlike 2D simulations, the 3D simulations performed have a smaller domain size due to the high computational cost. Consequently, the short separation between electrodes in the 3D model ($l_x = 80 \left[\mu m\right]$) induce faster deposition rates. Thus, future work may explore different meshing strategies to simulate larger 3D domains to achieve realistic simulation time scales.

In conclusion, we demonstrate the proposed phase-field modeling framework's effectiveness in performing 2D and 3D simulations of dendrite formation in lithium metal batteries with a reasonable computational cost. Beyond lithium electrodeposition, the framework has the potential to model other metal deposits in metal-anode batteries, such as zinc anode batteries. 

\section{Acknowledgments}
\label{section:acknow}

This work was supported by the Aberdeen-Curtin Alliance Scholarship. This publication was also made possible in part by the Professorial Chair in Computational Geoscience at Curtin University. This project has received funding from the European Union's Horizon 2020 research and innovation programme under the Marie Sklodowska-Curie grant agreement No 777778 (MATHROCKS). The Curtin Corrosion Centre and the Curtin Institute for Computation kindly provide ongoing support. 

\bibliographystyle{unsrt}
\bibliography{Reference}



\end{document}